\emergencystretch=\maxdimen
\documentclass[paper=a4, fontsize=12pt]{scrartcl} 

\usepackage[utf8]{inputenc} 
\usepackage[english]{babel}
\usepackage{mathtools,amsmath,amsfonts,amsthm,amssymb} 
\usepackage{bm} 
\usepackage{tikz-cd} 
\usetikzlibrary{decorations.markings} 
\usetikzlibrary{shapes} 
\usetikzlibrary{positioning, shapes.geometric} 
\usetikzlibrary{arrows, arrows.meta} 
\usetikzlibrary{decorations.pathreplacing,decorations.markings} 
\tikzset{   ->-/.style args={#1 #2 #3}{     decoration={       markings,         mark=         at position #1         with {\arrow{Stealth[#3,length=#2]}},     },       postaction=decorate, opacity=1.0   },     ->-/.default= 0.5 6pt black } 

\newcommand{\boundellipse}[3]
{(#1) ellipse (#2 and #3)
} 
\definecolor{ao}{rgb}{0.0, 0.5, 0.0} 
\usepackage[framemethod=tikz]{mdframed}

\newcommand{\coker}{\mathrm{coker}}
\newcommand{\im}{\mathrm{im}}

\definecolor{pur}{RGB}{186,146,162}
\definecolor{col}{RGB}{23,187,207}
\definecolor{gre}{RGB}{212,248,170}

\numberwithin{equation}{section} 
\usepackage{authblk} 
\usepackage{indentfirst} 
\usepackage{sectsty} 
\allsectionsfont{\centering \normalfont \scshape} 
\usepackage[margin=3cm]{geometry} 
\title{\LARGE Dynamical generalization of Yetter's model based on a crossed module of discrete groups}
\author{
Arkadiusz Bochniak\thanks{arkadiusz.bochniak@doctoral.uj.edu.pl} ,
Leszek Hadasz\thanks{leszek.hadasz@uj.edu.pl} \
and
B\l{}a\.zej Ruba\thanks{blazej.ruba@doctoral.uj.edu.pl, corresponding author}
}
\affil{\textit{Institute of Theoretical Physics, Jagiellonian University}}
\affil{\textit{prof. Łojasiewicza 11, 30-348 Kraków, Poland}}

\date{\today}

\begin{document}
\begin{titlepage}

\maketitle
\begin{abstract}
\centerline{\large\textbf{Abstract}}

\vskip 12pt

We construct a dynamical lattice model based on a crossed module of possibly non-abelian finite groups.
Its degrees of freedom are defined on links and plaquettes, while gauge transformations are based on vertices and links of the underlying lattice.
We specify the Hilbert space, define basic observables (including the Hamiltonian) and initiate a~discussion on the model's phase diagram.
The constructed model generalizes, and in appropriate limits reduces to, topological theories with symmetries described by groups and crossed modules,
lattice Yang-Mills theory and $2$-form electrodynamics. We conclude by reviewing classifying spaces of crossed modules, with an emphasis on the direct relation between their geometry and properties of gauge theories under consideration.
\end{abstract}

\thispagestyle{empty}
\end{titlepage}
\setcounter{page}{1}
\tableofcontents

\section{Introduction and summary}

One of the most fruitful ideas in the study of phase transitions is Landau's theory~\cite{landau}, which classifies phases of matter according to their symmetries. Despite this success, it is
currently known \cite{koesterlitz} that there exist transitions not driven by spontaneous symmetry breaking. In the case of gapped quantum systems, possibly with no symmetries, it has been proposed \cite{wen89} that phases may be distinguished by their topological orders, which were later interpreted in more physical terms \cite{wen10} as patterns of long range entanglement.

Topological aspects of many body quantum physics also turned out to play a~role in understanding the quantum Hall effect \cite{kanemele}, topological insulators \cite{fukanemele}, superconductors \cite{hansson} and other quantum phases of matter \cite{fradkin,chiu}. Several interesting applications arise in the study of geometry of Fermi surfaces \cite{wehling}. Topologically nontrivial observables are often robust against local perturbations, and hence have been suggested to possess potential to be used in fault-tolerant quantum computation \cite{kitaev2003,wang2010}.

A popular framework for description of topological order is that of Topological Quantum Field Theories (TQFTs) \cite{atiyah, witten}. Many well-known TQFTs are gauge theories, for example Chern-Simons \cite{witten}, $BF$ \cite{baezBF} or Dijkgraaf-Witten \cite{dw} theories. Several constructions, such as the Turaev-Viro \cite{turaev} or Crane-Yetter \cite{craneyetter} models, are based on quantum algebra, e.g. fusion categories. There is a~closely related line of study in which gapped lattice hamiltonian models are considered, such as in the Kitaev's quantum double model \cite{kitaev2003}, Levin-Wen string nets \cite{levinwen} or Walker-Wang model \cite{walker_wang}. One of advantages of this approach is that it encodes not only the space of ground states, but also its possible excitations. In this work we study a~generalized lattice gauge theory, which may be seen as a non-topological extension of the Yetter's $2$-type TQFT \cite{yetter}. It is shown that various TQFTs, as well as the lattice Yang-Mills theory with finite gauge group \cite{wegner, wilson, kogut} and $2$-form gauge theory may be obtained as limits of our model.

A crucial role in the analysis of physical systems is played by symmetries. It~has been suggested \cite{kapustin14,gksw} that so-called higher symmetries, which act on extended objects, also play a~significant role. An excellent example is provided by the center symmetry \cite{holland} in Yang-Mills theory (possibly with adjoint matter), which acts trivially on all local operators, but changes the value of Polyakov loops. Spontaneous breaking of this symmetry is responsible for a phase transition, which, however, is absent in QCD due to explicit breaking of the center symmetry.

Just as ordinary symmetries, higher symmetries may be used to derive selection rules on correlation functions. Moreover, they may be anomalous~\cite{benini}, which can be used to obtain theoretical constraints on the renormalization group flow. This is also related to the proposal \cite{kapustin17} of Symmetry Protected Topological (SPT) phases \cite{wen13} involving higher symmetries.

Higher symmetries may also be gauged, which leads to so-called higher gauge theories. One of the first models of this type, involving parallel transports over surfaces, was proposed in the context of string theory by Kalb and Ramond~\cite{kalbramond}. It~has been argued \cite{pform} that higher gauge theories are necessarily abelian, essentially because there is no meaningful notion of time ordering on objects of dimension higher than one. To some extent this conviction is defied by models inspired by higher category theory \cite{baez, pfeiffer, baez_huerta}. In this case parallel transports are indeed valued in an abelian group, but they are defined in terms of genuinely non-abelian degrees of freedom. Besides truly dynamical models, higher gauge fields appear also in TQFTs such as the Yetter's model \cite{yetter}, its generalizations \cite{porter,martins_porter,gpp08,kapustin17} and hamiltonian formulations \cite{bullivant17,delcamp,bullivant20}. Related models have also been proposed \cite{gukov,KT14,kapustin17} as effective descriptions of Yang-Mills theory vacua.

Conventional gauge theories depend on a choice of a gauge group. It has been proposed \cite{baez} that generalization of this notion suitable for theories with transports over surfaces \cite{mackaay,bs,sw} is a~$2$-group. There are several equivalent ways to define these objects \cite{baezlauda}. Here we choose to work with the formulation through crossed modules, whose definition will be recalled in the main text. We remark that these objects have also found applications in the classification of defects \cite{angprakash}, mathematically modeled as solitonic sectors of sigma models.

Models discussed in this paper allow non-abelian degrees of freedom associated to edges and faces of a spatial lattice. They are subject to a constraint called fake flatness. This enables to consistently define parallel transports over spheres, besides the more standard Wilson loops. As usually, there is a gauge freedom, which is however reduced with respect to that present in Yetter's model. This is necessary in order to preserve the dynamical (rather than purely topological) nature of the \mbox{$1$-form} gauge field, and hence to construct models generalizing the Yang-Mills theory. We~propose a suitable hamiltonian and discuss its symmetries and various special cases, including Yetter's theory. Ground states are described in several integrable limits, which allows to formulate initial conjectures concerning the phase diagram.

The organization of this paper is as follows. Section \ref{sec:basic} is introductory and sets the stage for subsequent developments. In~subsection \ref{sec:geo_setup} we review basic geometric notions used in the text, to some extent following \cite{bullivant20}. This allows to state precisely what is meant by field configurations valued in a crossed module. Interpretation of these fields is discussed in subsection \ref{sec:dofah}, where we define also the basic observables. Subsection \ref{sec:gt} is devoted to transformations of the configuration space, including a presentation of our motivation to restrict the group of gauge transformations. Examples in subsection \ref{sec:examples} illustrate several aspects of the subtle interplay between spatial topology and algebra of crossed modules, in which the fields are valued. In~section \ref{sec:ham} we complete the construction of our model and present first results about its dynamics. Then in subsection \ref{sec:construction} we specify the Hilbert space and define basic operators, including the hamiltonian. In order to make this more concrete, in~subsection \ref{sec:cubic_example} we carry out the construction explicitly in the case of a hypercubic lattice and a particular crossed module. Symmetries of proposed hamiltonians are discussed in subsection \ref{sec:sym}. Afterwards, in subsection~\ref{sec:vacua}, we~describe ground states of four integrable limits of our model and in each case relate it to some well-known TQFT. This is followed by subsection \ref{sec:dyn}, in which it is shown that in a certain region of the phase diagram, intermediate between TQFTs and the full model, one finds Yang-Mills theory or $2$-form gauge theory. Appendices \ref{sec:exact_seq} and \ref{sec:twisted} are devoted to a review of certain technical, albeit standard mathematical tools used in the main text. The more extensive appendix \ref{sec:classifying} is devoted to a discussion of classifying spaces of crossed modules. Relation of classifying spaces to gauge theories based on crossed modules is derived. This offers an interesting perspective on several properties of higher gauge theories. These results are not new, but we are not aware of an exposition which stresses relation with field theory. We hope this way of presentation will be helpful for some readers.

A natural next step would be to analyze the dynamics of proposed models in more detail, e.g. using perturbation theory. There exists a~natural candidate for a~state sum formulation of the model presented here, which could be studied using strong coupling expansion or Monte Carlo methods. Similar questions may also be asked about corresponding models with continuous spacetimes and crossed modules of Lie groups.

\section{Basic notions} \label{sec:basic}

\subsection{Geometric setup and field configurations}\label{sec:geo_setup}

Homotopy classes of (parametrized) paths in a topological space form a structure very similar to a group, since they can be composed in a way which is associative and admits multiplicative inverses. There is only one complication: composition $\gamma' \gamma$ exists only if the ``source'' of $\gamma'$ coincides with the ``target'' of $\gamma$. This is abstracted by the notion of a \textbf{grupoid}, whose definition we now recall. A grupoid consists of:
\begin{enumerate}
\item sets $G$ and $\mathrm{Ob}_G$, called the set of arrows and the set of objects, respectively,
\item functions $s,t : G \to \mathrm{Ob}_G$, called the source and the target map,
\item an associative binary operation on $G$, denoted by juxtaposition, with $\gamma' \gamma$ defined if $\gamma, \gamma' \in G$ are such that $s(\gamma')=t(\gamma)$.
\end{enumerate}
These data are subject to two axioms:
\begin{enumerate}
\item For every object $x$ there exists an arrow $\mathrm{id}_x$, with source and target $x$, such that $\gamma \, \mathrm{id}_x = \gamma$ and $\mathrm{id}_x \gamma = \gamma$ whenever these compositions are defined.
\item For every arrow $\gamma$ there exists an arrow $\gamma^{-1}$ with $s(\gamma^{-1})=t(\gamma)$, $t(\gamma^{-1})=s(\gamma)$, $\gamma^{-1}\gamma = \mathrm{id}_{s(\gamma)}$ and $\gamma \gamma^{-1} = \mathrm{id}_{t(\gamma)}$.
\end{enumerate}
In further discussion we will abuse the language by calling the set $G$ itself a~grupoid. 

We note that for any $x \in \mathrm{Ob}_G$ the set of all $\gamma \in G$ with $x=t(\gamma) = s(\gamma)$ is a~group. In particular, if $\mathrm{Ob}_G$ has exactly one element, then $G$ itself is~a~group.

If $B$ is a subspace of a topological space $A$, the \textbf{fundamental grupoid} $\pi_1(A;B)$ has $B$ as its set of objects and the set of homotopy classes of paths in $A$ with (fixed) endpoints in $B$ as the set of arrows. Source and target maps are obvious. A~composition $\gamma \gamma'$ is defined as $\gamma'$ followed by $\gamma$, which makes sense if $s(\gamma)=t(\gamma')$. We note that the fundamental group $\pi_1(A;b)$ of $A$ based at $b \in B$ may be described as $\{ \gamma \in \pi_1(A;B) \, | \, t(\gamma) = s (\gamma) = b \}$.

In our applications we shall consider connected spaces $X$ equipped with a lattice decomposition\footnote{For the most part it would be sufficient to consider smooth manifolds with a triangulation, though we prefer to allow also polytopes other than simplices in our decompositions. More generally, one may consider CW-complexes with cellular attaching maps for $2$-cells and $3$-cells. Here we have in mind the standard CW-decompositions of $S^1$ and $S^2$ with exactly two cells.}. In this situation we have a chain of inclusions
\begin{equation}
X_0 \subseteq X_1 \subseteq X_2 \subseteq ... \subseteq X_d = X,
\end{equation}
where $d$ is the dimension of $X$. Here $X_0$ is the set of vertices (also called lattice sites or $0$-cells), $X_1$ is constructed by gluing in edges (links or $1$-cells) to $X_0$, $X_2$ by gluing in faces (plaquettes or $2$-cells) to $X_1$ etc. $3$-cells will be referred to as balls.
We will make an extensive use of the groupoid $\pi_1(X_1;X_0)$. Its set of arrows may be described as the free grupoid generated by the edges of $X$. This means that any arrow is a product of some number of edges (identity arrows being understood as empty products), and that the only relations between two such products are those which follow from associativity of composition and identification of an edge $e$ with orientation reversed with the inverse of the edge $e$.

The above description of the fundamental grupoid is convenient for applications in field theory, since it is given in terms of local data. In some arguments another set of generators proves to be useful. Let us choose some $\ast \in X_0$. The fundamental group $\pi_1(X_1;\ast)$ is free \cite[p.~83]{hatcher}, i.e.\ there exists a set $L$ of generators satisfying no non-trivial relations, called a basis of loops. Secondly, we may choose a maximal tree $T$, i.e.\ a maximal set of edges with the property that there exists no non-trivial loop composed entirely of edges in $T$. Then $\pi_1(X_1;X_0)$ is freely generated by $L \cup T$.

\begin{figure}[ht]
\begin{minipage}[c]{0.4\textwidth}
\centering
\begin{tikzpicture}
\begin{scope}[very thick]
      \draw[decoration={markings, mark=at position 0.5 with {\arrow{stealth}}},postaction={decorate}] (0,2) coordinate (b1) -- node[left]{$e_2$} (0,0) coordinate (a1);
      \draw[decoration={markings, mark=at position 0.4 with {\arrow{stealth}}}, postaction={decorate},red] (1,1) coordinate (c1) -- node[right](q1){$e_1$} (0,2) coordinate (b1);
      \draw[decoration={markings, mark=at position 0.45 with {\arrow{stealth}}}, postaction={decorate},red] (0,0) coordinate (a1) -- node[right](q2){$e_3$} (1,1) coordinate (c1);
      \draw[decoration={markings, mark=at position 0.5 with {\arrow{stealth}}}, postaction={decorate},red] (1,1) coordinate (c1) -- node[below](q3){$e_0$} (3,1) coordinate (d1);
      \draw[decoration={markings, mark=at position 0.5 with {\arrow{stealth}}}, postaction={decorate},red] (3,1) coordinate (d1) -- node[below](q4){$e_4$} (4,0) coordinate (e1);
      \draw[decoration={markings, mark=at position 0.5 with {\arrow{stealth}}}, postaction={decorate}] (4,0) coordinate (e1) -- node[right](q5){$e_5$} (4,2) coordinate (f1);
      \draw[decoration={markings, mark=at position 0.5 with {\arrow{stealth}}}, postaction={decorate},red] (4,2) coordinate (f1) -- node[left](q6){$e_6$} (3,1) coordinate (d1);
      \node (c1) at (1,1) {{\color{red}$\bullet$}};
      \node[left] (c2) at (1,1) {$\bm{\ast}$};
      \draw[dashed, blue, decoration={markings, mark=at position 0.5 with {\arrow[scale=1.5]{stealth}}}, postaction={decorate}] \boundellipse{0.25,1}{0.9}{1.4};
      \draw[dashed, ao, decoration={markings, mark=at position 0.825 with {\arrow[scale=1.5]{stealth}}},,
        postaction={decorate}] \boundellipse{2.8,1}{2.4}{1.6};
      \node at (0,2.7) {{\color{blue}$l_1$}};
      \node at (5,1) {{\color{ao}$l_2$}};
\end{scope}
\end{tikzpicture}
\end{minipage}\hfill
 \begin{minipage}[c]{0.6\textwidth}
    \caption{
       Illustration of possible sets of generators of $\pi_1(X_1;X_0)$ for a certain space $X$. Edges are depictured by continuous lines, with a maximal tree distinguished by the red color. Two independent loops based at the point $\ast$ (the big dot) are indicated by dashed lines.
    } \label{fig:eight}
  \end{minipage}
\end{figure}
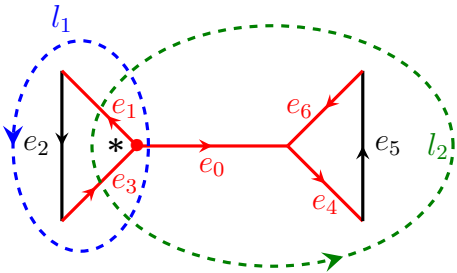

A simple example is in order. Consider the space illustrated on figure~\ref{fig:eight}. It~has seven edges $\{ e_i \}_{i=0}^6$. A basis of loops based at $\ast$ may be taken as $L = \{ l_1, l_2 \}$, where $l_1 = e_3e_2e_1$ and $l_2=e_0^{-1}e_6e_5e_4 e_0$. Set $T = \{ e_0, e_1, e_3, e_4, e_6 \}$ is a maximal tree. Grupoid $\pi_1(X_1;X_0)$ is generated by the loops $l_1, l_2$ and the edges in $T$. There are no non-trivial relations between these generators.

Now let $G,G'$ be grupoids. Map $F : G \to G'$ is called a homomorphism if:
\begin{enumerate}
\item there exists a map $F_0 : \mathrm{Ob}_G \to \mathrm{Ob}_{G'}$ such that $s \circ F = F_0 \circ s$ and $t \circ F = F_0 \circ t$,
\item $F(\gamma \gamma') = F(\gamma) F(\gamma')$ whenever $s(\gamma) = t(\gamma')$.
\end{enumerate}
We note that $F_0$ is uniquely determined by $F$ and that the first property guarantees that the second one makes sense. Secondly, if $G,G'$ are groups, $F$ is simply a~homomorphism of groups.

To give a concrete example: lattice gauge field on $X$ valued in a group $G$ may be defined as a homomorphism $\pi_1(X_1;X_0) \to G$. Since there are no relations between distinct edges, regarded as arrows of $\pi_1(X_1;X_0)$, defining a lattice gauge field amounts to specifying independently a group element $g_e \in G$ for every edge $e$. The element associated to a path $\gamma = e_n \ldots e_1$ is $g_{\gamma}=g_{e_n} \ldots g_{e_1}$. Alternatively, a~lattice gauge field may be specified by giving a homomorphism $\pi_1(X_1;\ast) \to G$ for some $\ast \in X_0$ and the values of $g_e$ for edges $e$ from any maximal tree $T$. These data can be chosen independently because there are no relations between generators of $\pi_1(X_1;\ast)$ and elements of $T$.
In order to capture two-dimensional aspects of geometry needed to formulate models considered in this work, we need to review another algebraic structure. A~\textbf{crossed module} of grupoids is a quadruple $(G,H, \rhd, \partial)$ consisting of:
\begin{enumerate}
\item grupoids $G,H$ with $\mathrm{Ob}_H = \mathrm{Ob}_G$ and $s(h)=t(h)$ for any $h \in H$,
\item homomorphism $\partial : H \to G$ with $\partial_0 : \mathrm{Ob}_H \to \mathrm{Ob}_G$ the identity map,
\item action $\rhd$ of $G$ on $H$: $g \rhd h \in H$ with $t(g \rhd h) = t(g)$ is defined for $g \in G$, $h \in G$ whenever $s(g) = t(h)$.
\end{enumerate}
These data are subject to the axioms:
\begin{enumerate}
\item $\mathrm{id}_{t(h)} \rhd h = h$ for any $h \in H$,
\item $(gg') \rhd h = g \rhd (g' \rhd h)$ whenever $s(g) = t(g')$ and $s(g')=t(h)$,
\item $g \rhd (hh') = (g \rhd h)(g \rhd h')$ whenever $s(g) = t(h)=t(h')$,
\item \textbf{1st Peiffer identity:} $\partial(g \rhd h) = g \partial(h) g^{-1}$ whenever $s(g) = t(h)$,
\item \textbf{2nd Peiffer identity:} $(\partial h) \rhd h' = hh'h^{-1}$ whenever $t(h)=t(h')$.
\end{enumerate}
Properties $1$--$3$ characterize the action $\rhd$, while Peiffer identites $4$ and $5$ are compatibility conditions between $\rhd$ and $\partial$. If $G$ has exactly one object, $(G,H,\partial, \rhd)$ is called a crossed module of groups. We proceed to motivate this lengthy definition by giving the example most important for our models.

\begin{figure}[ht]
\begin{minipage}[c]{0.5\textwidth}
\centering
\begin{tikzpicture}
      \draw[ultra thick]
       (0,0) coordinate (a1) -- node[left]     {$b$} (0,2) coordinate (d1)
       (2,0) coordinate (b1) -- node[right](q1){$b$} (2,2) coordinate (c1);
     \draw[thick] (d1) -- node[above]    {$B$}(c1);
     \draw[ultra thick] (a1) -- node[below](f1){$b$}(b1);
\end{tikzpicture}
\end{minipage}\hfill
 \begin{minipage}[c]{0.5\textwidth}
    \caption{
       Illustration of conditions satisfied by maps representing elements of the second relative homotopy group $\pi_2(A,B;b)$.
    } \label{fig:pi2_def}
  \end{minipage}
\end{figure}
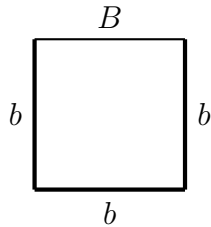

For a topological space $A$, its subspace $B$ and an element $b \in B$, the second relative homotopy group $\pi_2(A,B;b)$ of $A$ relative to $B$ and base $b$ is defined as the set of homotopy classes of maps $[0,1]^2 \to A$ such that $[0,1] \times \{ 1 \}$ is mapped to $B$ and $([0,1] \times \{ 0 \}) \cup (\{ 0, 1 \} \times [0,1]) $ is mapped to $b$. See figure \ref{fig:pi2_def} for a pictorial representation of these conditions.

\begin{figure}[ht]
\begin{minipage}[c]{0.5\textwidth}
\centering
$\sigma \sigma' \ = \ $
\begin{tikzpicture}[baseline=0.9cm]
      \draw[ultra thick]
       (0,0) coordinate (a1) -- node[left] {$b$} (0,2) coordinate (d1)
       (2,0) coordinate (b1) -- node[below] {$b$} (4,0) coordinate (e1)
       (2,0) coordinate (b1) -- node[right](q1){} (2,2) coordinate (c1);
     \draw[thick] (2,2) coordinate (c1) -- node[above] {$B$} (4,2) coordinate (f1);
     \draw[thick] (d1) -- node[above]    {$B$}(c1);
     \draw[ultra thick] (a1) -- node[below](g1){$b$}(b1);
     \draw[ultra thick] (e1) -- node[right](h1){$b$}(f1);
     \node (gg) at (1,1) {$\sigma$};
     \node (gg) at (3,1) {$\sigma'$};
\end{tikzpicture}
\end{minipage}\hfill
 \begin{minipage}[c]{0.5\textwidth}
    \caption{
       Definition of the product in~$\pi_2(A,B;b)$, which is given by horizontal concatenation.
    } \label{fig:pi2_concac}
  \end{minipage}
\end{figure}
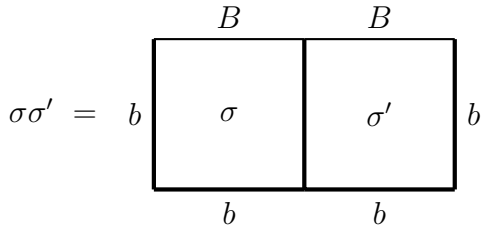

Multiplication of elements of $\pi_2(A,B;b)$ is given by horizontal concatenation -- see figure \ref{fig:pi2_concac}. One may also show \cite[p.~343]{hatcher} that elements of $\pi_2(A,B;b)$ describe homotopy classes of maps of a disc to $A$ which map the boundary to $B$ and a single point of the boundary to $b$. In the case $B= \{ b \}$ we abbreviate $\pi_2(A,B;b) = \pi_2(A;b)$. Elements of this group are homotopy classes of maps $S^2 \to A$, since a square with its boundary crushed to a point is a two-sphere.

\begin{figure}[ht]
\begin{minipage}[c]{0.5\textwidth}
\centering
\begin{tikzpicture}
      \draw[ultra thick]
       (0,0) coordinate (a1) -- node[left]     {$c$} (0,2) coordinate (d1)
       (2,0) coordinate (b1) -- node[right](q1){$c$} (2,2) coordinate (c1);
     \draw[thick, decoration={markings, mark=at position 0.5 with {\arrow{stealth}}},postaction={decorate}] (c1) -- node[above]    {$\partial \sigma$}(d1);
     \draw[ultra thick] (a1) -- node[below](f1){$c$}(b1);
     \node (gg) at (1,1) {$\sigma$};
\end{tikzpicture}
\end{minipage}\hfill
 \begin{minipage}[c]{0.5\textwidth}
    \caption{
       Definition of~$\partial$: homotopy class of the map given by a square $\sigma$ is~mapped to the loop given by its upper edge.
    } \label{fig:partial_def}
  \end{minipage}
\end{figure}
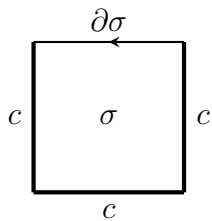

More generally, for a~subspace $C \subseteq B$ we let $\pi_2(A,B;C)$ be the groupoid with object set $C$ and the set of arrows from $c$ to $c'$ given by $\pi_2(A,B;c)$ if $c=c'$ and empty otherwise. Homomorphism $\partial : \pi_2(A,B;C) \to \pi_1(B;C)$ is defined by mapping the homotopy class of a map $\sigma$ to the homotopy class of $\left. \sigma \right|_{[0,1] \times \{ 1 \}}$, see figure \ref{fig:partial_def}.

\begin{figure}[ht]
\begin{minipage}[c]{0.45\textwidth}
\centering
$\gamma\rhd\sigma \ = \ $
\begin{tikzpicture}[baseline=1.4cm]
      \draw[ultra thick]
       (0,0) coordinate (a1) -- node[left]     {$c$} (0,3) coordinate (d1)
       (3,0) coordinate (b1) -- node[right](q4){$c$} (3,3) coordinate (c1);
     \draw[thick, decoration={markings, mark=at position 0.5 with {\arrow{stealth}}},postaction={decorate}] (1,3) --  (0,3)node[midway,above]{$\gamma$} ;
     \draw[thick, decoration={markings, mark=at position 0.5 with {\arrow{stealth}}},postaction={decorate}] (2,3) -- (1,3)node[midway,above]{$\partial\sigma$}  coordinate (x2);
     \draw[thick, decoration={markings, mark=at position 0.5 with {\arrow{stealth}}},postaction={decorate}]  (2,3) -- (3,3) node[midway,above]{$\gamma$} coordinate(c1);
     \draw[ultra thick] (a1) -- node[below](f1){$c$}(b1);
     \draw[thick]
     (1,1) coordinate (g1) -- (2,1)
     (1,1) coordinate (g2) -- (1,3)
     (2,3) coordinate (g3) -- (2,1);
     \draw[thick, decoration={markings, mark=at position 0.5 with {\arrow{stealth}}},postaction={decorate}](1,2) coordinate (g4) -- (0,2) node[midway,above] {$\gamma$};
     \draw[thick, decoration={markings, mark=at position 0.5 with {\arrow{stealth}}},postaction={decorate}](2,2) coordinate (g4) -- (3,2) node[midway,above] {$\gamma$};
     \draw[thick, decoration={markings, mark=at position 0.5 with {\arrow{stealth}}},postaction={decorate}]
     (1.5,1) coordinate (g4) -- (1.5,0) node[midway,right] {$\gamma$};
     \node (gg) at (1.5,2) {$\sigma$};
\end{tikzpicture}
\end{minipage}\hfill
 \begin{minipage}[c]{0.55\textwidth}
    \caption{
       Definition of~the~action of~$\pi_1(B;C)$ on $\pi_2(A,B;C)$. Here $\gamma$ is a path from $c'$ to $c$ and $\sigma$ belongs to $\pi_2(A,B;c')$.} \label{fig:action_def}
  \end{minipage}
\end{figure}
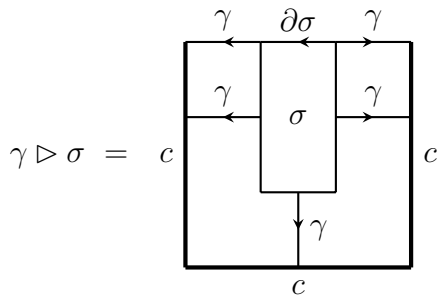

Last, but not least, an action of $\pi_1(B;C)$ on $\pi_2(A,B;C)$ is defined on figure \ref{fig:action_def}.

Inspection of figures \ref{fig:pi2_concac}--\ref{fig:action_def} reveals that $\Pi_{2}(A,B;C)=(\pi_1(B;C),\pi_2(A,B;C), \partial, \rhd)$ satisfies all axioms of a crossed module of grupoids.

A homomorphism of crossed modules of grupoids $(G,H,\partial, \rhd) \to (G',H',\partial',\rhd')$ is a pair of homomorphisms of grupoids, $E : G \to G'$ and $F : H \to H'$, such that $\partial' \circ F = E \circ \partial$ and $F(g \rhd h) = E(g) \rhd' F(h)$ whenever $s(g)=t(h)$. Now let $\mathbb G = (\mathcal E, \Phi, \Delta, \rhd)$ be a fixed crossed module of groups with finite $\mathcal E$ and $\Phi$. A~$\mathbb G$-valued lattice gauge field is defined as a homomorphism $\Pi_2(X_2,X_1;X_0) \to \mathbb G$. In~order to turn this concise definition into an~operational one, we need a description of the groupoid $\pi_2(X_2,X_1;X_0)$ in terms of explicit generators, preferably constructed in terms of local data. It is a nontrivial fact, which follows from the results of Whitehead \cite{whitehead01,brown03}, that this is indeed possible. This is what we will review next.

\begin{figure}[ht]
\begin{minipage}[c]{0.45\textwidth}
\centering
\begin{tikzpicture}
\node[draw, minimum size=3cm, regular polygon,
    regular polygon sides=5,
    label=side 1:$e_4$, label=side 2:$e_5$, label=side 3:$e_1$,
    label=side 4:$e_2$, label=side 5:$e_3$] (pol) {};

\foreach \x/\y in {2/3,3/4,5/1}
  \path[auto=right, ->-]
    (pol.corner \x)--(pol.corner \y);

\foreach \x/\y in {2/1,5/4}
  \path[auto=right, ->-]
    (pol.corner \x)--(pol.corner \y);

\node[below] (bf) at (pol.corner 3) {$b(f)$};
\node (point) at (pol.corner 3) {$\bullet$};

\node (reference) at (-0.6,-0.5) {};
\draw[->,>=stealth,semithick, blue] (reference) arc[radius=0.8, start angle=-140, end angle=160];
\node (f) at (0,0) {$f$};
\end{tikzpicture}
\end{minipage}\hfill
 \begin{minipage}[c]{0.55\textwidth}
    \caption{Pentagonal plaquette with a~chosen orientation of edges and of the face. In this case $\partial f = e_5 e_4^{-1}e_3 e_2^{-1}e_1$.} \label{fig:pentagon}
  \end{minipage}
\end{figure}
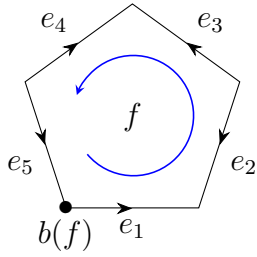

For every face $f$ we choose a basepoint $b(f)$ and an orientation. The~boundary of $f$ then forms a loop $\partial f$ based at $b(f)$. See figure \ref{fig:pentagon} for an example.

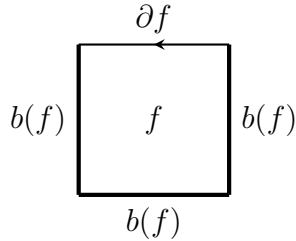
\begin{figure}[ht]
\begin{minipage}[c]{0.35\textwidth}
\centering
\begin{tikzpicture}
      \draw[ultra thick]
       (0,0) coordinate (a1) -- node[left]     {$b(f)$} (0,2) coordinate (d1)
       (2,0) coordinate (b1) -- node[right](q1){$b(f)$} (2,2) coordinate (c1);
     \draw[thick, decoration={markings, mark=at position 0.5 with {\arrow{stealth}}},postaction={decorate}] (c1) -- node[above]    {$\partial f$}(d1);
     \draw[ultra thick] (a1) -- node[below](f1){$b(f)$}(b1);
     \node (gg) at (1,1) {$f$};
\end{tikzpicture}
\end{minipage}\hfill
 \begin{minipage}[c]{0.65\textwidth}
    \caption{Schematic representation of~a~map representing the~element in $\pi_2(X_2,X_1;b(f))$ corresponding to a plaquette $f$.} \label{fig:plaquette_f}
  \end{minipage}
\end{figure}

There exists a corresponding element $f \in \pi_2(X_2,X_1; b(f))$, given by the homotopy class of any map of the schematic form depictured on figure \ref{fig:plaquette_f}.

By acting on faces with paths it is possible to obtain new elements, possibly based at different points. It turns out that the set of all $\gamma \rhd f$ with $s(\gamma)=b(f)$ generates the grupoid $\pi_2(X_2,X_1;b(f))$. The only non-trivial relations between these elements follow from Peiffer identities and are of the form
\begin{equation}
(\gamma \, \partial f \, \gamma^{-1}) \rhd (\gamma' \rhd f') = (\gamma \rhd f) \, (\gamma' \rhd f')\, (\gamma \rhd f)^{-1}
\label{eq:free_relation}
\end{equation}
for every $\gamma, \gamma', f$ and $f'$ such that $t(\gamma')=t(\gamma)$, $s(\gamma') = b(f')$ and $s(\gamma) = b(f)$.

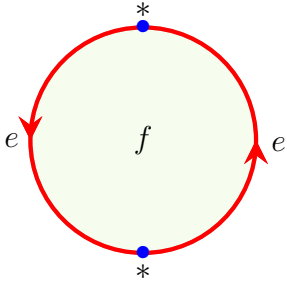
\begin{figure}[ht]
\begin{minipage}[c]{0.4\textwidth}
\centering
\begin{tikzpicture}
      \draw[ultra thick,fill=gre,opacity=0.2] (0,0) circle (1.5 cm);
      \draw[ultra thick, red, decoration={markings, mark=at position 0 with {\arrow[scale=1.5]{stealth}}}, postaction={decorate},decoration={markings, mark=at position 0.5 with {\arrow[scale=1.5]{stealth}}}, postaction={decorate}] (0,0) circle (1.5 cm);
      \node[below] (gg) at (0,-1.5cm) {$\ast$};
      \node at (0,-1.5cm) {{\color{blue}$\bullet$}};
      \node[above] (gh) at (0,1.5cm) {$\ast$};
      \node at (0,1.5cm) {{\color{blue}$\bullet$}};
      \node at (1.8cm, -0.05cm) {$e$};
      \node at (-1.75cm, 0cm) {$e$};
      \node at (0,0) {$f$};
\end{tikzpicture}
\end{minipage}\hfill
 \begin{minipage}[c]{0.6\textwidth}
    \caption{The real projective plane is a~disc with antipodal points of the bounding circle identified. It~can be constructed by attaching a~plaquette to a circle along a~map of winding number two.} \label{fig:rp2}
  \end{minipage}
\end{figure}

For the sake of example, we consider the real projective plane, $X=\mathbb{RP}^2$. It~admits a~decomposition with exactly one cell in every dimension up to $2$ -- see figure~\ref{fig:rp2}. In~this case the grupoid $\pi_1(X_1;X_0)$ has one object $\ast$ and one generator $e$. There is one plaquette $f$, with $\partial f = e^2$. The relative homotopy group $\pi_2(X_2,X_1;\ast)$ is generated by elements $f_n:=e^n\rhd f$, $n\in\mathbb{Z}$. The first Peiffer identity gives $\partial f_n = e^2$, so~relations \eqref{eq:free_relation} reduce to
\begin{equation}
f_{m+2}=f_nf_mf_n^{-1} \qquad \text{for all } n,m \in \mathbb Z.
\end{equation}
Evaluating this for $m=n$ gives $f_n=f_{n+2}$. Thus all generators can be expressed in terms of $f_0$ and $f_1$. Secondly, taking $n=0$, $m=1$ gives $f_0f_1=f_1f_0$. There are no other independent relations, so $\pi_2(X_2,X_1;\ast)\cong \mathbb{Z}^2$.

We are now ready to explain what are field configurations in the considered models. In order to define a homomorphism $\Pi_2(X_2,X_1;X_0) \to \mathbb G$ we have to assign an element $\epsilon_e \in \mathcal E$ to every edge $e$ and $\varphi_f \in \Phi$ to every face $f$. Since this assignment is to define a homomorphism $\pi_1(X_1; X_0) \to \mathcal E$, we map a path $\gamma= e_n \ldots e_1$ to $\epsilon_{\gamma} = \epsilon_{e_n} \ldots \epsilon_{e_1}$. Element $\gamma \rhd f \in \pi_2(X_2,X_1;t(\gamma))$ has to be sent to $\epsilon_{\gamma} \rhd \varphi_f$, by the definition of a homomorphism of crossed modules. Since any arrow in $\pi_2(X_2, X_1; X_0)$ is a product of arrows of this form, the element $\varphi_{\sigma} \in \Phi$ assigned to any $\sigma$ is determined. We still have to make sure that this is consistent. Firstly, the definition of a homomorphism asserts that we should have
\begin{equation}
\Delta \varphi_f = \epsilon_{\partial f} \qquad \text{for any face }f.
\label{eq:ff}
\end{equation}
This is a non-trivial constraint on the collections $\bm \epsilon = \{ \epsilon_e\}, \, \bm \varphi = \{ \varphi_f \}$, called \textbf{fake flatness}. We claim that there are no other constraints, since compatibility with the relation \eqref{eq:free_relation} is automatic. Indeed, equality
\begin{equation}
(\epsilon_{\gamma} \, \Delta \varphi_f \, \epsilon_{\gamma}^{-1} \, \epsilon_{\gamma'}) \rhd \varphi_{f'} = (\epsilon_{\gamma} \rhd \varphi_f)\, (\epsilon_{\gamma'} \rhd \varphi_{f'}) \,(\epsilon_{\gamma} \rhd \varphi_f)^{-1}
\end{equation}
follows from the fake flatness condition and Peiffer identities in $\mathbb G$.

To understand how field configurations look like in practice, consider the example of $X$ taken to be the pentagon presented on figure \ref{fig:pentagon}. A field configuration consists of elements $\epsilon_{e_1},...,\epsilon_{e_5} \in \mathcal E$ and $\varphi_f \in \Phi$ subject to the constraint
\begin{equation}
\Delta \varphi_f = \epsilon_{e_5} \epsilon_{e_4}^{-1} \epsilon_{e_3} \epsilon_{e_2}^{-1} \epsilon_{e_1}.
\end{equation}

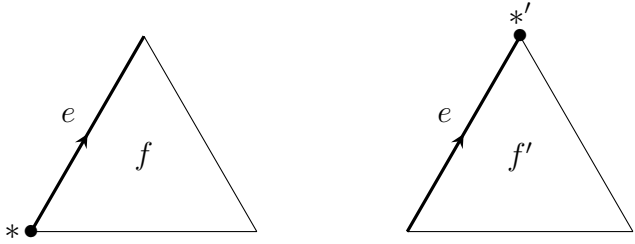
\begin{figure}[ht]
\begin{minipage}[c]{0.65\textwidth}
\centering
\begin{tikzpicture}
\draw
(0,0) coordinate (a1) -- ++(0:3) coordinate (a2)
(a2) -- ++(120:3) coordinate (a3)
(a3) -- (a1)
;

\draw
(5,0) coordinate (b1) -- ++(0:3) coordinate (b2)
(b2) -- ++(120:3) coordinate (b3)
(b3) -- (b1)
;

\draw[very thick, decoration={markings, mark=at position 0.5 with {\arrow{stealth}}},postaction={decorate}](a1) -- node[above left] {$e$}(a3);

\draw[very thick, decoration={markings, mark=at position 0.5 with {\arrow{stealth}}},postaction={decorate}](b1) -- node[above left] {$e$}(b3);

\node at (a1) {$\bullet$};
\node[left] at (a1) {$\ast$};

\node at (b3) {$\bullet$};
\node[above] at (b3) {$\ast'$};

\node at (1.5,1) {$f$};
\node at (6.5,1) {$f'$};

\end{tikzpicture}
\end{minipage}\hfill
 \begin{minipage}[c]{0.35\textwidth}
    \caption{Illustration of the change of a base point. In this case $f'=e\rhd f$.}
    \label{fig:triangles2}
  \end{minipage}
\end{figure}

In the above discussion we have been forced to choose base points and orientations for the elementary plaquettes. Distinct choices correspond to distinct choices of generators of the same algebraic structure. We close this section with an explanation how generators are transformed upon a change of these choices:
\begin{enumerate}
\item Change of orientation of a plaquette maps the element $f$ to $f^{-1}$.
\item Change of the base point from $\ast$ to $\ast'$ (with both elements belonging to the boundary of $f$) changes $f$ to $\gamma \rhd f$, where $\gamma$ is a path from $\ast$ to $\ast'$ along the boundary of $f$. For an example see figure \ref{fig:triangles2}. If $f$ is simply-connected, the element $\gamma \rhd f$ does not depend on the choice of $\gamma$. Indeed, in this case any other allowed path takes the form $\gamma' = \gamma \, (\partial f)^n$ for some $n$ and $\partial f \rhd f =f$, by~the second Peiffer identity.
\end{enumerate}

\subsection{Degrees of freedom and holonomies}
\label{sec:dofah}

In models based on crossed modules there are, besides holonomies along loops (built out of degrees of freedom located on edges), also holonomies along surfaces (built out of degrees of freedom located on edges and faces). In order to explain their construction we first need to discuss certain basic properties of crossed modules.

Let $(G,H,\partial, \rhd)$ be a crossed module of groups. We note two important consequences of the first Peiffer identity:
\begin{enumerate}
\item The image $\mathrm{im}(\partial)$ of $\partial$ is a normal subgroup of $G$. Thus there is a group structure on the quotient space $\coker(\partial)=G/\mathrm{im}(\partial)$.
\item The action of $G$ on $H$ preserves $\mathrm{ker}(\partial)$, the kernel of $\partial$:
\begin{equation}
h \in \mathrm{ker} (\partial) \implies \forall g \in G \ \ g \rhd h \in \mathrm{ker}(\partial),
\label{eq:ker_pres}
\end{equation}
\end{enumerate}
and two conclusions from the second Peiffer identity:
\begin{enumerate}
\item $\mathrm{ker}(\partial)$ is a central subgroup of $H$. In particular $\mathrm{ker}(\partial)$ is abelian.
\item Elements of $\mathrm{im}(\partial)$ act trivially on $\mathrm{ker}(\partial)$, i.e.\ $g \rhd h = h$ for all  $g\in\mathrm{im}(\partial)$ and $h\in\ker\partial$. Thus there is an induced action of the group $\coker(\partial)$ on $\mathrm{ker}(\partial)$.
\end{enumerate}
Furthermore, if $(E, F) : (G,H,\partial, \rhd) \to (G',H',\partial',\rhd')$ is a homomorphism of crossed modules of groups, then:
\begin{enumerate}
\item $E (\mathrm{im}(\partial)) \subseteq \mathrm{im}(\partial')$, so there is an induced map $\overline E : \coker(\partial) \to \coker(\partial')$.
\item $F$ maps $\ker(\partial)$ to $\mathrm{ker}(\partial')$. We denote the induced map $\ker(\partial) \to \ker(\partial')$ by $\overline F$.
\end{enumerate}
For future use we remark that if $\overline E$ and $\overline F$ are group isomorphisms, $(E,F)$ is said to be a \textbf{weak isomorphism}. Existence of a weak isomorphism $\mathbb G \to \mathbb G'$ does not imply\footnote{For the sake of example, consider the crossed module with $G=\mathbb Z_2$ and trivial $H$, $\partial$ and $\rhd$. Secondly, let us take $G'=\mathbb Z_4$, $H'=\mathbb Z_2$, $\partial'$ given by reduction modulo two and trivial $\rhd'$. Standard embedding $\mathbb Z_2 \to \mathbb Z_4$, together with the trivial homomorphism $H \to H'$, is a weak isomorphism. It is easy to check that there exists no weak isomorphism in the opposite direction.} that there is a weak isomorphism $\mathbb G' \to \mathbb G$. Thus in order for this notion to yield an equivalence relation, one declares two crossed modules $\mathbb G$ and $\mathbb G'$ to be \textbf{weakly equivalent} if there exist a~family of crossed modules $\mathbb G_1,...,\mathbb G_n$ and a zig-zag sequence of weak isomorphisms of the form
\begin{equation}
    \mathbb G\xrightarrow{\hspace*{0.6cm}}\mathbb G_1\xleftarrow{\hspace*{0.6cm}}\mathbb G_2\xrightarrow{\hspace*{0.6cm}}...\xleftarrow{\hspace*{0.6cm}}\mathbb G_n\xrightarrow{\hspace*{0.6cm}}\mathbb G'.
\end{equation}
In other words, weak equivalence is the coarsest equivalence relation such that weakly isomorphic crossed modules are equivalent.

Let us now specialize to the crossed module $\Pi_2(X_2,X_1;\ast)$ for some $\ast \in X_0$. As~reviewed in the appendix \ref{sec:exact_seq}, $\coker(\partial)$ and $\ker(\partial)$ are the fundamental group of $X$ and the second homotopy group of $X_2$, respectively. Therefore any field configuration induces homomorphisms $\pi_1(X;\ast) \to \coker(\Delta)$ and $\pi_2(X_2;\ast) \to \mathrm{ker}(\Delta)$. We will now explain their significance.

Consider a field configuration given by $\bm \epsilon$ and $\bm \varphi$. Element $\epsilon_{\gamma} \in \mathcal E$ assigned to a~path $\gamma$ has the interpretation of a parallel transport from $s(\gamma)$ to $t(\gamma)$ along $\gamma$. Parallel transports along closed paths ($s(\gamma)=t(\gamma)$) will be called \textbf{$1$-holonomies}, to~distinguish them from \textbf{$2$-holonomies}, to be considered soon.

We define $\overline \epsilon_{\gamma}$ as the reduction of $\epsilon_{\gamma}$ modulo $\mathrm{im}(\Delta)$. Assignment $\gamma \mapsto \overline{\epsilon}_{\gamma}$ defines an~ordinary $\coker(\Delta)$-valued lattice gauge field $\overline{\bm \epsilon}$. Its definition is motivated by inspecting the fake flatness condition (\ref{eq:ff}) reduced modulo $\mathrm{im}(\Delta)$:
\begin{equation}
\overline \epsilon_{\partial f} =1 \qquad \text{for any face }f.
\label{eq:ff_reduced}
\end{equation}
This is the statement that $\overline{\bm \epsilon} = \{ \overline \epsilon_e \}$ is a flat gauge field: the holonomy along any loop in $X$ which bounds a surface is trivial\footnote{This follows directly from the condition above only for surfaces in $X_2$. Here we are also using the fact that attaching $3$-cells to a space does not change its fundamental group \cite[Prop.~1.26(b)]{hatcher}.}, so holonomies along homotopic loops are equal. In other words, $\overline{\bm \epsilon}$ defines a homomorphism $\pi_1(X;X_0) \to \coker(\Delta)$.

To further understand the fake flatness condition, consider the problem of finding its solutions $\bm \epsilon$, $\bm \varphi$ for a fixed flat $\overline{\bm \epsilon}$. First, each $\epsilon_e$ is determined by $\overline \epsilon_e$ up to multiplication by $\Delta \psi_e$ for some $\psi_e \in \Phi$. Having chosen any particular $\bm \epsilon$, we are guaranteed by flatness of $\overline{\bm \epsilon}$ that each $\epsilon_{\partial f}$ belongs to $\mathrm{im}(\Delta)$: there exists some $\varphi_f \in \Phi$, unique up to multiplication by any $\chi_f \in \mathrm{ker}(\Delta)$, such that $\Delta \varphi_f = \epsilon_{\partial f}$.

The above discussion may be summarized as follows. Gauge field valued in a~crossed module may be though of as consisting of three components:
\begin{enumerate}
\item $\coker(\Delta)$-valued field located on edges, constrained by (\ref{eq:ff_reduced}) and hence defining a flat gauge field $\overline{\bm \epsilon}$,
\item $\mathrm{im}(\Delta)$-valued degrees of freedom located on edges, responsible for the freedom in the choice of $\bm \epsilon$ for a given $\overline{\bm \epsilon}$,
\item $\mathrm{ker}(\Delta)$-valued degrees of freedom located on faces, responsible for the freedom in the choice of $\bm \varphi$ for a given $\bm \epsilon$.
\end{enumerate}
As in ordinary gauge theory, some degrees of freedom are eventually removed by introducing a ``gauge equivalence'' relation on the set of field configurations. This will be discussed in subsection \ref{sec:gt}.

Typical observables sensitive to degrees of freedom of the third type are (functions~of) $2$-holonomies, i.e.\ elements $\varphi_{\sigma} \in \Phi$ assigned to $\sigma \in \mathrm{ker}(\partial)$. Notice that $\varphi_{\sigma}$ are automatically in $\mathrm{ker}(\Delta)$. Indeed, $\Delta \varphi_{\sigma} = \varphi_{\partial \sigma} = 1$. This gives the promised homomorphism $\pi_2(X_2;\ast) \to \ker(\Delta)$. It~may be interpreted as a~two-dimensional analogue of parallel transport along closed paths, with loops replaced by spheres embedded in $X_2$.

To summarize the above discussion, $\mathrm{ker}(\Delta)$-valued holonomy along any sphere in $X_2$ is defined. We will now demonstrate how to compute it in some simple examples.

\begin{figure}[ht]
\begin{minipage}[c]{0.5\textwidth}
\centering
\begin{tikzpicture}
\begin{scope}[very thick]

     \coordinate (a) at (0,0);
     \coordinate (b) at (2.5,1.5);;
     \coordinate (c) at (-1.5,2);
     \coordinate (d) at (1.2,4);

     \node (0,0) {$\bullet$};

     \draw[thick, decoration={markings, mark=at position 0.5 with {\arrow{stealth}}},postaction={decorate}](a) -- (b) node[midway,below right] {$e_{1}$};

     \draw[dashed, decoration={markings, mark=at position 0.65 with {\arrow{stealth}}},postaction={decorate}](c) -- (b) node[pos=0.65,above] {$e_{6}$};

     \draw[thick, decoration={markings, mark=at position 0.5 with {\arrow{stealth}}},postaction={decorate}](a) -- (c) node[midway,left] {$e_{3}$};

 \draw[thick, decoration={markings, mark=at position 0.65 with {\arrow{stealth}}},postaction={decorate}](a) -- (d) node[pos=0.65,left] {$e_{2}$};

 \draw[thick, decoration={markings, mark=at position 0.5 with {\arrow{stealth}}},postaction={decorate}](d) -- (b) node[pos=0.40,right] {$e_{4}$};

     \draw[thick, decoration={markings, mark=at position 0.5 with {\arrow{stealth}}},postaction={decorate}](c) -- (d) node[midway,above left] {$e_{5}$};

\fill[gre,opacity=0] (b)--(c)--(d)--cycle;
\fill[gre,opacity=0.2] (a)--(b)--(d)--cycle;
\fill[gre,opacity=0.2] (a)--(c)--(d)--cycle;
\fill[gre,opacity=0] (a)--(b)--(c)--cycle;

\node[below] (0,0) {$\ast$};

\node (f1) at (4,1.5) {$f_1$};
\node (f1a) at (1.5,2) {};

\node (f2) at (-3,1.5) {$f_2$};
\node (f2a) at (0.2,2) {};

\node (f3) at (3,-0.5) {$f_3$};
\node (f3a) at (0.5,0.3) {};

\node (f4) at (-2,4) {$f_4$};
\node (f4a) at (0.5,3.5) {};

\sffamily

\draw[-latex,blue] (f1) to[out=110,in=0] node[midway,font=\scriptsize,above] {} (f1a);
\draw[-latex,blue] (f2) to[out=-60,in=160] node[midway,font=\scriptsize,above] {} (f2a);
\draw[-latex,blue] (f3) to[out=90,in=-70] node[midway,font=\scriptsize,above] {} (f3a);
\draw[-latex,blue] (f4) to[out=-90,in=90] node[midway,font=\scriptsize,above] {} (f4a);

\end{scope}
\end{tikzpicture}
\end{minipage}\hfill
 \begin{minipage}[c]{0.5\textwidth}
    \caption{Tetrahedron as an example of~a~triangulation of a~$2$-sphere. Chosen orientations of edges are indicated by arrows.} \label{fig:tetrahedron}
  \end{minipage}
\end{figure}
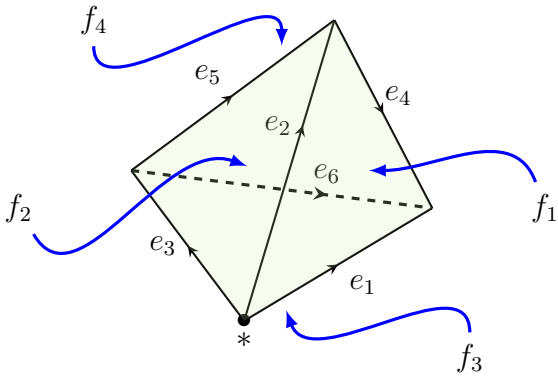

Consider the triangulation of a two-sphere presented on figure \ref{fig:tetrahedron}. We~choose $\ast$ as the base point of $f_1,f_2,f_3$ and $t(e_1)$ as the base point of $f_4$. Faces $f_1,\ldots,f_4$ are oriented so that
\begin{equation}
\label{tetrahedron:faces:orientations}
\partial f_1 = e_2^{-1}e_4^{-1}e_1,
\hskip 5mm
\partial f_2 = e_3^{-1}e_5^{-1}e_2,
\hskip 5mm
\partial f_3 = e_1^{-1}e_6e_3,
\hskip 5mm
\partial f_4 = e_4e_5 e_6^{-1}.
\end{equation}

\begin{figure}[ht]
\centering
$\sigma =$
\begin{tikzpicture}[baseline=1.0cm]
      \draw[ultra thick]
       (0,0) coordinate (a1) -- node[left]     {} (2.5,0) coordinate (a2)
       (2.5,0) coordinate (a2) -- node[left]     {} (5,0) coordinate (a3)
       (7,0) coordinate (a3) -- node[left]     {} (7.5,0) coordinate (a4)
       (7.5,0) coordinate (a4) -- node[left]     {} (10,0) coordinate (a5)
       (0,0) coordinate (a1) -- node[left]     {} (0,2) coordinate (b1)
       (2.5,0) coordinate (a2) -- node[left]     {} (2.5,2) coordinate (b2)
       (5,0) coordinate (a3) -- node[left]     {} (5,2) coordinate (b3)
       (7.5,0) coordinate (a4) -- node[left]     {} (7.5,2) coordinate (b4)
       (10,0) coordinate (a5) -- node[left]     {} (10,2) coordinate (b5);

     \draw[thick, decoration={markings, mark=at position 0.5 with {\arrow{stealth}}},postaction={decorate}] (b2) -- node[above]    {$e_1^{-1}e_4e_5e_6^{-1}e_1$}(b1);
     \draw[thick, decoration={markings, mark=at position 0.5 with {\arrow{stealth}}},postaction={decorate}] (b3) -- node[above]    {$e_1^{-1}e_6e_3$}(b2);
	 \draw[thick, decoration={markings, mark=at position 0.5 with {\arrow{stealth}}},postaction={decorate}] (b4) -- node[above]    {$e_3^{-1}e_5^{-1}e_2$}(b3);
     \draw[thick, decoration={markings, mark=at position 0.5 with {\arrow{stealth}}},postaction={decorate}] (b5) -- node[above]    {$e_2^{-1}e_4^{-1}e_1$}(b4);

     \draw[ultra thick] (a3) -- node[below]{}(a4);

     \node (f1) at (1.25,1) {$e_1^{-1}\rhd f_4$};
     \node (f1) at (3.75,1) {$f_3$};
     \node (f1) at (6.25,1) {$f_2$};
     \node (f1) at (8.75,1) {$f_1$};
\end{tikzpicture} \\
$\sigma' =$
\begin{tikzpicture}[baseline=1.0cm]
      \draw[ultra thick]
       (0,0) coordinate (a1) -- node[left]     {} (2.5,0) coordinate (a2)
       (2.5,0) coordinate (a2) -- node[left]     {} (5,0) coordinate (a3)
       (7,0) coordinate (a3) -- node[left]     {} (7.5,0) coordinate (a4)
       (7.5,0) coordinate (a4) -- node[left]     {} (10,0) coordinate (a5)
       (0,0) coordinate (a1) -- node[left]     {} (0,2) coordinate (b1)
       (2.5,0) coordinate (a2) -- node[left]     {} (2.5,2) coordinate (b2)
       (5,0) coordinate (a3) -- node[left]     {} (5,2) coordinate (b3)
       (7.5,0) coordinate (a4) -- node[left]     {} (7.5,2) coordinate (b4)
       (10,0) coordinate (a5) -- node[left]     {} (10,2) coordinate (b5);

     \draw[thick, decoration={markings, mark=at position 0.5 with {\arrow{stealth}}},postaction={decorate}] (b2) -- node[above]    {$e_2^{-1}e_5e_6^{-1}e_4e_2$}(b1);
     \draw[thick, decoration={markings, mark=at position 0.5 with {\arrow{stealth}}},postaction={decorate}] (b3) -- node[above]    {$e_2^{-1}e_4^{-1}e_1$}(b2);
	 \draw[thick, decoration={markings, mark=at position 0.5 with {\arrow{stealth}}},postaction={decorate}] (b4) -- node[above]    {$e_1^{-1}e_6e_3$}(b3);
     \draw[thick, decoration={markings, mark=at position 0.5 with {\arrow{stealth}}},postaction={decorate}] (b5) -- node[above]    {$e_3^{-1}e_5^{-1}e_2$}(b4);

     \draw[ultra thick] (a3) -- node[below]{}(a4);

     \node (f1) at (1.25,1) {$(e_4e_2)^{-1}\!\rhd\! f_4$};
     \node (f1) at (3.75,1) {$f_1$};
     \node (f1) at (6.25,1) {$f_3$};
     \node (f1) at (8.75,1) {$f_2$};
\end{tikzpicture}
\caption{Graphical representation of two ways to construct a~generator of~$\pi_2(X_2,X_1;\ast)$ for the tetrahedron from figure \ref{fig:tetrahedron}: $\sigma = (e_1^{-1} \rhd f_4)f_3 f_2 f_1$ and $\sigma' =( (e_4 e_2)^{-1} \rhd f_4) f_1 f_3 f_2$. In fact we have $\sigma = \sigma'$.} \label{fig:ho_squares_01}
\end{figure}
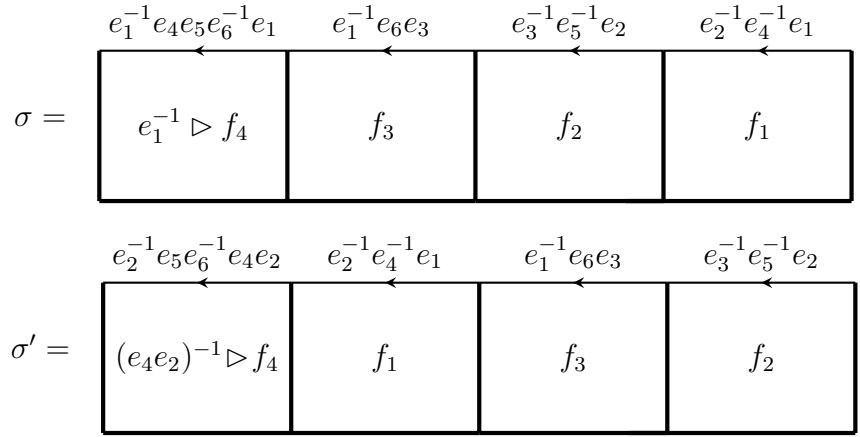

It is well known that the second homotopy group of $S^2$ is infinite cyclic. Choice of one of two possible generators of this group is equivalent to a~choice of orientation. One may construct a generator by multiplying the four faces (all transported to a~common base point by acting with edges) in such a way that an element with trivial boundary is obtained. There is more than one way to do this, as shown on figure \ref{fig:ho_squares_01}. It is not difficult to convince oneself that the two elements $\sigma, \sigma'$ presented on figure \ref{fig:ho_squares_01} represent the same orientation. Thus they must be equal. We will now check this by a direct computation:
\begin{equation}
\sigma' = \left( \underbrace{(e_4 e_2)^{-1} e_1}_{\partial f_1} e_1^{-1} \rhd f_4 \right) f_1 f_3 f_2 = f_1 (e_1^{-1} \rhd f_4 ) f_3 f_2 = f_1 \sigma f_1^{-1} = \sigma,
\end{equation}
where we applied the second Peiffer identity, inserted the definition of $\sigma$ and used the fact that $\sigma$ is central in the second, third and fourth equalities, respectively.

Having constructed the element $\sigma \in \pi_2(X_2,X_1;\ast)$, we compute $\varphi_{\sigma}$ simply by replacing each edge $e$ (resp. plaquette $f$) in the definition of $\sigma$ by the corresponding $\epsilon_e$ (resp. $\varphi_f$). Thus (compare with figure \ref{fig:ho_squares_01}):
\begin{equation}
\varphi_{\sigma} = (\epsilon_{e_1}^{-1} \rhd \varphi_{f_4}) \, \varphi_{f_3}\, \varphi_{f_2}\, \varphi_{f_1}.
\end{equation}
By construction, $\varphi_{\sigma}$ may also be computed as $ (  (\epsilon_{e_4} \, \epsilon_{e_2})^{-1} \rhd \varphi_{f_4}) \, \varphi_{f_1} \, \varphi_{f_3} \, \varphi_{f_2}$.

More generally, element $\sigma$ may always be constructed in an essentially unique way for any decomposition of a two-sphere with a chosen base point and orientation (possibly embedded in a larger space). The case particularly important for us is that of a sphere bounding a ball $q$, oriented and based at a point $b(q) \in X_0$. In~this case we denote the corresponding element $\sigma \in \pi_2(X_2;b(q))$ by $\partial q$.


We close this section with remarks about $\Pi_2(X_2,X_1;X_0)$ in the case when $X$ is (a decomposition of) a disc. Then the first and the second (non-relative) homotopy groups are trivial. Thus $\partial$ has trivial kernel and cokernel, i.e.\ it is an isomorphism. This means that a polygon embedded in $X$ bounded by a loop $l$ corresponds to the uniquely determined element $\partial^{-1}(l) \in \pi_2(X_2,X_1;s(l))$. There is more than one way to construct the element $\partial^{-1}(l)$ out of elementary plaquettes, but they are all equal due to Peiffer identities. Of course there is still some arbitrariness in the choice of the base point $s(l)$, but groups $\pi_2(X_2,X_1;x)$ with distinct $x \in X_0$ are canonically isomorphic. Remarks of this paragraph are also applicable to calculations performed in completely general geometries $X$, as long as only elements constructed out of edges and plaquettes in a contractible subcomplex of $X_2$ are involved.

\begin{figure}[ht]
\begin{minipage}[c]{0.55\textwidth}
\centering
\begin{tikzpicture}
\draw
(0,0) coordinate (a1) -- ++(-30:3) coordinate (a2)
(a2) -- ++(30:3) coordinate (a3)
(a3) -- ++(150:3) coordinate (a4)
(a4) -- (a1)
(a4) -- (a2)
;

\draw[very thick, decoration={markings, mark=at position 0.5 with {\arrow{stealth}}},postaction={decorate}](a1) -- node[below left] {$e_1$}(a2);
\draw[very thick, decoration={markings, mark=at position 0.5 with {\arrow{stealth}}},postaction={decorate}](a2) -- node[below right] {$e_2$}(a3);
\draw[very thick, decoration={markings, mark=at position 0.5 with {\arrow{stealth}}},postaction={decorate}](a3) -- node[above right] {$e_3$}(a4);
\draw[very thick, decoration={markings, mark=at position 0.5 with {\arrow{stealth}}},postaction={decorate}](a4) -- node[above left] {$e_4$}(a1);
\draw[very thick, decoration={markings, mark=at position 0.5 with {\arrow{stealth}}},postaction={decorate}](a2) -- node[right] {$e_5$}(a4);

\node at (a1) {$\bullet$};
\node[left] at (a1) {$\ast$};
\node at (1.7,0) {$f_1$};
\node at (3.5,0) {$f_2$};

\end{tikzpicture}
\end{minipage}\hfill
 \begin{minipage}[c]{0.45\textwidth}
    \caption{Triangulation of a~disc. We take $\ast$ and $s(e_2)$ as the base points of $f_1$ and $f_2$, respectively. Both faces are oriented counterclockwise.}
    \label{fig:double}
  \end{minipage}
\end{figure}
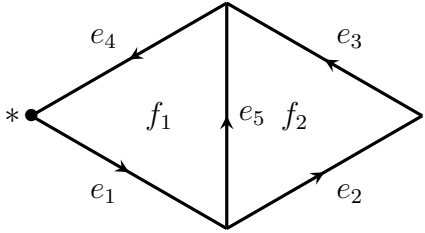

As an example, let us consider the triangulation of a disc depictured on the figure~\ref{fig:double}. With the chosen base points and orientations of faces we have
\begin{equation}
\partial f_1 = e_4e_5e_1, \hskip 1cm \partial f_2 = e_5^{-1}e_3e_2.
\end{equation}
We will construct the element corresponding to the whole disc out of elementary plaquettes and edges. We choose the counterclockwise orientation and pick $\ast$ as the base point. Then the bounding loop is $l=e_4 e_3 e_2 e_1$. We observe that
\begin{equation}
l = \underbrace{e_4 e_5 e_1}_{\partial f_1} e_1^{-1} \underbrace{e_5^{-1} e_3 e_2}_{\partial f_2} e_1 = \partial \left( f_1 \, (e_1^{-1} \rhd f_2) \right),
\end{equation}
so $\partial^{-1} (l) = f_1 \, (e_1^{-1} \rhd f_2)$. On the other hand, we also have $l = \partial \left( (e_4 e_5 \rhd f_2) \, f_1 \right)$. Thus $f_1 \, (e_1^{-1} \rhd f_2)=(e_4 e_5 \rhd f_2) \, f_1$. Indeed, this is easy to verify directly:
\begin{equation}
f_1 \, (e_1^{-1} \rhd f_2) = \underbrace{f_1 \, (e_1^{-1} \rhd f_2)\, f_1^{-1}}_{\text{Peiffer}}\, f_1 = ((\partial f_1 e_1^{-1}) \rhd f_2) \, f_1 = (e_4 e_5 \rhd f_2) \, f_1.
\end{equation}

\subsection{Gauge and electric transformations} \label{sec:gt}

As in ordinary gauge theory, there exist two particularly important broad classes of transformations of the set of field configurations. Firstly, we have gauge transformations. They describe redundancies in the description of the system, since configurations related by gauge transformations are regarded as physically indistinguishable. Secondly, there are transformations which are used to define higher analogues of the electric field operators in the quantized theory. Here we will discuss both types at the same time, as they are closely related\footnote{One example of this fact is that in conventional gauge theory quantized in temporal gauge, time-independent gauge transformations are generated by the divergence of the electric field. Secondly, the center $1$-form symmetry operators are also of electric type: they shift the gauge field by a center-valued cocycle. This operation reduces to a gauge transformation for cocycles of trivial cohomology class.}.

Another distinction between various transformation arises from geometric considerations: $p$-form transformations are parametrized by data associated to geometric objects of dimension $p$. Here we will consider vertex ($0$-form) transformations, regarded as gauge redundancies, analogous to those present in the ordinary gauge theory. Secondly, there will be edge ($1$-form) transformations. Declaring them to be gauge transformations is necessary to obtain the Yetter's topological field theory and its twisted versions. We will discuss the possibility to restrict the group of gauge transformations. This increases the number of physical degrees of freedom and hence allows to construct models with richer dynamics. Finally, we~will introduce plaquette ($2$-form) transformations. They play the role of electric transformations and are very analogous to corresponding transformations in abelian $2$-form gauge theory.

\begin{figure}[ht]
\begin{minipage}[c]{0.5\textwidth}
\centering
\begin{tikzcd}[column sep=2cm, row sep=2cm](\bm \epsilon,\bm \varphi)\arrow[d,"\hspace{-25pt}\{\xi_v\}"]\arrow[rd, "\{\xi_v'\,\xi_v\}"]& \\
(\bm{\epsilon'},\bm{\varphi'})\arrow[r, "\{\xi_v'\}"]& (\bm{\epsilon''},\bm{\varphi''})
\end{tikzcd}
\end{minipage}\hfill
 \begin{minipage}[c]{0.5\textwidth}
    \caption{
       Composition rule for vertex transformations: $\{ \xi_v \}$ followed by $\{ \xi'_v \}$ coincides with $\{ \xi_v'\, \xi_v \}$.
    } \label{fig:gauge_triangle}
  \end{minipage}
\end{figure}
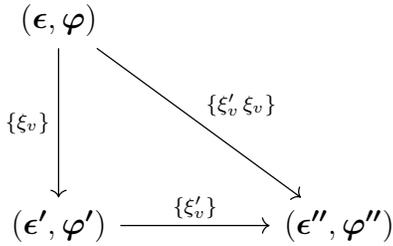

We begin with the discussion of \textbf{vertex transformations}. They are parametrized by collections $\bm \xi = \{ \xi_v \}$ of elements of $\mathcal E$ indexed by lattice sites. Their action on $\bm \epsilon$ is as for usual gauge fields, while $ \varphi_f$ transforms as a matter field placed on the lattice site $b(f)$:
\begin{equation}
\epsilon_e' = \xi_{t(e)} \, \epsilon_e \, \xi_{s(e)}^{-1}, \qquad \varphi_f' = \xi_{b(f)} \rhd \varphi_f.
\end{equation}
We will call them vertex transformations. They preserve the fake flatness, since
\begin{eqnarray}
\Delta\varphi'_{f}
 =
\Delta\left(\xi_{b(f)}\rhd\varphi_{f}\right)
\;\stackrel{\mathrm{Peiffer}}{=} \;
\xi_{b(f)} \, \Delta \varphi_{f}  \, \xi^{-1}_{b(f)}
\;\stackrel{\mathrm{f.f.}}{=} \;
\xi_{b(f)} \, \epsilon_{\partial f} \, \xi^{-1}_{b(f)}
\; = \;
\epsilon'_{\partial f}.
\end{eqnarray}
Thus they define a left action of the group $\mathcal E_{X}^{(0)}$ of all collections $\bm{\xi}$ (with vertex-wise multiplication, see~figure~\ref{fig:gauge_triangle}) on the set of field configurations. All~transformations in this group will be regarded as gauge redundancies. 

Secondly, an \textbf{edge transformation} is parametrized by a collection $\bm \psi = \{ \psi_e \}$ of elements of~$\Phi$. It changes $\bm \epsilon$ according to
\begin{equation}
\epsilon_e' = \Delta \psi_e \, \epsilon_e.
\label{eq:eps_psi_trans}
\end{equation}
Before we give the transformation law for $\bm \varphi$, let us inspect how $\epsilon_{\gamma}$ changes for general~$\gamma$. We observe that $\overline \epsilon_e' = \overline \epsilon_e$ implies that $\overline \epsilon_{\gamma}' = \overline \epsilon_{\gamma} $ for any path $\gamma$. Thus
\begin{equation}
\epsilon_{\gamma}' = \Delta \psi_{\gamma}^{(\epsilon)} \, \epsilon_{\gamma}
\label{eq:general_eps_transformation}
\end{equation}
for some $\psi_{\gamma}^{(\epsilon)}$, which depends on $\bm \psi$ as well as on $\bm \epsilon$. This equation determines each $\psi_{\gamma}^{(\epsilon)}$ up to multiplication by an element of $\ker(\Delta)$. As a step towards an unambiguous definition, we consider a composite path $\gamma \gamma'$ and evaluate $\epsilon'_{\gamma \gamma'}$ in two different ways. Firstly, by equation \eqref{eq:general_eps_transformation}, it is equal to $\Delta \psi_{\gamma \gamma'}^{(\epsilon)} \, \epsilon_{\gamma \gamma'}$. On the other hand we have $\epsilon_{\gamma \gamma'}'= \epsilon_{\gamma}'\, \epsilon_{\gamma'}'$. Applying \eqref{eq:general_eps_transformation} to the two terms separately we obtain
\begin{equation}
\epsilon_{\gamma \gamma'}' =  \Delta \psi_{\gamma}^{(\epsilon)} \, \underbrace{\epsilon_{\gamma}  \, \Delta \psi_{\gamma'}^{(\epsilon)} \, \epsilon_{\gamma}^{-1}}_{\text{Peiffer}} \, \epsilon_{\gamma}\, \epsilon_{\gamma'} = \Delta \left( \psi_{\gamma}^{(\epsilon)} \, \left( \epsilon_{\gamma} \rhd \psi_{\gamma'}^{(\epsilon)} \right) \right) \, \epsilon_{\gamma \gamma'}.
\end{equation}
Comparison of the two results yields
\begin{equation}
\Delta \psi_{\gamma \gamma'}^{(\epsilon)} \, \epsilon_{\gamma \gamma'} = \Delta \left( \psi_{\gamma}^{(\epsilon)} \, \left( \epsilon_{\gamma} \rhd \psi_{\gamma'}^{(\epsilon)} \right) \right) \, \epsilon_{\gamma \gamma'}.
\end{equation}
This formula has the consequence that, perhaps up to multiplication of the right hand side by an element of $\ker(\Delta)$,
\begin{equation}
\psi_{\gamma \gamma'}^{(\epsilon)} = \psi_{\gamma}^{(\epsilon)} \, \left( \epsilon_{\gamma} \rhd \psi_{\gamma'}^{(\epsilon)} \right).
\label{eq:psi_comp_path}
\end{equation}
It is convenient to \textit{define} $\psi_{\gamma}^{(\epsilon)}$ for general paths $\gamma$ by demanding that this relation is satisfied exactly (rather than merely up to elements from $\ker(\Delta)$) and that $\psi_{e}^{(\epsilon)} = \psi_e$ for any edge $e$. Freeness of the grupoid $\pi_1(X_1;X_0)$ guarantees that this definition is well-posed. More explicitly, for a path $\gamma = e_n e_{n-1} \ldots e_1$ it gives
\begin{equation}
\psi_{\gamma}^{(\epsilon)} = \psi_{e_n} \, \left( \epsilon_{e_n} \rhd \psi_{e_{n-1}} \right)\, \ldots \, \left( \epsilon_{e_n} \ldots \epsilon_{e_2} \rhd \psi_{e_1} \right).
\end{equation}
By induction on $n$, formulas \eqref{eq:eps_psi_trans} and \eqref{eq:psi_comp_path} imply that with this definition of $\psi_{\gamma}^{(\epsilon)}$, transformation law \eqref{eq:general_eps_transformation} is indeed satisfied for any $\gamma$. We also note that the composition rule \eqref{eq:psi_comp_path} yields the inversion formula
\begin{equation}
\psi_{\gamma^{-1}}^{(\epsilon)} = \left( \epsilon_{\gamma}^{-1} \rhd \psi_{\gamma}^{(\epsilon)} \right)^{-1}.
\end{equation}

Let us now return to the problem of defining an action of edge transformations on $\bm \varphi$. The guiding principle is the preservation of the fake flatness condition. Thus we must have
\begin{equation}
\Delta \varphi_f' = \epsilon_{\partial f}' = \Delta \psi_{\partial f}^{(\epsilon)} \, \epsilon_{\partial f}= \Delta \left( \psi_{\partial f}^{(\epsilon)} \, \varphi_f \right) .
\end{equation}
The simplest way to satisfy this condition is to declare
\begin{equation}
\varphi_f'= \psi_{\partial f}^{(\epsilon)} \, \varphi_f.
\label{eq:edge_phi_trans}
\end{equation}

We illustrate the above definitions by considering a field configuration on the geometry depictured on figure \ref{fig:double}. Such configuration consists of five elements $\epsilon_{e_i} \in \mathcal E$ and two $\varphi_{f_i} \in \Phi$. Vertex transformation given by the collection $\{ \psi_{e_i} \}$ maps $\epsilon_{e_i}$ to $\Delta \psi_{e_i} \, \epsilon_{e_i}$. Action on $\varphi$ variables is given by
\begin{subequations}
\begin{align}
\varphi_{f_1} & \mapsto \psi_{e_4} \, \left( \epsilon_{e_4} \rhd \psi_{e_5} \right) \, \left( \epsilon_{e_4} \epsilon_{e_5} \rhd \psi_{e_1} \right) \, \varphi_{f_1}, \\
\varphi_{f_2} & \mapsto \underbrace{\left( \epsilon_{e_5}^{-1} \rhd \psi_{e_5}^{-1} \right)}_{\psi_{e_5^{-1}}} \, \left( \epsilon_{e_5}^{-1} \rhd \psi_{e_3} \right) \, \left( \epsilon_{e_5}^{-1} \epsilon_{e_3} \rhd \psi_{e_2} \right) \, \varphi_{f_2}.
\end{align}
\end{subequations}

Definition \eqref{eq:edge_phi_trans} implies the following transformation law for $\varphi_{\sigma}$ for arbitrary $\sigma$:
\begin{equation}
\varphi_{\sigma}' = \psi^{(\epsilon)}_{\partial \sigma} \, \varphi_{\sigma}.
\label{eq:2hol_trans}
\end{equation}
This can be proven as follows. First we note that, by definition, it holds for $\sigma=f$ for any face $f$. Secondly, if $\sigma$ and $\sigma'$ share the base point and are such that \eqref{eq:2hol_trans} holds, then the same is true for the product $\sigma \sigma'$:
\begin{equation}
\begin{split}
\varphi_{\sigma}\, \varphi_{\sigma'}\xmapsto{\hspace{20pt}}&\, \psi_{\partial\sigma}^{(\epsilon)}\,\varphi_\sigma \, \psi_{\partial\sigma'}^{(\epsilon)}\, \varphi_{\sigma'}=\psi^{(\epsilon)}_{\partial(\sigma\sigma')}\,\left(\epsilon_{\partial\sigma}\rhd {\psi_{\partial \sigma'}^{(\epsilon)}}\right)^{-1}\, \varphi_\sigma\, \psi_{\partial\sigma'}^{(\epsilon)}\,\varphi_{\sigma'}\\
=\,& \psi^{(\epsilon)}_{\partial(\sigma\sigma')}\, \left(\epsilon_{\partial\sigma}\rhd {\psi_{\partial \sigma'}^{(\epsilon)}}\right)^{-1}\, \varphi_\sigma\,\psi_{\partial\sigma'}^{(\epsilon)}\, \varphi_\sigma^{-1}\, \varphi_{\sigma\sigma'}\\
=\,&\psi^{(\epsilon)}_{\partial(\sigma\sigma')}\, \left(\epsilon_{\partial\sigma}\rhd {\psi_{\partial \sigma'}^{(\epsilon)}}\right)^{-1}\, \left(\Delta\varphi_\sigma\rhd \psi^{(\epsilon)}_{\partial\sigma'}\right)\, \varphi_{\sigma\sigma'}\\
\stackrel{\mathrm{f.f.}}{=}\,&\psi^{(\epsilon)}_{\partial(\sigma\sigma')}\, \left(\epsilon_{\partial\sigma}\rhd {\psi_{\partial \sigma'}^{(\epsilon)}}\right)^{-1}\, \left(\epsilon_{\partial \sigma}\rhd \psi^{(\epsilon)}_{\partial\sigma'}\right)\, \varphi_{\sigma\sigma'}=\psi^{(\epsilon)}_{\partial(\sigma\sigma')}\, \varphi_{\sigma\sigma'}.
\end{split}
\end{equation}
Next we show that if $\sigma$ is such that \eqref{eq:2hol_trans} holds and $b(\sigma) = s(\gamma)$, then \eqref{eq:2hol_trans} holds also for $\gamma \rhd \sigma$. Indeed, in this situation we have
\begin{equation}
\varphi_\sigma\xmapsto{\hspace{20pt}} \psi_{\partial\sigma}^{(\epsilon)}\, \varphi_\sigma, \qquad \epsilon_\gamma\xmapsto{\hspace{20pt}} \Delta\psi_\gamma^{(\epsilon)}\, \epsilon_\gamma.
\end{equation}
Since $\partial(\gamma\rhd \sigma)=\gamma\, \partial\sigma \,\gamma^{-1}$, we have to check that
\begin{equation}
\epsilon_\gamma\rhd \varphi_\sigma \xmapsto{\hspace{20pt}}\psi^{(\epsilon)}_{\gamma\, \partial\sigma \, \gamma^{-1}}\, \left(\epsilon_{\gamma}\rhd \varphi_{\sigma}\right).
\end{equation}
This is indeed the case, since $\epsilon_\gamma\rhd \varphi_\sigma$ transforms as:
\begin{equation}
\begin{split}
\epsilon_\gamma\rhd \varphi_\sigma \xmapsto{\hspace{20pt}}&\left(\Delta\psi_\gamma^{(\epsilon)}\, \epsilon_\gamma\right)\rhd \left(\psi_{\partial\sigma}\, \varphi_\sigma\right)=\psi_\gamma^{(\epsilon)}\, \left(\epsilon_\gamma\rhd \left(\psi^{(\epsilon)}_{\partial\sigma}\,\varphi_\sigma\right)\right)\, {\psi_{\gamma}^{(\epsilon)}}^{-1}\\
=\,&\psi_{\gamma}^{(\epsilon)}\, \left(\epsilon_\gamma\rhd \psi^{(\epsilon)}_{\partial\sigma}\right)\, \left(\epsilon_\gamma\rhd \varphi_\sigma\right)\, {\psi_\gamma^{(\epsilon)}}^{-1}\, \left(\epsilon_\gamma\rhd \varphi_\sigma\right)^{-1}\, \left(\epsilon_\gamma\rhd \varphi_\sigma\right)\\
=\,&\psi^{(\epsilon)}_{\gamma\,\partial\sigma}\, \left(\epsilon_\gamma\rhd \varphi_\sigma\right)\, {\psi_\gamma^{(\epsilon)}}^{-1}\, \left(\epsilon_\gamma\rhd \varphi_\sigma\right)^{-1}\, \left(\epsilon_\gamma\rhd \varphi_\sigma\right)\\
=\,&\psi^{(\epsilon)}_{\gamma\,\partial\sigma}\,\left(\Delta\left(\epsilon_\gamma\rhd\varphi_\sigma\right)\rhd{\psi^{(\epsilon)}_\gamma}^{-1}\right)\, \left(\epsilon_\gamma\rhd \varphi_\sigma\right)\\
=\,&\psi^{(\epsilon)}_{\gamma\,\partial\sigma}\, \left(\left(\epsilon_\gamma\, \Delta\varphi_\sigma\, \epsilon_\gamma^{-1}\right)\rhd {\psi_\gamma^{(\epsilon)}}^{-1}\right) \, \left(\epsilon_\gamma\rhd \varphi_\sigma\right)\\
=\,&\psi^{(\epsilon)}_{\gamma\,\partial\sigma}\, \left(\epsilon_{\gamma\, \partial\sigma \, \gamma^{-1}}\rhd{\psi_\gamma^{(\epsilon)}}^{-1}\right)\,\left(\epsilon_\gamma\rhd \varphi_\sigma\right)=\psi^{(\epsilon)}_{\gamma\, \partial\sigma \, \gamma^{-1}}\, \left(\epsilon_{\gamma}\rhd \varphi_{\sigma}\right).
\end{split}
\end{equation}
This concludes the proof, since any $\sigma$ may be written as a product of some number of elements of the form $\gamma \rhd f$.

The following special case of the above result is worth to be mentioned separately: if $\sigma$ has trivial boundary ($\partial \sigma=1$), then $\varphi_{\sigma}$ is invariant with respect to edge transformations.

\begin{figure}[ht]
\begin{minipage}[c]{0.55\textwidth}
\centering
\begin{tikzcd}[column sep=2cm, row sep=2cm](\bm \epsilon,\bm \varphi)\arrow[d,"\hspace{-25pt}\{\psi_e\}"]\arrow[rd, "\{\psi_e'\,\psi_e\}"]& \\
(\bm{\epsilon'},\bm{\varphi'})\arrow[r, "\{\psi_e' \}"]& (\bm{\epsilon''},\bm{\varphi''})
\end{tikzcd}
\end{minipage}\hfill
 \begin{minipage}[c]{0.45\textwidth}
    \caption{
       Edge transformation $\{ \psi_e \}$ followed by $\{ \psi'_e \}$ coincides with $\{ \psi_e'\, \psi_e \}$.
    } \label{fig:gauge_triangle_2}
  \end{minipage}
\end{figure}
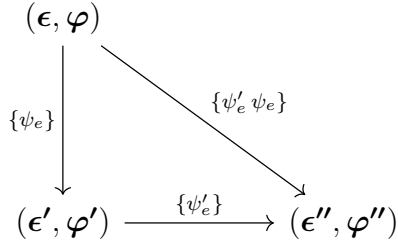

Edge transformations form a group $\Phi^{(1)}_X$, with composition computed edge-wise (see~figure \ref{fig:gauge_triangle_2}): transformation $\{ \psi_e \}$ followed by $\{ \psi_e' \}$ coincides with $\{ \psi_e'' \}$, where $\psi''_e = \psi'_e \,\psi_e$. Indeed, for configurations as on figure \ref{fig:gauge_triangle_2} we have
\begin{subequations}
\begin{align}
\epsilon''_e &= \Delta \psi_{e}' \, \epsilon'_e = \Delta \psi'_e \, \Delta \psi_e \, \epsilon_e = \Delta \psi''_e \, \epsilon_e, \\
\varphi''_f &= \psi'^{(\epsilon')}_{\partial f} \, \varphi'_f = \psi'^{(\epsilon')}_{\partial f} \, \psi^{(\epsilon)}_{\partial f} \, \varphi_f.
\end{align}
\end{subequations}
Thus in order to prove the claimed composition law it only remains to show that $\psi''^{(\epsilon)}_{\partial f} = \psi'^{(\epsilon')}_{\partial f} \, \psi^{(\epsilon)}_{\partial f} $. In fact even more is true: for any path $\gamma$ we have
\begin{equation}
\psi''^{(\epsilon)}_{\gamma} = \psi'^{(\epsilon')}_{\gamma} \, \psi^{(\epsilon)}_{\gamma}
\end{equation}
Indeed, by the induction principle, it is sufficient to demonstrate that the above equality is satisfied for a composite path $\gamma \gamma'$ provided that it holds for $\gamma$ and $\gamma'$ separately. To this end we use \eqref{eq:psi_comp_path} and apply the inductive hypothesis:
\begin{align}
\psi''^{(\epsilon)}_{\gamma \gamma'} &= \psi''^{(\epsilon)}_{\gamma} \, \left( \epsilon_{\gamma} \rhd  \psi''^{(\epsilon)}_{\gamma'} \right) = \psi'^{(\epsilon')}_{\gamma} \, \psi^{(\epsilon)}_{\gamma} \, \left( \epsilon_{\gamma} \rhd \left( \psi'^{(\epsilon')}_{\gamma'} \, \psi^{(\epsilon)}_{\gamma'} \right) \right) \nonumber \\
&= \psi'^{(\epsilon')}_{\gamma} \, \underbrace{\psi^{(\epsilon)}_{\gamma} \, \left( \epsilon_{\gamma} \rhd \psi'^{(\epsilon')}_{\gamma'} \right) \, \psi^{(\epsilon) -1}_{\gamma}}_{\text{Peiffer}} \, \underbrace{\psi^{(\epsilon)}_{\gamma} \, \left( \epsilon_{\gamma} \rhd \psi_{\gamma'}^{(\epsilon)} \right)}_{\psi^{(\epsilon)}_{\gamma \gamma'}} \\
&= \psi'^{(\epsilon')}_{\gamma} \, \left( \left( \Delta \psi^{(\epsilon)}_{\gamma} \, \epsilon_{\gamma} \right) \rhd \psi'^{(\epsilon')}_{\gamma'} \right) \, \psi^{(\epsilon)}_{\gamma \gamma'} = \psi'^{(\epsilon')}_{\gamma \gamma'} \, \psi^{(\epsilon)}_{\gamma \gamma'}. \nonumber
\end{align}

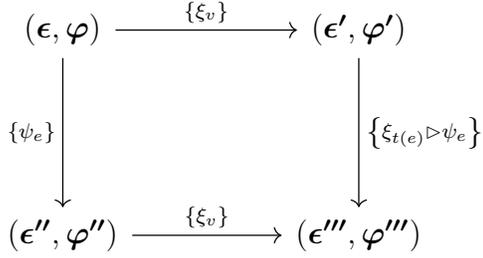
\begin{figure}[ht]
\begin{minipage}[c]{0.5\textwidth}
\centering
\begin{tikzcd}[column sep=2cm, row sep=2cm]
(\bm{\epsilon},\bm{\varphi})\arrow[d,"\hspace{-25pt} \{\psi_e\}"]\arrow[r, "\{\xi_v\}"]&(\bm{\epsilon'},\bm{\varphi'})\arrow[d, "\left\{\xi_{t(e)}\rhd \psi_e\right\}"] \\
(\bm{\epsilon''},\bm{\varphi''})\arrow[r, "\{\xi_v\}"]& (\bm{\epsilon'''},\bm{\varphi'''})
\end{tikzcd}
\end{minipage}\hfill
 \begin{minipage}[c]{0.5\textwidth}
    \caption{Illustration of the semi-direct product structure of the group $\mathcal E_X^{(0)} \ltimes \Phi_X^{(1)}$ generated by vertex and edge transformations.}
     \label{fig:gauge_square}
  \end{minipage}
\end{figure}

We remark also that conjugation of an edge transformation with a vertex transformation gives another edge transformation, see figure \ref{fig:gauge_square}. This means that vertex transformations together with edge transformations form a semi-direct product structure $\mathcal E^{(0)}_{X} \ltimes \Phi^{(1)}_X$.

Next we define \textbf{plaquette transformations}. They are labeled by $\mathrm{ker}(\Delta)$-valued collections $\bm \chi = \{ \chi_f \}$ indexed by faces. The action on fields is given by
\begin{equation}
\epsilon_e' = \epsilon_e, \qquad \varphi_f' = \chi_f \, \varphi_f.
\end{equation}
It is clear that the fake flatness condition is preserved.

As announced at the beginning of this subsection, in topological field theories based on crossed modules all edge transformations are regarded as gauge transformations. We will now list some important consequences of this choice:
\begin{enumerate}
\item Up to a gauge transformation, $\bm \epsilon$ is uniquely determined by $\overline{\bm \epsilon}$. Thus the only gauge invariant functions constructed entirely of $\bm \epsilon$ are the (conjugacy classes~of) holonomies of $\overline{\bm \epsilon}$, which are topological observables.
\item Apart from topological degrees of freedom present in $\bm{\overline \epsilon}$, there remain $\ker(\Delta)$-valued degrees of freedom in $\bm \varphi$. These can be made topological by introducing additional flatness constraint: $\varphi_{\partial q}=1$ for every ball $q$.
\item There exists a space $B\mathbb G$, called the \textbf{classifying space} of $\mathbb G$, with the property that gauge equivalence classes of flat field configurations are in one-to-one correspondence with homotopy classes of maps from $X$ to $B \mathbb G$. In~particular the set of gauge orbits of flat gauge fields is a homotopy invariant of $X$. We~review this in appendix \ref{sec:classifying}.
\item Despite the fact that models under consideration are formulated in terms of the crossed module $\mathbb G$, they depend only on its weak equivalence class. We will obtain this fact as a corollary from considerations in section \ref{sec:vacua}. Furthermore, we give its second, logically independent proof in appendix \ref{appendix:cswecm}.
\end{enumerate}

Here we would like to focus on an alternative possibility and regard only edge transformations with $\psi_e \in \ker(\Delta)$ for each edge $e$ as gauge redundancies. With this definition it is possible to formulate dynamical models with $1$-form and $2$-form gauge fields interacting in an interesting way. Indeed, the conjugacy classes of holonomies of $\bm{\epsilon}$ (rather than merely their reductions modulo $\im(\Delta)$) become gauge invariant. These holonomies are not necessarily trivial for contractible loops, so some non-topological degrees of freedom may be present in the field $\bm{\epsilon}$.

Topological quantum field theories briefly discussed above may still be recovered in a certain limit, by enforcing invariance with respect to all edge transformations and flatness of the $\bm {\varphi}$ field dynamically. Furthermore, two other well-known models may be obtained as special cases:
\begin{itemize}
\item If $\mathcal E$ is taken to be trivial, $\Phi$ can still be any abelian group. In this case one recovers $2$-form lattice gauge theory valued in $\Phi$.
\item Taking $\Phi = \mathcal E$, homomorphism $\Delta$ to be the identity map and the action of $\mathcal E$ on $\Phi$ given by conjugation we recover the standard lattice gauge theory.
\end{itemize}
There are two other special cases which correspond to slight variations of the above:
\begin{itemize}
\item Given any $\mathcal E$ and an abelian group $\Phi$ on which $\mathcal E$ acts one can form a crossed module by letting $\Delta$ be the trivial homomorphism. Then fake flatness implies that $\bm \epsilon$ is flat, so it carries no local gauge-invariant degrees of freedom. The effect of nontrivial holonomies of $\bm \epsilon$ along non-contractible loops may be loosely described as imposing twisted boundary conditions for the field $\bm \varphi$. Models of this type may be obtained from $2$-form gauge theories by gauging a global symmetry of the form $\varphi_f \mapsto \epsilon \rhd \varphi_f$.
\item Taking a crossed module with injective $\Delta$ one obtains lattice gauge theory with gauge group $\mathcal E$ in which the curvature is constrained to be valued in the normal subgroup $\mathrm{im}(\Delta) \cong \Phi$.
\end{itemize}

We have shown that proposed models unify and at the same time generalize several interesting classes of gauge theories involving $1$-form and $2$-form gauge fields, which provides compelling motivation to study them.

We close this subsection with a technical lemma, to be used later. Consider the effect of a vertex transformation $\bm \xi$ valued in $\mathrm{im}(\Delta)$ on a configuration $(\bm \epsilon, \bm \varphi)$. For~each vertex $v$ we write $\xi_v = \Delta (\rho_v)$ for some $\rho_v \in \Phi$. Then we have
\begin{eqnarray}
\nonumber
\epsilon'_e
& = &
\Delta(\rho_{t(e)})\, \epsilon_e \, \,\Delta(\rho_{s(e)}^{-1})
\; = \;
\Delta(\rho_{t(e)}) \, \overbrace{\epsilon_e \,\,\Delta(\rho_{s(e)}^{-1})\, \epsilon_e^{-1}}^{\mathrm{Peiffer}}\, \epsilon_e
\\
& = &
\Delta(\rho_{t(e)})\,\Delta\big(\epsilon_e\rhd \rho^{-1}_{s(e)}\big)\,\epsilon_e
\; = \;
\Delta(\psi_e)\,\epsilon_e,
\label{eq:G_im_step}
\end{eqnarray}
where $\psi_e = \rho_{t(e)} \, \left( \epsilon_e \rhd \rho_{s(e)}^{-1} \right)$. We claim that one also has $
\varphi_f' = \psi_{\partial_f}^{(\epsilon)} \, \varphi_f$, so~the pertinent vertex transformation is equivalent to an edge transformation with some $\bm \psi$ depending on $\bm \rho$ and $\bm \epsilon$. Indeed, first observe that we have
\begin{align}
\varphi_f' &= \xi_{b(f)} \rhd \varphi_f = \Delta \rho_{b(f)} \rhd \varphi_f \stackrel{\mathrm{Peiffer}}{=}  \rho_{b(f)} \, \overbrace{\varphi_f \, \rho_{b(f)}^{-1}  \, \varphi_f^{-1}}^{\mathrm{Peiffer}} \, \varphi_f \nonumber \\
 & = \rho_{b(f)} \, \left( \Delta \varphi_f \rhd \rho_{b(f)}^{-1} \right) \, \varphi_f \stackrel{\mathrm{f.f.}}{=} \rho_{b(f)} \, \left( \epsilon_{\partial_f} \rhd \rho_{b(f)}^{-1} \right) \, \varphi_f.
\end{align}
It only remains to show that $\psi_{\partial f}^{(\epsilon)} = \rho_{b(f)} \, \left( \epsilon_{\partial f} \rhd \rho_{b(f)}^{-1} \right)$. In fact even more is true: $\psi^{(\epsilon)}_{\gamma} = \rho_{t(\gamma)} \, \left( \epsilon_{\gamma} \rhd \rho_{s(\gamma)}^{-1} \right)$ for any path $\gamma$. This follows from \eqref{eq:psi_comp_path} and the definition of $\bm \psi$, by induction on the number of edges in $\gamma$.

\subsection{Interesting field configurations - examples} \label{sec:examples}

In this subsection we present examples of field configurations illustrating certain phenomena that will play important roles in the further discussion.

\begin{figure}[ht]
\begin{minipage}[c]{0.4\textwidth}
\centering
\begin{tikzpicture}
\draw
(0,0) coordinate (a1) -- ++(0:2) coordinate (a2)
(a2) -- ++(90:2) coordinate (a3)
(a3) -- ++(180:2) coordinate (a4)
(a4) -- (a1)
;
\draw[very thick, decoration={markings, mark=at position 0.5 with {\arrow{stealth}}},postaction={decorate}](a1) -- node[below] {$a$}(a2);
\draw[very thick, decoration={markings, mark=at position 0.5 with {\arrow{stealth}}},postaction={decorate}](a2) -- node[right] {$b$}(a3);
\draw[very thick, decoration={markings, mark=at position 0.5 with {\arrow{stealth}}},postaction={decorate}](a3) -- node[above] {$a^{-1}$}(a4);
\draw[very thick, decoration={markings, mark=at position 0.5 with {\arrow{stealth}}},postaction={decorate}](a4) -- node[left] {$b^{-1}$}(a1);
\node at (a1) {$\bullet$};
\node at (a2) {$\bullet$};
\node at (a3) {$\bullet$};
\node at (a4) {$\bullet$};
\node[below left] at (a1) {$\ast$};
\node[below right] at (a2) {$\ast$};
\node[above right] at (a3) {$\ast$};
\node[above left] at (a4) {$\ast$};
\node at (1,1) {$f$};
\end{tikzpicture}
\end{minipage}\hfill
 \begin{minipage}[c]{0.6\textwidth}
    \caption{A decomposition of the~$2$-torus. In this case we have $\partial f=b^{-1}a^{-1}ba$.} \label{fig:torus_gauge}
  \end{minipage}
\end{figure}
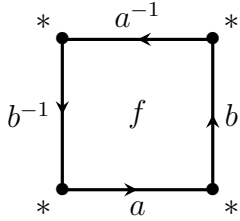

Firstly, we would like to point out that flatness of ${\overline{\bm\epsilon}}$ does not guarantee that one can find a corresponding \textit{flat} $\bm{\epsilon}$. To illustrate this feature we consider the decomposition of a $2$-torus depictured on figure \ref{fig:torus_gauge}.

We take $\mathcal{E}$ to be the dihedral group $D_4$. It is generated by elements $x,y,z$, which are subject to relations
\begin{equation}
 x^2=y^2=z^2=1, \quad  xz=zx, \quad yz=zy, \quad  xy=zyx .
\end{equation}
The group $\Phi$ is taken to be $\{1,z\}$, with $\Delta$ the inclusion map. Action of $\mathcal{E}$ on $\Phi$ is trivial. As a result, $\ker(\Delta)$ is trivial and $\mathrm{coker}(\Delta)=\mathbb{Z}_2\times\mathbb{Z}_2$.

Now consider a configuration with $\overline{\epsilon_a}=\overline{x}$ and $\overline{\epsilon_b}=\overline{y}$. Since $\partial f=b^{-1}a^{-1}ba$ and $\mathrm{coker}(\Delta)$ is abelian, we have $\overline{ \epsilon_{\partial f}}=1$. On the other hand we must have $\epsilon_{a} \in \{ x, zx \}$. Similarly, $\epsilon_b \in \{ y, zy \}$. In each of the four possibilities we get $\epsilon_{\partial f} = z$. Therefore $\bm \epsilon$ cannot be flat for the given $\overline{\bm \epsilon}$.

Next we would like to show that there may exist field configurations which are not gauge equivalent, even though all holonomies coincide\footnote{This is true regardless of the choice whether edge transformations not valued in $\ker(\Delta)$ are regarded as gauge transformations.}. To this end we continue to consider the $2$-torus, but we no longer restrict ourselves to the specific choice of the crossed module.

We choose $\epsilon_a = \epsilon_b = 1$. Notice that this condition is preserved by vertex and edge transformations with arbitrary $\xi_{\ast}$ and $\psi_a$, $\psi_b$. Furthermore, $\varphi_f$ may be any element from the kernel of $\Delta$. Under a gauge transformation, this element changes according to the formula

\begin{equation}
\begin{split}
\varphi_f\mapsto\xi_\ast & \rhd \left(\psi_{b^{-1}a^{-1}ba}\, \varphi_f\right).
\end{split}
\label{eq:varphi_tran_ex}
\end{equation}
Next, since $\ker(\Delta)$ is abelian,
\begin{equation}
\psi_{b^{-1}a^{-1}ba}\stackrel{\epsilon=1}{=}\psi_{b^{-1}}\, \psi_{a^{-1}}\, \psi_b\, \psi_a\stackrel{\epsilon=1}{=}\psi_b^{-1}\, \psi_a^{-1}\, \psi_b\, \psi_a = 1.
\end{equation}
Therefore the formula \eqref{eq:varphi_tran_ex} simplifies to
\begin{equation}
\varphi_f\mapsto \xi_\ast \rhd \varphi_f,
\end{equation}
which does not depend on the choice of $\psi_a$ and $\psi_b$ in $\ker(\Delta)$. This means that configurations with $\bm \epsilon = \bm 1$ and $\varphi_f$ in different orbits of $\mathcal E$ are not related by a gauge transformation. Thus there will be at least two such non-equivalent configurations if $\ker(\Delta)$ is nontrivial. On the other hand, all these configurations have the same values of all holonomies. Indeed, $1$-holonomies are all equal $1$ and there are no nontrivial $2$-holonomies, since the second homotopy group of a torus vanishes. This indicates existence of gauge invariant observables associated to non-spherical surfaces. This is indeed true, but they are slightly tricky to define. We will not consider this problem here. An interesting discussion in the context of state sum formulation of topological higher gauge theories was given in \cite{kapustin17}.


In the final example of this subsection we shall show that some $2$-holonomies may be determined already by $\overline{\bm \epsilon}$. In particular, it may happen that for some $\overline {\bm \epsilon}$ it is not possible to choose $\bm \epsilon$ and $\bm \varphi$ so that $\varphi_{\partial q}=1$ for every ball $q$.

Let us consider the decomposition of the projective plane presented on figure~\ref{fig:rp2}. We take the crossed module with
\begin{equation}
\mathcal{E}=\Phi=\mathbb{Z}_4, \quad \Delta(n)=2n, \quad m\rhd n=(-1)^m n,
\end{equation}
where we use additive notation. In~this case $\ker(\Delta) \cong \mathrm{coker}(\Delta) \cong \mathbb{Z}_2$.

In the present example, fake flatness does not impose any conditions on $\overline {\epsilon_e}$. Thus we can set it to be the nonzero element of $\mathbb Z_2$. Then $\epsilon_e\in\{1,3\}$, so $\epsilon_{\partial f} = \epsilon_{e^2}=2$. Then fake flatness gives $\Delta(\varphi_f)=2$, so $\varphi_f\in\{1,3\}$.

Recall now that the second homotopy group of $\mathbb{RP}^2$ is generated by the element $\sigma=(e\rhd f)\, f^{-1}$. Evaluation of the $2$-holonomy along this generator gives
\begin{equation}
\varphi_\sigma=(\epsilon_e\rhd \varphi_f)-\varphi_f=2\varphi_f=2,
\end{equation}
regardless of which of the two possible values of $\epsilon_e$ and $\varphi_f$ have been chosen. Similarly one can show that if $\overline{\epsilon_e}$ is trivial, then $\varphi_{\sigma}=0$.

One can also embed $\mathbb{RP}^2$ in $\mathbb {RP}^3$ by attaching an additional $3$-cell $q$ along the generator of $\pi_2(\mathbb{RP}^2)$. In other words, we may decompose $\mathbb{RP}^3$ into $\mathbb {RP}^2$ and an extra ball $q$ with $\partial q = \sigma$. Field configurations discussed above make sense also as configurations on $\mathbb{RP}^3$. Hence we see that in this case if $\overline {\bm\epsilon}$ is nontrivial, then $\bm \varphi$ cannot be chosen to be flat. As~reviewed in the appendix \ref{appendix:postnikov}, this phenomenon is controlled by the so-called \textbf{Postnikov class}.


\section{Hamiltonian models}\label{sec:ham}

\subsection{Construction}\label{sec:construction}

In this section we present the proper construction of our models. We work in the hamiltonian formulation of quantum mechanics. As the first step we construct the Hilbert space ${\cal H}$ and local operators resembling electric field operators from the usual gauge theory. Hamiltonian $\mathsf H_{\scriptscriptstyle\mathrm{E}}$ is defined in terms of these electric operators. It may be though of as a kinetic term. The full hamiltonian $\mathsf{H}$ involves also a magnetic term $\mathsf{H}_{\scriptscriptstyle\mathrm{M}}$. Each of $\mathsf H_{\scriptscriptstyle\mathrm{E}}$ and $\mathsf{H}_{\scriptscriptstyle\mathrm{M}}$ is separately solvable (being a sum of commuting local terms), but its action exchanges states which are stationary for the other. Thus the sum is expected to describe interesting dynamics.

Let us consider the Hilbert space $\mathcal{H}_0$ with an orthonormal basis whose elements are labeled by collections $\bm{\epsilon}=\{\epsilon_e\}$, $\bm{\varphi}=\{\varphi_f\}$ of elements of $\mathcal E$ and $\Phi$,
\begin{equation}
{\cal H}_0 \cong \Big(\bigotimes\limits_{e} L^2({\cal E})\Big)\otimes \Big(\bigotimes\limits_{f} L^2(\Phi)\Big).
\end{equation}
The Hilbert space of the constructed model will be the subspace  ${\cal H}  \subset{\cal H}_0$ spanned by those  $\left|\bm{\epsilon},\bm{\varphi}\right\rangle$ for which the fake flatness condition
$\Delta(\varphi_f) = \epsilon_{\partial f}$ is satisfied. This Hilbert space is not the tensor product of local Hilbert spaces associated to edges and faces, but it does admit a basis consisting of product states.

Several interesting classes of operators may be defined on ${\cal H}$:
\begin{itemize}
\item
For a collection $\bm{\xi}=\{\xi_v\}$ of elements of ${\cal E}$ we define $\mathsf G\big(\bm{\xi}\big)$ by
\begin{equation}
\mathsf G\big(\bm{\xi}\big)\left|\bm{\epsilon},\bm{\varphi}\right\rangle = \big|\{\xi_{t(e)}\,\epsilon_e\, \xi_{s(e)}^{-1}\},\{\xi_{b(f)}\rhd\varphi_f\}\big\rangle.
\end{equation}
\item
For a collection $\bm{\psi}=\{\psi_e\}$ of elements of $\Phi$ we let $\mathsf V\big(\bm{\psi}\big)$ be
\begin{equation}
\mathsf V\big(\bm{\psi}\big)\left|\bm{\epsilon},\bm{\varphi}\right\rangle = \left|\{\Delta(\psi_e)\, \epsilon_e\},\{\psi_{\partial f}^{(\epsilon)}\, \varphi_f\}\right\rangle.
\end{equation}
\item
For a collection $\bm{\chi}=\{\chi_f\}$ of elements of $\mathrm{ker}(\Delta)$ we introduce $\mathsf W\big(\bm{\chi}\big)$ by putting
\begin{equation}
\mathsf W\big(\bm{\chi}\big) \left|\bm{\epsilon},\bm{\varphi}\right\rangle = \left|\bm{\epsilon},\{\chi_f\,\varphi_f\}\right\rangle.
\end{equation}
\end{itemize}

Operators $\mathsf G\big(\bm{\xi}\big)$ form a~representation of the group of vertex transformations $\mathcal{E}_X^{(0)}$ on ${\cal H}$. We will call them vertex Gauss' operators. Only elements of ${\cal H}$ which satisfy the vertex Gauss' law, i.e.\ are invariant with respect to the action of all $\mathsf G\big(\bm{\xi}\big)$, will be regarded as physical states.

Operators $\mathsf V\big(\bm{\psi}\big)$ form a representation of the group $\Phi_X^{(1)}$ of edge transformations. As discussed in subsection \ref{sec:gt}, its subgroup $\mathrm{ker}(\Delta)_X^{(1)}$ describes (a part of the) gauge redundancy of the constructed model. Therefore we will call $\mathsf V\big(\bm{\psi}\big)$ with $\psi_e \in \mathrm{ker}(\Delta)$ the edge Gauss' operators. The final requirement for an element of ${ \cal H }$ to be regarded as a~physical state is that it should satisfy the edge Gauss' law, i.e.\ be invariant with respect to the action of all edge Gauss' operators.

We will now construct electric operators associated to edges. These are required to be gauge invariant, i.e.\ to commute with Gauss' operators of both types. Let us denote by $\mathsf V_e(\psi)$ the operator $\mathsf V\big(\{\psi_{e'}\}\big)$
with $\psi_{e'} = \psi$ for $e'= e$ and $\psi_{e'} =1$ otherwise. Recall that $\ker(\Delta)$ is a central subgroup of $\Phi$, so operators $\mathsf V_e(\psi)$ do commute with all edge Gauss' operators. However, they are not invariant with respect to vertex gauge transformations. Instead we have
\begin{equation}
\mathsf G\big(\bm{\xi}\big)\mathsf V_e(\psi)\mathsf G\big(\bm{\xi}\big)^{-1} = \mathsf V_e(\xi_{t(e)}\rhd\psi).
\end{equation}
This means that to obtain a gauge invariant operator it is sufficient to sum $V_e(\psi)$ over $\psi$ with any $\mathcal E$-invariant weight $\mu : \Phi \to \mathbb C$. Explicitly, we define
\begin{equation}
\mathsf V_{e,\mu} = \sum\limits_{\psi\in\Phi}\mu(\psi)\mathsf V_e(\psi).
\end{equation}
This operator commutes with $\mathsf G\big(\bm{\xi}\big)$ provided that $\mu(\xi\rhd\psi) = \mu(\psi)$ for all $\xi \in {\cal E}$.

If $\mu$ and $\mu'$ are two $\mathcal E$-invariant functions, operators $\mathsf V_{e,\mu}$ and $\mathsf V_{e',\mu'}$ commute. This is obvious for $e'\neq e,$ while:
\begin{equation}
\begin{split}
\mathsf V_{e,\mu}\mathsf V_{e,\mu'}&=\sum\limits_{\psi,\psi'\in\Phi}\mu(\psi)\mu'(\psi')\mathsf V_{e}(\psi\psi')
=\sum\limits_{\psi,\psi'\in\Phi}\mu(\psi)\mu'(\psi')\mathsf V_{e}(\underbrace{\psi \psi' \psi^{-1}}_{\psi''})\mathsf V_e(\psi)\\
&=\sum\limits_{\psi,\psi''\in\Phi}\mu(\psi)\mu'\big(\psi^{-1} \psi'' \psi \big)\mathsf V_{e}(\psi'')\mathsf V_e(\psi)=\mathsf V_{e,\mu'}\mathsf V_{e,\mu}.
\end{split}
\label{V:e:alpha"commutativity}
\end{equation}

Operators $\mathsf W\big(\bm{\chi}\big)$ form a representation of the abelian group $\mathrm{ker}(\Delta)^{(2)}_X$ and, in particular, commute with each other. Moreover, they commute with all $\mathsf V\big(\bm{\psi}\big)$. We~will use them to construct electric operators associated to faces.

Let us denote by $\mathsf W_f(\chi)$ the operator $\mathsf W\big(\{\chi_f'\}\big)$
with $\chi_f' = \chi$ for $f' = f$ and $\chi_f' = 1$ for $f'\neq f$. For a function $\nu:\ \mathrm{ker}(\Delta) \to {\mathbb C}$ we put
\begin{equation}
\mathsf W_{f,\nu} = \sum_{\chi \in \mathrm{ker}(\Delta)}\nu(\chi)\mathsf W_f(\chi).
\end{equation}

The gauge transformation law for $\mathsf W_f(\chi)$ operators takes the form
\begin{equation}
\mathsf G\big(\bm{\xi}\big)\mathsf W_f(\chi)\mathsf G\big(\bm{\xi}\big)^{-1} = \mathsf W_f(\xi_{b(f)}\rhd\chi),
\end{equation}
hence $\mathsf W_{f,\nu}$ commutes with $\mathsf G\big({\bm\xi}\big)$ if and only if $\nu(\xi\rhd\chi) = \nu(\chi)$ for all $\xi \in {\cal E}$.


As our candidate for the electric hamiltonian we take
\begin{equation}
\mathsf{H}_{\scriptscriptstyle\mathrm{E}} =\mathsf{H}_V+\mathsf{H}_W, \qquad \text{where} \qquad \mathsf H_V=\sum\limits_{e}\mathsf V_{e,\mu_e}, \qquad \mathsf H_W= \sum\limits_{f} \mathsf W_{f,\nu_f}
\label{electric:Hamiltonian}
\end{equation}
with a priori different functions $\mu_e$ and $\nu_f$ for different edges $e$ and faces $f$, since the spatial lattice is not necessarily assumed to admit any symmetries. By~construction, $\mathsf{H}_{\scriptscriptstyle\mathrm{E}}$ commutes with all Gauss' operators
and thus is a well-defined operator on the physical subspace of ${\cal H}$. In order for $\mathsf{H}_{\scriptscriptstyle\mathrm{E}}$ to be self-adjoint we have to take functions $\mu, \nu$ to satisfy $\mu (\psi^{-1}) = \overline{\mu(\psi)}$ and $\nu(\chi^{-1}) = \overline{\nu(\chi)}$. Furthermore, we would like $\mathsf{H}_{\scriptscriptstyle\mathrm{E}}$ to admit either a unique ground state, or at most a finite number of ground states, dependent only on the topology. This can be achieved by assuming that all functions $\mu_e$ and $\nu_f$ are such that their Fourier transforms vanish at the trivial representation and are positive otherwise\footnote{It can be shown that functions satisfying this condition as well as the required invariance properties always exist.}, as will be demonstrated in subsection \ref{sec:vacua}.

Following a common terminology we shall call operators, which are diagonal in the adapted basis of ${\cal H}$, ``magnetic''. The first important class of operators of this type are those constructed out of $1$-holonomies. Consider a function $\eta : \mathcal E \to \mathbb C$ and a path $\gamma$. We define an operator $\mathsf A_{\gamma, \eta}$ by
\begin{equation}
\mathsf A_{\gamma, \eta} \big|\bm{\epsilon},\bm{\varphi}\big\rangle = \eta(\epsilon_\gamma) \big|\bm{\epsilon},\bm{\varphi}\big\rangle.
\end{equation}
This operator is gauge invariant if and only if the endpoints of $\gamma$ coincide and $\eta$ is a~class function, i.e.\  $\eta( \xi \, \epsilon \, \xi^{-1}) = \eta(\epsilon)$ for any $\xi, \epsilon \in \mathcal E$. Our magnetic hamiltonian will involve only terms $\mathsf A_{\partial f, \eta}$ for faces $f$, as in standard lattice gauge theory. In~this case the function $\eta$ needs to be defined only on the subgroup $\mathrm{im}(\Delta) \subseteq \mathcal E$, since $\epsilon_{\partial f} \in \mathrm{im}(\Delta)$ by fake flatness.

Analogously, let $\theta$ be a complex function on $\Phi$ and let $\sigma \in \pi_2(X_2,X_1;x)$ for some base point $x \in X_0$. We define an operator $\mathsf B_{\sigma, \theta}$ by
\begin{equation}
\mathsf B_{\sigma, \theta} \big|\bm{\epsilon},\bm{\varphi}\big\rangle =\theta(\varphi_\sigma) \big|\bm{\epsilon},\bm{\varphi}\big\rangle.
\end{equation}
Recall that for $\sigma$ with trivial boundary the $2$-holonomy along $\sigma$ is invariant with respect to edge gauge transformations. Thus $\mathsf B_{\sigma, \theta}$ commutes with each $\mathsf V(\bm \psi) $.

We~note also that if $\partial \sigma=1$, function $\theta$ needs to be defined only on $\ker(\Delta)$, since then $\Delta(\varphi_{\sigma})=1$ by fake flatness.

Recall that $2$-holonomies transform nontrivially under vertex transformation. This has the consequence that $\mathsf B_{\sigma, \eta}$ commutes with all $\mathsf G\big(\bm{\xi}\big)$ (and thus defines an~operator
on the physical subspace of ${\cal H}$) provided that $\eta$ satisfies
\begin{equation}
\eta(\xi\rhd\varphi) = \eta(\varphi) \hskip 5mm \hbox{for all} \hskip 4mm \varphi \in \mathrm{ker}(\Delta),\ \xi \in {\cal E}.
\end{equation}


We are ready to propose our candidate for the magnetic hamiltonian:
\begin{equation}
\mathsf{H}_{\scriptscriptstyle\mathrm{M}} =\mathsf H_A +\mathsf{H}_B, \qquad \text{where} \qquad \mathsf H_A=\sum\limits_{f} \mathsf A_{\partial f, \eta_f}, \qquad \mathsf H_B= \sum\limits_{q} \mathsf B_{\partial q, \theta_q}
\label{magnetic:Hamiltonian}
\end{equation}
with a priori different functions $\eta_f$ and $\theta_q$ for different faces $f$ and balls $q$. We shall assume that functions $\eta_f$ and $\theta_q$ are non-negative, with value zero attained only for the neutral element. Thus the magnetic hamiltonian penalizes configurations with nontrivial holonomies along contractible loops and surfaces.

We close this subsection with a brief discussion on how the above construction needs to be modified if all edge transformations (i.e.\ with $\psi_e$ not necessarily in $\ker(\Delta)$) are regarded as gauge transformations. In this case the term in $\mathsf{H}_{\scriptscriptstyle\mathrm{M}}$ involving $\mathsf A$ operators has to be dropped, as it is no longer gauge invariant. Furthermore, all $\mathsf V$ operators in $\mathsf{H}_{\scriptscriptstyle\mathrm{E}}$ may be dropped, since they act trivially on the space of physical states. Thus the hamiltonian reduces to
$\mathsf{H}_B+\mathsf{H}_W$.

\subsection{An explicit example}\label{sec:cubic_example}

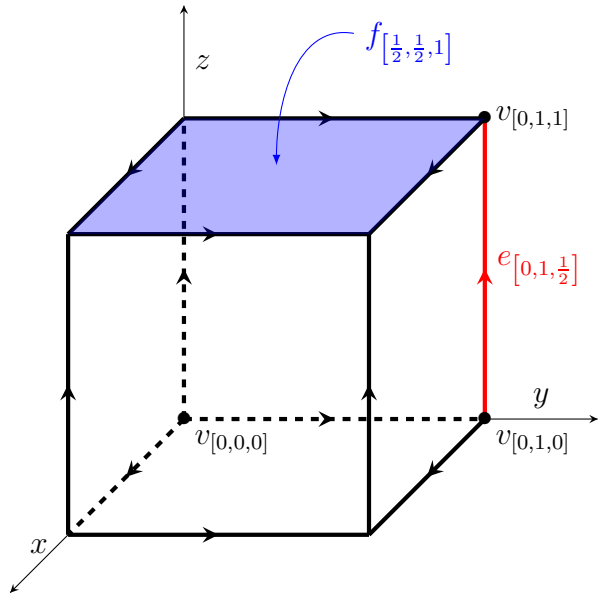
\begin{figure}[ht]
\centering
\begin{tikzpicture}
  \draw[ultra thick, dashed, decoration={markings, mark=at position 0.5 with {\arrow{stealth}}},postaction={decorate}]
  (0,0,0) coordinate (O) -- (4,0,0) coordinate (A1);
  \draw[ultra thick, dashed, decoration={markings, mark=at position 0.5 with {\arrow{stealth}}},postaction={decorate}]
  (O) -- (0,4,0) coordinate (A2);
  \draw[ultra thick, dashed, decoration={markings, mark=at position 0.5 with {\arrow{stealth}}},postaction={decorate}]
  (O) -- (0,0,4) coordinate (A3);
\draw[ultra thick, decoration={markings, mark=at position 0.5 with {\arrow{stealth}}},postaction={decorate},color=red]
  (A1) -- node[right] {{\color{red}$e_{\left[0,1,\frac{1}{2}\right]}$}} (4,4,0) coordinate (A4);
\draw[ultra thick, decoration={markings, mark=at position 0.5 with {\arrow{stealth}}},postaction={decorate}]
  (A1) -- (4,0,4) coordinate (A5);
  \draw[ultra thick, decoration={markings, mark=at position 0.5 with {\arrow{stealth}}},postaction={decorate}]
  (A2) -- (0,4,4) coordinate (A6);
\draw[ultra thick, decoration={markings, mark=at position 0.5 with {\arrow{stealth}}},postaction={decorate}]
  (A3) -- (A6);
\draw[ultra thick, decoration={markings, mark=at position 0.5 with {\arrow{stealth}}},postaction={decorate}]
  (A2) -- (A4);
  \draw[ultra thick, decoration={markings, mark=at position 0.5 with {\arrow{stealth}}},postaction={decorate}]
  (A3) -- (A5);
  \draw[ultra thick, decoration={markings, mark=at position 0.5 with {\arrow{stealth}}},postaction={decorate}]
  (A4) -- (4,4,4) coordinate (A7);
  \draw[ultra thick, decoration={markings, mark=at position 0.5 with {\arrow{stealth}}},postaction={decorate}]
  (A5) -- (A7);
  \draw[ultra thick, decoration={markings, mark=at position 0.5 with {\arrow{stealth}}},postaction={decorate}]
  (A6) -- (A7);
  \draw [decoration={markings, mark=at position 1.0 with {\arrow{stealth}}},postaction={decorate}] (A1) -- +(1.5pt,0,0) node [midway,above] {$y$};
  \draw [decoration={markings, mark=at position 1.0 with {\arrow{stealth}}},postaction={decorate}] (A2) -- +(0,1.5pt,0) node [midway,right] {$z$};
  \draw [decoration={markings, mark=at position 1.0 with {\arrow{stealth}}},postaction={decorate}] (A3) -- +(0,0,2pt) node [midway,above] {$x$};
  \node at (O) {$\bullet$};
  \node[below right] at (O) {$v_{\left[0,0,0\right]}$};
  \node at (A1) {$\bullet$};
  \node[below right] at (A1) {$v_{\left[0,1,0\right]}$};
  \node at (A4) {$\bullet$};
  \node[right] at (A4) {$v_{\left[0,1,1\right]}$};

\node (f1) at (3,5,0) {{\color{blue}$f_{\left[\frac{1}{2},\frac{1}{2},1\right]}$}};
\node (f1a) at (2,4,2) {};
\draw[-latex,blue] (f1) to[out=170,in=90] node[midway,font=\scriptsize,above] {} (f1a);
\fill[blue,opacity=0.3] (A2) -- (A6) -- (A7) -- (A4) -- cycle;

\end{tikzpicture}
    \caption{The cubic lattice with chosen orientations.} \label{fig:cube}
\end{figure}

In order to illustrate general features discussed so far, we consider here an~example constructed on a cubic lattice with a particular crossed module chosen. We shall parametrize the set of vertices by ordered triples of integers $[j_1,j_2,j_3]$, edges by ordered triples consisting of two integers and one half-integer, while for faces we use ordered triples consisting of an~integer and two half-integers - see figure \ref{fig:cube}.

Due to the translational symmetry we can restrict attention to one elementary cell. We will introduce the necessary notation and conventions based upon this cell, in order to avoid tedious expressions.

The orientation of edges is chosen as follows:
\begin{eqnarray}
&&
s\big(e_{[{\scriptscriptstyle\frac12},0,0]}\big) = v_{[0,0,0]},
\hskip 5mm
t\big(e_{[{\scriptscriptstyle\frac12},0,0]}\big) = v_{[1,0,0]},
\end{eqnarray}
and similarly for other edges, while faces are oriented so that:
\begin{equation}
\begin{aligned}
\partial f_{[{\scriptscriptstyle\frac12},{\scriptscriptstyle\frac12},0]} & = & e^{-1}_{[0,{\scriptscriptstyle\frac12},0]}\, e^{-1}_{[{\scriptscriptstyle\frac12},1,0]}\, e_{[1,{\scriptscriptstyle\frac12},0]} \, e_{[{\scriptscriptstyle\frac12},0,0]},
\end{aligned}
\end{equation}
and analogously for other faces. We illustrate this on the figure \ref{fig:cube}. The basepoints are chosen so that
$
b\big(f_{[{\scriptscriptstyle\frac12},{\scriptscriptstyle\frac12},0]}\big) =
b\big(f_{[{\scriptscriptstyle\frac12},0,{\scriptscriptstyle\frac12}]}\big) =
b\big(f_{[0, {\scriptscriptstyle\frac12},{\scriptscriptstyle\frac12}]}\big) = v_{[0,0,0]}
$, and so on.

Finally, each $3$-cell $q$ will be parameterized by an ordered triple of half-integers and oriented so that the orientation of
$\partial q_{[{\scriptscriptstyle\frac12},{\scriptscriptstyle\frac12},{\scriptscriptstyle\frac12}]}$
agrees with the orientations of
$f_{[1,{\scriptscriptstyle\frac12},{\scriptscriptstyle\frac12}]}, f_{[{\scriptscriptstyle\frac12},1,{\scriptscriptstyle\frac12}]}$ and $f_{[{\scriptscriptstyle\frac12},{\scriptscriptstyle\frac12},1]}$
and disagrees with
$f_{[0,{\scriptscriptstyle\frac12},{\scriptscriptstyle\frac12}]}, f_{[{\scriptscriptstyle\frac12},0,{\scriptscriptstyle\frac12}]}$ and $f_{[{\scriptscriptstyle\frac12},{\scriptscriptstyle\frac12},0]}$.

Let us now consider the crossed module ${\mathbb G}_{44} = (\Phi,{\cal E},\Delta,\rhd)$
with $\Phi \cong {\cal E} \cong {\mathbb Z}_4,$ $\Delta(n) = 2n$ for $n \in \Phi$ and $m\rhd n = (-1)^m \, n$ for  $n \in \Phi$ and $m \in {\cal E}.$
This is an~example of a crossed module with nontrivial\footnote{In this case $\ker(\Delta)\cong \mathrm{coker}(\Delta)\cong {\mathbb Z}_2$. Moreover, the Postnikov class, whose definition is given in the appendix \ref{appendix:postnikov}, is the nonzero element of $H^3( B \mathbb Z_2, \mathbb Z_2) \cong \mathbb Z_2$.} $\ker(\Delta)$ and $\mathrm{coker}(\Delta)$. Furthermore, ${\cal E}$ acts non-trivially on $\Phi$. However, it is still relatively simple, since $\coker(\Delta)$ is abelian and acts trivially on $\ker(\Delta)$.

Our present goal is to write down an explicit formula for the proposed Hamiltonian for the above system. We shall denote basis states in the Hilbert space ${\cal H}$ as $\left|\hbox{\boldmath $m$},\hbox{\boldmath $n$}\right\rangle$, where $\bm m = \{ m_e \}$ and $\bm n = \{ n_f \}$ are collections of integers modulo four. An operator which for a fixed edge $e$ shifts $m_e \to m_e+1$ will be denoted by ${\mathsf T}_{\hskip -1pt e}$. The definition of ${\mathsf T}_{\hskip -2pt f}$ is analogous. Furthermore, we let:
\begin{equation}
{\mathsf U}_e \left| \bm m, \bm n \right\rangle =  e^{\frac{i\pi m_e}{2}}\left| \bm m, \bm n \right\rangle,
\hskip 1cm
{\mathsf U}_f \left| \bm m, \bm n \right\rangle = e^{\frac{i\pi n_f}{2}}\left| \bm m, \bm n \right\rangle.
\end{equation}
More generally, we define $\mathsf U_{\gamma}$ and $\mathsf U_{\sigma}$ in the self-evident way.

Fake flatness constraint takes the form
\begin{equation}
     \sum_{e \in \partial f} m_e = 2 n_f,
\end{equation}
where summation is taken over all edges contained in the boundary of the face $f$, regardless of orientations. In other words, we restrict attention to states invariant under operators $\mathsf U_f^2 \prod\limits_{e \in \partial f} \mathsf U_e$. Secondly, there is the constraint of invariance with respect to vertex Gauss' operators, which can be written in the form
\begin{equation}
    \mathsf G_v=\left(\prod_{e:\,v=t(e)}\mathsf T_e \right)\left(\prod_{e:\,v=s(e)}\mathsf T_e^3 \right)\prod_{f:\, v=b(f)}\left(\mathsf T_f^2\frac{1-\mathsf U_f^2}{2}+\frac{1+\mathsf U_f^2}{2}\right).
\end{equation}
Finally, there are edge Gauss' operators, which take the form $\prod\limits_{f:\, e\in\partial f}\mathsf T_f^2$.

It can be shown that operators ${\mathsf W}_{f,\nu_f}$ satisfying all conditions given in subsection \ref{sec:construction} are essentially uniquely determined to be of the form
\begin{equation}
{\mathsf W}_{f,\nu_f} = \nu_{f,0}\left( 1 - {\mathsf T}_{\hskip -2pt f}^2\right), \qquad \text{with some } \nu_{f,0}>0.
\end{equation}
On the other hand, the operators $\mathsf V_{e,\mu_e}$ are given by
\begin{equation}
{\mathsf V}_{e,\mu_e} = \mu_{e,1}\left(2-{\mathsf T}_e^2 \prod\limits_{f:\, e \in \partial f}\left({\mathsf T}_f+{\mathsf T}_f^3\right)\right), \qquad \text{with some } \mu_{e,1}>0.
\end{equation}

Imposing further translational invariance of the system we are forced to put all parameters $\nu_{f,0}$ and $\mu_{e,1}$ to be constant, i.e.\ independent of $f$ and $e$, respectively. Therefore we have
\begin{equation}
    \mathsf H_W = \nu \sum_{f} \left( 1 - \mathsf T_f^2 \right), \qquad \mathsf H_V = \mu \sum_e \left( 2 - \mathsf T_e^2 \prod_{f:\,e \in \partial f} \left( \mathsf T_f + \mathsf T_f^3 \right) \right)
\end{equation}
with some $\mu ,\nu >0$.

There is some freedom in the definition of $\mathsf H_A$. One good choice is given by
\begin{equation}
\mathsf H_A = \eta \sum_{f} \left( 2 - \mathsf U_{\partial f} - \mathsf U_{\partial f}^3 \right), \qquad \text{with } \eta >0.
\end{equation}

Next we consider a ball $q$ and denote by $f_1, f_2$ and $f_3$ faces with $b(f_i) = b(q)$ (they are necessarily contained in $\partial q$). The three remaining faces of $q$ will be denoted by $f_j$, with $j=4,5,6$. For each of these three faces we take $e_j$ be the edge such that $s(e_j) = b(q)$ and $t(e_j) = b(f_j)$. With this notation we have:
\begin{equation}
    {\mathsf U}_{\partial q} = \prod\limits_{j=4}^6 {\mathsf U}_{e_j^{-1}\rhd f_j} \prod\limits_{i=1}^3 {\mathsf U}_{f_i}^\dagger,
\end{equation}
where ${\mathsf U}_{e_j^{-1}\rhd f_j}$ can be expressed in terms of elementary $\mathsf U_e$ and $\mathsf U_f$ operators as
\begin{equation}
{\mathsf U}_{e_j^{-1}\rhd f_j} = \mathsf U_f \frac{1 + \mathsf U_e^2}{2} + \mathsf U_f^{3} \frac{1 - \mathsf U_e^2}{2}.
\end{equation}
Hamiltonian $\mathsf H_B$ is essentially uniquely determined to be
\begin{equation}
\mathsf H_B = \theta \sum_q \left( 1 - \mathsf U_{\partial q} \right), \qquad \text{with } \theta >0.
\end{equation}

\subsection{Symmetries} \label{sec:sym}

We will now describe symmetries of our model. Firstly, any field configuration determines a flat gauge field $\overline{\bm \epsilon}$ valued in $\coker(\Delta)$. Such field is described up to gauge transformations by an element $[\bm {\overline \epsilon}] \in \mathrm{Hom}(\pi_1(X;\ast),\coker(\Delta)) {/\hskip -2.5pt/} \coker(\Delta)$, with double slash denoting the quotient with respect to a group action (in this case given by conjugation), and $\ast$ being an arbitrarily chosen base point in $X_0$. One may regard $[\bm {\overline \epsilon}]$ as an essentially classical quantity, because it is unchanged by the action of all operators introduced in our model thus far. In particular, the subspace of ${\cal H}$ spanned by all $| \bm \epsilon, \bm \varphi \rangle$ corresponding to a single $[\bm {\overline \epsilon}]$ is $\mathsf H$-invariant.

Secondly, there exist so-called electric symmetries. In order to introduce them, we~need to define the following two groups: $\mathcal E_0$ is the subgroup of $\mathcal E$ consisting of all elements which commute with the whole $\mathcal E$ and act trivially on $\Phi$, while $\Phi_0$ is the subgroup of $ \ker(\Delta)$ of all elements invariant under the action of whole $\mathcal E$.

Now let $\bm \zeta = \{ \zeta_e \}$ be a collection of elements of $\mathcal E_0$ such that\footnote{$\zeta_{\gamma}$ is defined for arbitrary path $\gamma$ by demanding that $\zeta_{\gamma \gamma'} = \zeta_{\gamma} \zeta_{\gamma'}$ whenever $s(\gamma) = t (\gamma')$.} $\zeta_{\partial f}=1$ for each $f$. We define an operator $\mathsf L_1(\bm \zeta)$ by the formula
\begin{equation}
\mathsf L_1 (\bm \zeta) | \bm \epsilon , \bm \varphi \rangle = | \{ \zeta_e \, \epsilon_e \} , \bm \varphi \rangle.
\label{eq:L1_operator}
\end{equation}
It is straightforward to check that this preserves fake flatness and that $\mathsf L_1(\bm \zeta)$ commutes with $\mathsf H$. Now suppose that $\bm \zeta$ is of the special form $\zeta_e = \lambda_{t(e)} \lambda_{s(e)}^{-1}$ for some collection $\bm \lambda = \{ \lambda_v \}$ valued in $\mathcal E_0$. In this case we have $\mathsf L_1( \bm \zeta ) = \mathsf G(\bm \lambda)$, which acts trivially on physical states.

We conclude from the preceding discussion that operators $\mathsf L_1$ define a~representation of the cohomology group $H^1(X, \mathcal E_0)$ on the space of physical states. This is a $1$-form symmetry with symmetry group $\mathcal E_0$.

Secondly, let $\bm \kappa = \{ \kappa_f \}$ be a collection of elements of $\Phi_0$ such that\footnote{Element $\kappa_{\partial q}$ is defined in terms of the $\kappa_f$ in the same way as $\varphi_{\partial q}$ is defined in terms of $\varphi_f$. It~does not depend on $\bm \epsilon$ because elements $\kappa_f$ are $\mathcal E$-invariant. Similar notations will be used in the remaining part of this subsection without further explanations.} $\kappa_{\partial q}=1$ for every ball $q$. Then we can introduce
\begin{equation}
\mathsf L_2(\bm \kappa) | \bm \epsilon, \bm \varphi \rangle = | \bm \epsilon, \{ \kappa_f \, \varphi_f \} \rangle.
\label{eq:L2_operator}
\end{equation}
Again, this operation preserves fake flatness and commutes with $\mathsf H$. For a collection $\bm \kappa$ of the form $\kappa_f = \omega_{\partial f}$ for some $\bm \omega = \{ \omega_e \}$ valued in $\Phi_0$ we have $\mathsf L_2(\bm \kappa) = \mathsf V(\bm \omega)$. Hence on the space of physical states a representation of $H^2(X, \Phi_0)$ is defined. This is a $2$-form symmetry with symmetry group $\Phi_0$.

The last type of symmetries we discuss here is related with automorphisms of crossed modules. An automorphism of $\mathbb G$ is a homomorphism $(E,F) : \mathbb G \to \mathbb G$ such that $E$ and $F$ are group automorphisms. An automorphism of $\mathbb G$ is said to be inner if it is of the form $E(\epsilon) = \xi \, \epsilon \, \xi^{-1}$, $F(\varphi) = \xi \rhd \varphi$ for some $\xi$ in $\mathcal E$.

Automorphisms of $\mathbb G$ form a group $\mathrm{Aut}(\mathbb G)$, with inner automorphisms being its normal subgroup. The~quotient group is called the group of outer automorphisms and denoted by $\mathrm{Out}(\mathbb G)$. We remark that the name is potentially misleading, because elements of $\mathrm{Out}(\mathbb G)$ are merely equivalence classes of automorphisms (typically there exists no embedding of $\mathrm{Out}(\mathbb G)$ as a subgroup of $\mathrm{Aut}(\mathbb G)$).

Now let $(E,F)$ be an automorphism of $\mathbb G$. We define an operator $\mathsf K(E,F)$ by
\begin{equation}
\mathsf K(E,F) | \bm \epsilon, \bm \varphi \rangle = | \{ E(\epsilon_e) \}, \{ F(\varphi_f) \} \rangle.
\end{equation}
Clearly this defines a representation of $\mathrm{Aut}(\mathbb G)$ on $\mathcal H$. Now let us observe that for an inner automorphism given by an element $\xi \in \mathcal E$ we have
\begin{equation}
\mathsf K(E,F) | \bm \epsilon, \bm \varphi \rangle = | \{ \xi \, \epsilon_e \, \xi^{-1} \} , \{ \xi \rhd \varphi \} \rangle = \mathsf G ( \{ \xi_v \} ) | \bm \epsilon , \bm \varphi \rangle
\end{equation}
with the constant collection $\xi_v = \xi$ for every $v$. As this is a gauge transformation, $\mathsf K(E,F)$ acts trivially on physical states in this case. Hence on the space of physical states a representation of the group $\mathrm{Out}(\mathbb G)$ is defined. Hamiltonian $\mathsf H$ may or may not be invariant under the action of these transformations, depending on the choice of "coupling constant" functions $\{ \mu_e \}$, $\{ \nu_f \}$, $\{ \eta_f \}$ and $\{ \theta_q \}$ in its definition. It is always possible to choose these functions so that whole $\mathrm{Out}(\mathbb G)$ is realized as a global symmetry group.

\subsection{Vacuum states} \label{sec:vacua}

One of the most interesting goals in the study of models described by hamiltonians $\mathsf{H} = \mathsf{H}_{\scriptscriptstyle\mathrm{M}} + \mathsf{H}_{\scriptscriptstyle\mathrm{E}}$ would be to describe their possible phases. We will now make a~small step in this direction by describing the space of ground states of $\mathsf H$ in various limits in which diagonalization can be performed exactly. In each case we have found that the lowest energy subspace:
\begin{itemize}
    \item is the space of states of a certain topological field theory,
    \item admits a basis whose elements are in one-to-one correspondence with homotopy classes of maps from $X$ to some other space.
\end{itemize}
These results are summarized in the table \ref{tab:models}. All proofs are given in the remainder of this section. For each hamiltonian we provide a more explicit description of the basis ground states, not involving classifying spaces.

We~speculate that some features found in the discussed limits may be generic for certain regions in the phase diagram of our model:
\begin{itemize}
    \item Ground states of $\mathsf{H}_{\scriptscriptstyle\mathrm{E}}$ are characterized by strong fluctuations of holonomies. Similar behaviour is expected to be exhibited also by ground states of the full hamiltonian in the regime in which $\mathsf{H}_{\scriptscriptstyle\mathrm{E}}$ dominates over $\mathsf{H}_{\scriptscriptstyle\mathrm{M}}$. Such phase, if it indeed exists, would likely be characterized by an area law for $1$-holonomies and a volume law for $2$-holonomies.
    \item The putative phase approximately described by ground states of $\mathsf H_{AW}$ would be characterized by a perimeter law for $1$-holonomies and a volume law for $2$-holonomies.
\item For ground states of $\mathsf{H}_{\scriptscriptstyle\mathrm{M}}$ slightly perturbed by the electric hamiltonian $\mathsf{H}_{\scriptscriptstyle\mathrm{E}}$, we~expect a perimer law for $1$-holonomies and an area law for $2$-holonomies.
\item In a phase continuously connected to dynamics of $\mathsf H_{BV}$ we expect an area law for $1$-holonomies as well as for $2$-holonomies.
\end{itemize}
This will be further corroborated by the discussion in subsection \ref{sec:dyn}, where we consider certain still simple, but already not purely topological limits of our model. They are shown to reduce to more standard purely $1$-form or $2$-form gauge theories, which are believed to exhibit behaviour consistent with the description above.

\begin{table}[ht!]
\begin{center}
\begin{tabular}{ |c||c|}
 \hline
 Hamiltonian of the model & Basis of ground states\\
 \hline
\hline
&\\
$\mathsf H_E=\mathsf H_V+\mathsf H_W$ & $\left[X,B\mathrm{coker}(\Delta)\right]$ \\
 &\\
 \hline
 &\\
 $\mathsf H_{AW}=\mathsf H_A+\mathsf H_W$ & $\left[X,B\mathcal{E}\right]$\\

 &\\
 \hline
 &\\
 $\mathsf H_M=\mathsf H_A+\mathsf H_B$ & $\left[X,B\mathbb G'\right]$ \\
 &\\
 \hline
 &\\
  $\mathsf H_{BV}=\mathsf H_B+\mathsf H_V$ & $\left[X,B\mathbb G\right]$\\
 &\\
 \hline
\end{tabular}
\end{center}
\caption{The ground states for four models described by integrable hamiltonians containing two out of four terms of $\mathsf H$. Here $[X,Y]$ is the set of homotopy classes of maps $X \to Y$. For the first two entries, the relevant spaces $Y$ are classifying spaces of groups. In the last two entries classifying spaces of crossed modules are meant. $\mathbb G'$ is the crossed module consisting of the trivial homomorphism $\ker(\Delta) \to \mathcal E$ and action of $\mathcal E$ on $\ker(\Delta)$ inherited from the crossed module $\mathbb G$.}
\label{tab:models}
\end{table}

\newpage
Let us start by considering the ground states of $\mathsf{H}_{\scriptscriptstyle\mathrm{E}}$. We will minimize every term in \eqref{electric:Hamiltonian} at the same time, which clearly minimizes the whole $\mathsf{H}_{\scriptscriptstyle\mathrm{E}}$. First let us observe that for every face $f$ we have a representation of the group $\ker (\Delta)$ by operators $\mathsf W_f(\chi)$. It is possible to diagonalize all of them at the same time. Eigenvectors are labeled by element $\widehat {\chi}$ of the Pontryagin dual of $\ker(\Delta)$, i.e.\ the group $\widehat{\ker(\Delta)}$ of homomorphisms $\ker(\Delta) \to \mathrm U(1)$. Such eigenvector, here labeled by $| \widehat \chi \rangle$, satisfies
\begin{equation}
\mathsf W_f(\chi) | \widehat \chi \rangle = \widehat \chi (\chi) | \widehat \chi \rangle \qquad \text{for every } \chi \in \mathrm{ker}(\Delta).
\end{equation}
It then follows that we have
\begin{equation}
\mathsf W_{f, \nu} | \widehat \chi \rangle = \left( \sum_{\chi} \nu(\chi) \widehat \chi (\chi) \right) | \widehat \chi \rangle.
\end{equation}
The quantity in the parenthesis is, by definition, $\widehat \nu (\widehat \chi)$ - the Fourier transform of $\nu$ evaluated at $\widehat \chi$. We have assumed that functions $\nu_f$ defining terms $\mathsf W_{f, \nu_f}$ in $\mathsf{H}_{\scriptscriptstyle\mathrm{E}}$ are such that $\widehat {\nu_f}(\widehat \chi) \geq 0 $, with the equality if and only if $\widehat \chi = 1$. Then zero is the smallest eigenvalue of $\mathsf W_{f, \nu_f}$ and one has $\mathsf W_{f, \nu_f} | \widehat \chi \rangle =0$ only for $\widehat \chi = 1$. This means that vectors minimizing $\mathsf{H}_{\scriptscriptstyle\mathrm{E}}$ are invariant with respect to all $\mathsf W( \bm \chi)$. An analogous analysis, involving Fourier analysis for the non-abelian group (see e.g. \cite[Part II]{terras}) $\Phi$~instead of Pontryagin duality, shows that they have to be invariant also with respect to all $\mathsf V(\bm \psi)$. Finally, we require also invariance with respect to Gauss' operators $\mathsf G(\bm \xi)$. States satisfying all these requirements may be obtained by summing $| \bm \epsilon , \bm \varphi \rangle$ over an orbit of the group generated by all vertex, edge and plaquette transformations, say
\begin{equation}
\sum_{\bm \xi , \bm \psi, \bm \chi} \mathsf G(\bm \xi) \mathsf V(\bm \psi) \mathsf W(\bm \chi) | \bm \epsilon , \bm \varphi \rangle
\end{equation}
for some reference $(\bm \epsilon, \bm \varphi)$. The next step is to understand the space of orbits.

It is clear that two configurations with the same $\bm \epsilon$ are related by a plaquette transformation. Furthermore, for two configurations with the same $\bm {\overline \epsilon}$ one can perform an edge transformation to make $\bm \epsilon$ equal. Collection $\bm {\overline \epsilon}$ itself is not changed by edge and plaquette transformations, but it transforms with respect to vertex transformations in the way usual for a gauge field.

We conclude that there is a basis of ground states of $\mathsf{H}_{\scriptscriptstyle\mathrm{E}}$ indexed by elements of $\mathrm{Hom}(\pi_1(X;\ast), \coker(\Delta)) {/\hskip -2.5pt/} \coker(\Delta)$. Distinct ground states may be distinguished by values of $1$-holonomies along nontrivial loops in $X$. In other words, we have found the space of states of a topological gauge theory with gauge group $\coker(\Delta)$.

Secondly, we discuss the space of ground states of $\mathsf H_{AW}$. It~admits a~basis consisting of vectors of the form
\begin{equation}
\sum_{\bm \xi, \bm \varphi} \mathsf G (\bm \xi) | \bm \epsilon, \bm \varphi \rangle,
\label{eq:H2_gs}
\end{equation}
where $\bm \epsilon$ is any collection with $\epsilon_{\partial f}=1$ for each $f$. The sum over $\bm \varphi$ runs over collections with $\varphi_f$ in $\ker(\Delta)$, by fake flatness. Distinct vectors of the form \eqref{eq:H2_gs} are labeled by elements of $\mathrm{Hom}(\pi_1(X;\ast), \mathcal E) {/\hskip -2.5pt/} \mathcal E$ determined by $\bm \epsilon$. Hence we find the space of states of a topological gauge theory with gauge group $\mathcal E$.

Next we consider the ground states of $\mathsf{H}_{\scriptscriptstyle\mathrm{M}}$. This is facilitated by the fact that holonomy operators act diagonally. To minimize all terms in \eqref{magnetic:Hamiltonian} at the same time we have to restrict attention to configurations $(\bm \epsilon, \bm \varphi)$ satisfying flatness conditions
\begin{subequations}
\begin{align}
\epsilon_{\partial f} &=1 \qquad \text{for every face } f, \\
\varphi_{\partial q} &= 1 \qquad \text{for every ball } q.  \label{eq:phi_flatness}
\end{align}
\label{eq:magnetic_constraints}
\end{subequations}
The first condition implies that each $\varphi_f$ is in $\ker(\Delta)$, by fake flatness. Besides these constraints, only gauge invariant states are allowed. Such state may be constructed by summing over the gauge orbit of some configuration $(\bm \epsilon, \bm \varphi)$ satisfying \eqref{eq:magnetic_constraints}:
\begin{equation}
| [ \bm \epsilon, \bm \varphi ] \rangle = \sum_{(\bm \epsilon', \bm \varphi') \sim (\bm \epsilon, \bm \varphi)} | \bm \epsilon', \bm \varphi' \rangle,
\end{equation}
where we write $(\bm \epsilon', \bm \varphi') \sim (\bm \epsilon, \bm \varphi)$ if $(\bm \epsilon', \bm \varphi')$ and $(\bm \epsilon, \bm \varphi)$ are related by a gauge transformation. Thus there is a basis of the space of ground states whose elements are in one-to-one correspondence with gauge orbits of configurations $(\bm \epsilon, \bm \varphi)$ subject to conditions (\ref{eq:magnetic_constraints}). We will now describe this space of orbits.

An admissible collection $\bm \epsilon$ defines a flat gauge field on $X$ valued in $\mathcal E$. For every conjugacy class of homomorphisms $\pi_1(X;\ast) \to \mathcal E$ we focus on one representative $\bm \epsilon$. Having fixed $\bm \epsilon$, we consider the allowed $\bm \varphi$. They have to satisfy \eqref{eq:phi_flatness}. Furthermore, we have to identify collections related by
\begin{equation}
\varphi_f' = \psi_{\partial f}^{(\epsilon)} \, \varphi_f
\end{equation}
for any collection $\bm \psi$ of elements of $\ker(\Delta)$. We note the fact that elements $\psi_{\partial f}^{(\epsilon)}$ actually depend on $\bm \epsilon$ only through $\overline{\bm \epsilon}$, since the image of $\Delta$ acts trivially on $\ker(\Delta)$.

The space of equivalence classes of admissible collections $\bm \varphi$ is the twisted cohomology group $H^2(X, \ker(\Delta), \overline{\bm \epsilon} )$, as recalled in the apppendix \ref{sec:twisted}. It is not true in general that distinct cohomology classes correspond to different gauge orbits. This is because there might exist vertex transformations $\bm \xi$ which preserve $\bm \epsilon$:
\begin{equation}
\xi_{t(e)} \, \epsilon_e \, \xi_{s(e)}^{-1} = \epsilon_e \qquad \text{for each edge } e.
\label{eq:vertex_stab}
\end{equation}
This formula implies that $\xi_\ast$ commutes with $\epsilon_{\gamma}$ for every loop based at $\ast$. Secondly, given such $\xi_\ast$ it is possible to uniquely determine $\xi_v$ for every vertex $v$ from the above relation. In summary, the group $\mathrm{Stab}_V(\bm \epsilon)$ of vertex transformations preserving $\bm \epsilon$ is isomorphic to the group of elements of $\mathcal E$ whose adjoint action preserves the homomorphism $ \pi_1(X,\ast) \to \mathcal E$ determined by $\bm \epsilon$. It acts on $H^2(X, \ker(\Delta), \overline{\bm \epsilon})$ by (abelian) group homomorphisms, so the quotient space is also a group.

We conclude that the set of gauge orbits of flat configurations is a disjoint union of groups $H^2(X, \ker(\Delta), \overline{\bm \epsilon}) {/\hskip -2.5pt/} \mathrm{Stab}_V(\bm \epsilon)$, with $\bm \epsilon$ running through a set of representatives of elements of $\mathrm{Hom}(\pi_1(X;\ast),\mathcal E) {/\hskip -2.5pt/} \mathcal E$. We remark that this is also the space of states of a topological gauge theory based on the crossed module $\mathbb G'$ which consists of the trivial homomorphism $\ker(\Delta) \to \mathcal E$ and action of $\mathcal E$ on $\ker(\Delta)$ inherited from~$\mathbb G$. Indeed, condition $\epsilon_{\partial f}=1$ for each $f$ is precisely the fake flatness constraint for~$\mathbb G'$. Moreover groups, in which fields $\bm \epsilon, \bm \varphi$ as well as gauge transformations are valued, coincide.

Vacuum states corresponding to non-equivalent $\bm \epsilon$ may always be distinguished by values of $1$-holonomy operators along nontrivial loops. It is not always true that states with the same $\bm \epsilon$ but non-equivalent $\bm \varphi$ can be discriminated by evaluating $2$-holonomies, as illustrated by one of examples in subsection \ref{sec:examples}.

Last, but not least, we consider the problem of minimization of $\mathsf H_{BV}$. There exists a~basis of ground states indexed by homotopy classes of maps $X \to B \mathbb G$. This implies that ground states of this hamiltonian form the space of states of the Yetter's model. Below we prove this statement and give a more explicit description of the space of ground states.

In order to minimize operators $\mathsf B_{\partial q, \theta_q}$ we have to restrict attention to configurations obeying $\varphi_{\partial q} = 1$ for every ball $q$. Given any such configuration $(\bm \epsilon, \bm \varphi)$ we obtain a ground state by forming the superposition
\begin{equation}
\sum_{\bm \xi, \bm \psi} \mathsf G(\bm \xi) \mathsf V (\bm \psi) | \bm \epsilon, \bm \varphi \rangle.
\label{eq:H1_gs}
\end{equation}
A basis of the space of ground states is formed by vectors of this form, one for each orbit of the group of vertex and edge transformations in the set of admissible configurations. The fact that these orbits are in one-to-one correspondence with homotopy classes of maps $X \to B \mathbb G$ has been reviewed in the appendix \ref{appendix:cscmp}. We~proceed to give an alternative description of the set of orbits.

Since in the present analysis configurations related by edge transformations with arbitrary $\bm \psi$ are identified, the only invariant datum specified by $\bm \epsilon$ is the corresponding element of $\mathrm{Hom}(\pi_1(X;\ast), \coker(\Delta)) {/\hskip -2.5pt/} \coker(\Delta)$. For every element of this set we choose one representative $\overline{\bm \epsilon}$ and lift it to some $\bm \epsilon$. It is not always possible to choose $\bm \epsilon$ which is itself flat, as shown in examples in subsection \ref{sec:examples}.

Next we consider the set of allowed $\bm \varphi$ for the given $\bm \epsilon$. As illustrated in subsection \ref{sec:examples}, for $\overline{\bm \epsilon}$ having nontrivial holonomies it may happen that no $\bm \varphi$ satisfying the flatness condition $\varphi_{\partial q}=1$ exists. Let us consider the case in which some flat $\bm \varphi$ does exist. Then any other flat $\bm \varphi'$ is of the form
\begin{equation}
\varphi_f' = \chi_f \, \varphi_f
\label{eq:phi_fixed_eps}
\end{equation}
for some twisted cocycle $\bm \chi$ (see~apppendix \ref{sec:twisted}). The more stringent condition that $\bm \chi$ is a~twisted coboundary holds if and only if configurations $(\bm \epsilon, \bm \varphi)$ and $(\bm \epsilon, \bm \varphi')$ are related by an edge transformation with $\psi_e$ in $\ker(\Delta)$ for each $e$. Therefore the set $\mathcal F(\bm \epsilon)$ of equivalence classes of flat $\bm \varphi$ (with the given $\bm \epsilon$) modulo $\ker(\Delta)$-valued edge transformations is an affine space over $H^2(X, \ker(\Delta), \overline{\bm \epsilon})$. That is not the end of the story, because the group $\mathrm{Stab}_{V,E}(\bm \epsilon)$ of combined vertex and edge transformations preserving $\bm \epsilon$ acts on $\mathcal F(\bm \epsilon)$. We will now show that this action factors through the smaller group $\mathrm{Stab}_V(\overline{\bm \epsilon})$ of $\coker(\Delta)$-valued vertex transformations, i.e.\ collections $\overline{\bm \xi}=\{ \overline \xi_v \}$ of elements of $\coker(\Delta)$ such that
\begin{equation}
\overline \epsilon_{\gamma} = \overline \xi_{t{(\gamma)}} \, \overline \epsilon_{\gamma}\, \overline \xi_{s(\gamma)}^{-1} \qquad \text{for every path } \gamma.
\label{eq:stab_eps_bar}
\end{equation}

Transformations in $\mathrm{Stab}_{V,E}(\bm \epsilon)$ are represented by operators of the form $\mathsf G(\bm \xi) \mathsf V(\bm \psi)$, where the pair $(\bm \xi, \bm \psi)$ is such that
\begin{equation}
\epsilon_{\gamma} = \xi_{t(\gamma)} \, \Delta \left( \psi_{\gamma}^{(\epsilon)} \right) \, \epsilon_{\gamma} \, \xi_{s(\gamma)}^{-1} \qquad \text{for any path } \gamma.
\end{equation}
Let us first consider a pair $(\bm \xi , \bm \psi)$ such that in addition $\xi_v = \Delta \rho_v$ for each $v$. The discussion around equation \eqref{eq:G_im_step} gives
\begin{eqnarray}
\mathsf G(\bm \xi) \mathsf V(\bm \psi) | \bm \epsilon, \bm \varphi \rangle =  \mathsf V(\widetilde {\bm \psi}) \mathsf V(\bm \psi) | \bm \epsilon, \bm \varphi \rangle = \mathsf V( \widetilde{\bm \psi} \bm \psi) | \bm \epsilon, \bm \varphi \rangle
\end{eqnarray}
for some $\Phi$-valued collection $\widetilde{\bm \psi}$. By the definition of $\mathrm{Stab}_{V,E}(\bm \epsilon)$, we must have \hbox{$\widetilde \psi_e \psi_e \in \ker(\Delta)$} for each $e$. Thus the action of $(\bm \xi , \bm \psi)$ on the set of allowed $\bm \varphi$ for the given $\bm \epsilon$ reduces to a $\ker(\Delta)$-valued edge transformation. Hence $(\bm \xi , \bm \psi)$ acts trivially on the set $\mathcal F(\bm \epsilon)$.

Next let us observe that the map $\mathrm{Stab}_{V,E}(\bm \epsilon) \ni (\bm \xi, \bm \psi) \mapsto \overline{\bm \xi} \in \mathrm{Stab}_V(\overline{\bm \epsilon})$ is a~homomorphism. Preceding discussion shows that its kernel acts trivially on $\mathcal F(\bm \epsilon)$, so to complete the proof it is sufficient to show that this homomorphism is surjective. Thus we choose some $\overline{\bm \epsilon}$ obeying \eqref{eq:stab_eps_bar} and lift it to an $\mathcal E$-valued collection $\bm \epsilon$ arbitrarily. By construction, we have that for each path $\gamma$ the element
\begin{equation}
    \mu_{\gamma} = \xi_{t(\gamma)}^{-1} \, \epsilon_{\gamma} \, \xi_{s(\gamma)} \, \epsilon_{\gamma}^{-1}
    \label{eq:mu_def}
\end{equation}
belongs to $\mathrm{im}(\Delta)$. Directly from the definition of $\mu_{\gamma}$ we have that
\begin{equation}
\mu_{\gamma' \gamma} = \mu_{\gamma'} \, \epsilon_{\gamma'} \, \mu_{\gamma} \, \epsilon_{\gamma'}^{-1}
\end{equation}
is satisfied for any composite path $\gamma' \gamma$. Now for every edge $e$ we choose some $\psi_e$ such that $\mu_e = \Delta \psi_e$. Then $\mu_{\gamma}$ coincides with $\Delta \psi_{\gamma}^{(\epsilon)}$ whenever $\gamma$ is a single edge, and furthermore these two collections satisfy the same composition rule for concatenated paths. Thus $\mu_{\gamma} =\Delta \psi_{\gamma}^{(\epsilon)}$ for every $\gamma$. Plugging this into \eqref{eq:mu_def} we obtain
\begin{equation}
\xi_{t(\gamma)} \, \Delta \psi_{\gamma}^{(\epsilon)} \, \epsilon_{\gamma} \, \xi_{s(\gamma)}^{-1} = \epsilon_{\gamma},
\end{equation}
and hence the claim is proven.

In summary, vectors of the form \eqref{eq:H1_gs} form a basis of ground states of $\mathsf H_{BV}$. They can be labeled by elements of the disjoint union of sets $\mathcal F (\bm \epsilon ) {/\hskip -2.5pt/} \mathrm{Stab}_{V}(\overline{\bm \epsilon})$ with $\bm \epsilon$ running through representatives of elements of $\mathrm{Hom}(\pi_1(X;\ast),\coker(\Delta)) {/\hskip -2.5pt/} \coker(\Delta)$.

We are now ready to deduce that the space of ground states of $\mathsf H_{BV}$, as well as the space of states invariant under all vertex and edge transformations, but not necessarily with flat $\bm \varphi$, depends on $\mathbb G$ only through its weak equivalence class. Clearly it is sufficient to consider a weak isomorphism $(E,F) : \mathbb G \to \mathbb G' = (\mathcal E', \Phi', \Delta', \rhd')$. We~let $T$ be the map which sends a $\mathbb G$-valued configuration $(\bm \epsilon, \bm \varphi)$ to $(E(\bm \epsilon), F(\bm \varphi))$, where $E(\bm \epsilon)= \{ E(\epsilon_e) \}$ and $F(\bm \varphi) = \{ F(\varphi_f) \}$. This intertwines between $\mathbb G$ and $\mathbb G'$-valued vertex and edge transformations, so there is an induced mapping on the space of orbits. The latter is in both cases a disjoint union of subsets labeled~by
\begin{equation*}
    \mathrm{Hom}(\pi_1(X;\ast),\coker(\Delta)) {/\hskip -2.5pt/} \coker(\Delta) \cong \mathrm{Hom}(\pi_1(X;\ast),\coker(\Delta')) {/\hskip -2.5pt/} \coker(\Delta'),
\end{equation*}
where we used the fact that $\overline E : \coker(\Delta) \to \coker(\Delta')$ is an isomorphism. Mapping $T$ preserves this decomposition. Thus it is sufficient to consider configurations of the form $(\bm \epsilon , \bm \varphi)$ and $(E(\bm \epsilon), \bm \varphi')$ for one $\bm \epsilon$, for now with no constraint on $\bm \varphi$. Let $\mathcal C(\bm \epsilon)$ and $\mathcal C'(E(\bm \epsilon))$ be the sets of all allowed $\bm \varphi$ and $\bm \varphi'$ modulo $\ker(\Delta)$- (resp. $\ker(\Delta')$-)valued edge transformations. They are affine over $H^2(X_2, \ker(\Delta), \bm {\overline \epsilon}) \cong H^2(X_2, \ker(\Delta'), \overline E(\bm {\overline \epsilon}) )$, because a flat configuration on $X_2$ is the same as an arbitrary configuration on $X$ (constraint $\varphi_{\partial q}=1$ for every ball $q$ being vacuous if balls are absent). The map $T$ intertwines between the affine structures, so~we have $\mathcal C(\bm \epsilon) \cong \mathcal C'(E(\bm \epsilon))$. Furthermore, we clearly have $\mathrm{Stab}_V(\bm{\overline \epsilon}) \cong \mathrm{Stab}_V( \overline E (\bm{\overline \epsilon}))$, and again $T$ preserves actions of these groups. Thus
\begin{equation*}
\mathcal C(\bm \epsilon) {/\hskip -2.5pt/} \mathrm{Stab}_V(\bm{\overline \epsilon})  \cong \mathcal C'(E(\bm \epsilon)) {/\hskip -2.5pt/} \mathrm{Stab}_V(E(\bm{\overline \epsilon})),
\end{equation*}
which proves that $T$ is a bijection. Finally, let us observe that $F(\bm \varphi)_{\partial q} = \overline F(\varphi_{\partial q})$, since $\varphi_{\partial q} \in \ker(\Delta)$. Since $\overline F$ is an isomorphism, this implies that flatness of $F(\bm \varphi)$ is equivalent to flatness of $\bm \varphi$. Hence $T$ is bijective also after restricting to flat configurations.

Another interesting point to be raised here is that there is an explicit topological criterion to determine when the set $\mathcal F(\bm \epsilon)$ is nonempty. Here we give a~short summary, with a more detailed description postponed to the appendix \ref{appendix:postnikov}. Field $\bm{\overline \epsilon}$ determines (up to homotopy) a map $h_{\bm{\overline \epsilon}}$ from $X$ to $B \coker(\Delta)$, the classifying space of the group $\coker(\Delta)$. There is a distiniguished twisted cohomology class $\beta$ on $B \coker(\Delta)$, called the Postnikov class. The set $\mathcal F(\bm \epsilon)$  is nonempty if and only if the pullback $h_{\bm{\overline \epsilon}}^{\ast} \, \beta$ is the trivial cohomology class on $X$.

We close this section with the remark that it has been shown in \cite{kapustin17} that the topological field theory describing ground states of $\mathsf H_{VB}$ may be formulated using fields valued in groups $\ker(\Delta)$ and $\coker(\Delta)$ only. In this approach crossed modules do not have to be invoked explicitly. One has to merely specify the action of $\coker(\Delta)$ on $\ker(\Delta)$ and the Postnikov class $\beta$. These are precisely the data that determine the crossed module up to weak isomorphisms \cite{maclane_whitehead}, in accord with the fact that the model possesses weak isomorphism invariance.

\subsection{A peek at dynamics} \label{sec:dyn}

\begin{figure}[ht]
\begin{minipage}[c]{0.56\textwidth}
\centering
\begin{tikzpicture}
\begin{scope}[very thick]

      \draw[red,fill=red,opacity=.2] \boundellipse{1,1}{1}{0.5};
      \node (a1) at (1,1) {{\color{blue}$\mathsf H_V+\mathsf H_B$}};
      \draw[red,fill=red,opacity=.2] \boundellipse{6,1}{1}{0.5};
      \node (a2) at (6,1) {{\color{blue}$\mathsf H_A+\mathsf H_B$}};
      \draw[red,fill=red,opacity=.2] \boundellipse{1,-4}{1}{0.5};
      \node (a3) at (1,-4) {{\color{blue}$\mathsf H_V+\mathsf H_W$}};
      \draw[red,fill=red,opacity=.2] \boundellipse{6,-4}{1}{0.5};
      \node (a4) at (6,-4) {{\color{blue}$\mathsf H_A+\mathsf H_W$}};

      \node (b1d) at (1,0.5) {};
      \node (b1r) at (2,1) {};

      \node (b2d) at (6,0.5) {};
      \node (b2l) at (5,1) {};

      \node (b3u) at (1,-3.5) {};
      \node (b3r) at (2,-4) {};

      \node (b4u) at (6,-3.5) {};
      \node (b4l) at (5,-4) {};
      {\color{black}
      \draw (b1r)--(b2l) node [midway,below] {$\mathsf H_{ABV}$};
      \draw (b3r)--(b4l) node [midway,above] {$\mathsf H_{AVW}$};
      \draw (b1d)--(b3u) node [midway,right] {$\mathsf H_{BVW}$};
      \draw (b2d)--(b4u) node [midway,left] {$\mathsf H_{ABW}$};

      \draw (b1r)--(b2l) node [midway,above] {Yang-Mills};
      \draw (b3r)--(b4l) node [midway,below] {Yang-Mills};
      \draw (b3u)--(b1d) node [midway,sloped,above] {$2$-form gauge theory};
      \draw (b2d)--(b4u) node [midway,sloped,above] {$2$-form gauge theory};
      }
\end{scope}
\end{tikzpicture}
\end{minipage}\hfill
 \begin{minipage}[c]{0.44\textwidth}
    \caption{
       A diagram representing four possible models with hamiltonians consisting of three out of four terms.
    } \label{fig:diagram}
  \end{minipage}
\end{figure}
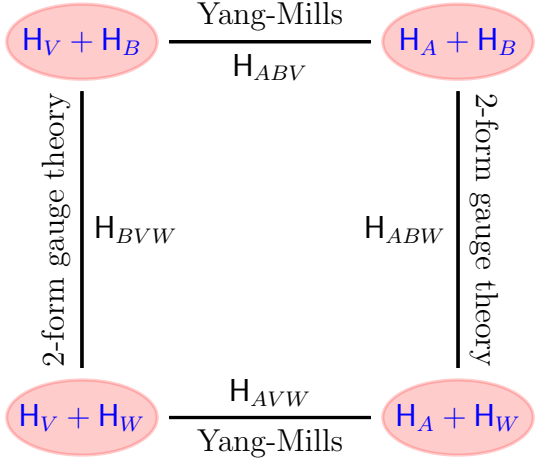

In this subsection we discuss models described by hamiltonians in which three out of four terms of $\mathsf H$ are present. See figure \ref{fig:diagram} for an illustration of the four possibilities. In each case dynamics reduces to that of some simpler theory. Therefore we can understand the dynamics generated by $\mathsf H$ along the boundary of its phase diagram.

Several topological aspects of our models have been discussed in subsection \ref{sec:vacua}. In this step we would like to focus instead on local dynamics. Therefore we assume now a topologically trivial situation, i.e.\ that the first two homotopy groups of $X$ vanish. In this case we can always fix gauge $\overline{\epsilon_e}=1$. In other words, the $\bm \epsilon$ field can be regarded as valued in $\mathrm{im}(\Delta)$. True physical states may be obtained in the end by summing over vertex transformations. Thus in the further analysis it is necessary to explicitly take into account only those vertex transformations $\bm \xi$ which preserve the gauge condition $\overline{\epsilon_e}= 1$, i.e.\ those with constant $\overline {\bm \xi}$.

Let us begin with the case $\mathsf H_{ABW}=\mathsf H_A + \mathsf H_B + \mathsf H_W$. Term $\mathsf H_A$ commutes with the other two, so it may be minimized exactly\footnote{Strictly speaking, it could happen that an eigenvector of $\mathsf H_B + \mathsf H_W$ to a much lower eigenvalue could be found in excited subspaces of $\mathsf H_A$. Thus presented analysis is valid exactly only under the additional assumption that $\mathsf H_A$ dominates over the other two terms.}. Therefore we may restrict attention to field configurations with $\epsilon_{\partial f}=1$ for each $f$. Each gauge equivalence class of fields with this property admits a representative with $\epsilon_e=1$ for each edge $e$. For these representatives the fake flatness constraint implies that $\varphi_f\in\ker(\Delta)$. As a result, the only physical degree of freedom is a $\ker(\Delta)$-valued $2$-form field. Residual gauge freedom consists of transformations of two types: edge transformations valued in $\ker(\Delta)$, which play the role of standard gauge transformations for the $2$-form field, and vertex transformations with constant $\xi$. Explicitly, the latter acts according to the formula  $\varphi_\sigma \longmapsto \xi\rhd \varphi_\sigma$ for every $\sigma$. From the point of view of the $2$-form theory this is a global symmetry. Summarizing, the ground states of $\mathsf{H}_{ABW}$ coincide with ground states of a $2$-form gauge theory valued in $\ker(\Delta)$, restricted to the singlet sector of a certain global symmetry.

For the hamiltonian $\mathsf H_{AVW}$, the analysis is analogous. Field $\varphi$ is effectively removed by exactly minimizing $\mathsf H_W$, which enforces that for any $\bm \epsilon$ all configurations $(\bm \epsilon, \bm \varphi)$ allowed by fake flatness enter with an equal amplitude. The final conclusion is that ground states are the same as in a Yang-Mills theory with gauge group $\mathrm{im}(\Delta)$, restricted to the singlet sector of a global symmetry.

Next, consider the model with hamiltonian $\mathsf H_{ABV}$. In this case we impose the constraint $\varphi_{\partial q}=1$ for each $q$. There exists a unique such $\bm \varphi$, up to edge transformations valued in $\ker(\Delta)$, for every $\bm \epsilon$. Therefore the field $\bm \varphi$ is effectively removed from the theory. In the end we obtain the singlet sector of lattice Yang-Mills theory with gauge group $\mathrm{im}(\Delta)$, as in the case of $\mathsf H_{AVW}$.

It remains to analyze the theory with $\mathsf H_{BVW}$ as a hamiltonian. In this case we minimize exactly the $\mathsf H_V$ term. Therefore ground states may be written as superpositions of states of the form
\begin{equation}
\sum_{\bm \psi} \mathsf V(\bm \psi) | \bm 1 , \bm \varphi \rangle = \sum_{\bm \psi} | \{ \Delta \psi_e \} , \{ \psi_{\partial f} \, \varphi_f  \} \rangle,
\end{equation}
which are labeled by collections $\bm \varphi$ valued in $\ker(\Delta)$, modulo $2$-form gauge transformations $\varphi_f \mapsto \psi_{\partial f} \, \varphi_f$
with $\ker(\Delta)$-valued $\bm \psi$. Thus we obtain the space of states of a $2$-form gauge theory. Vertex gauge transformations with $\bm \xi$ valued in $\mathrm{im}(\Delta)$ act trivially, because they reduce to edge transformations, which were already taken care of. There remains only the condition of invariance with respect to vertex transformations with constant $\bm \xi$, which again can be interpreted as a~global symmetry.

Finally, we would like to emphasize that, in spite of the preceding discussion, models found on opposite edges of the diagram on figure \ref{fig:diagram} are not identical. They differ in their global properties once we start considering spaces $X$ with nontrivial homotopy groups. Firstly, let us compare hamiltonians $\mathsf H_{ABV}$ and $\mathsf H_{AVW}$. In the first case low-lying states have flat $\bm \varphi$, but can be distinguished by $2$-holonomies along non-contractible spheres in $X$. There is a possibility of ground state degeneracy due to existence of several non-equivalent flat $\bm \varphi$ for a given $\bm \epsilon$. Thus the $2$-form electric symmetry may be broken. On the other hand for the hamiltonian $\mathsf H_{AVW}$ field $\bm \varphi$ is effectively absent. Since ground states are invariant under all $\mathsf W$ operators, the $2$-form symmetry is unbroken. Comparison of $\mathsf H_{ABW}$ and $\mathsf H_{BVW}$ is similar. In the former case fields $\bm \epsilon$ are flat, but they may still have nontrivial $1$-holonomies. Thus the $1$-form electric symmetry may be broken. On the other hand for ground states of $\mathsf H_{BVW}$ holonomies of $\bm \epsilon$ are undefined, since they are not invariant with respect to edge transformations (which are symmetries of the states).

\subsection*{Acknowledgements}

LH would like to thank Martin Ro\v{c}ek and Rikard von Unge for a discussion which initiated the current project. We also thank A.~Czarnecki for a~discussion. BR~was supported by the MNS donation for PhD students and young scientists N17/MNS/000040. The work of LH was supported by the TEAM programme of the Foundation for
Polish Science co-financed by the European Union under the European Regional
Development Fund (POIR.04.04.00-00-5C55/17-00).


\appendix

\section{Kernel and cokernel of $\partial$} \label{sec:exact_seq}

Given a topological space $A$, subspace $B$ and a base point $\ast\in B$, one has an~exact sequence of groups \cite[Thm.~4.3]{hatcher}
\begin{equation}
    \pi_1(A;\ast)\xleftarrow{\hspace*{0.6cm}}\pi_1(B;\ast)\xleftarrow{\hspace*{0.3cm}\partial\hspace*{0.3cm}}\pi_2(A,B;\ast)\xleftarrow{\hspace*{0.6cm}}\pi_2(A;\ast)\xleftarrow{\hspace*{0.6cm}}\pi_2(B;\ast).
\end{equation}
It follows that $\ker(\partial)$ may be identified with the quotient of $\pi_2(A;\ast)$ by the image of the homomorphism $\pi_2(B;\ast)\rightarrow\pi_2(A;\ast)$.

Furthermore, notice that if the map $\pi_1(B;\ast)\xrightarrow{}\pi_1(A;\ast)$ is surjective, the cokernel of $\partial$ is isomorphic to $\pi_1(A;\ast)$. This is true in particular for $A=X$, $B=X_1$, by the cellular approximation theorem \cite[Thm.~4.8]{hatcher}. Secondly, the universal cover a~one-dimensional CW-complex is contractible. Thus $\pi_2(X_1; \ast)$ is trivial, so we have an identification $\ker(\partial)\cong \pi_2(X;\ast)$.

\section{Twisted cohomology} \label{sec:twisted}

In this appendix we give a definition of twisted cohomology as it arises directly in calculations done in this paper. We refer to \cite[p.~255--290]{whitehead_book} for a more complete treatment. We shall use relative homotopy groups $\pi_n(A,B ; \ast)$ with any $n \geq 2$, as well as the action of $\pi_1(B; \ast)$ on these groups. Their definition is entirely analogous to the case $n=2$ and can be found e.g. in \cite[p.~343]{hatcher}. They are abelian for $n \geq 3$. As~for $n=2$, there is a homomorphism $\partial : \pi_n(A,B;\ast) \xrightarrow{\partial} \pi_{n-1}(B; \ast)$, whose kernel coincides with the image of the self-evident map $\pi_n(A; \ast) \to \pi_n(A,B ; \ast)$. Furthermore, a map $A \to A'$ which takes $B$ to $B' \subseteq A'$ induces a homomorphism $\pi_n(A,B;\ast) \to \pi_n(A',B';\ast)$, which is unchanged by homotopic deformations preserving the condition that $B$ is mapped to $B'$ at all intermediate stages. All that generalizes to a grupoid version $\pi_n(A,B;C)$, for which a~whole set $C \subseteq B$ of base points is allowed, in a way analogous to the case $n=2$.

In our applications we need the above structure with $A=X_n$, $B=X_{n-1}$ and $C=X_0$. Thus $\pi_1(B;C) = \pi_1(X_1;X_0)$ if $n=2$ and $\pi_1(B;C) = \pi_1(X;X_0)$ for $n \geq 3$. Since the latter group is a quotient of $\pi_1(X_1;X_0)$, we have an action of $\pi_1(X_1;X_0)$ on $\pi_n(X_n,X_{n-1};X_0)$ in each case. Groups $\pi_n(X_n, X_{n-1};x)$ with $x \in X_0$ and $n \geq 3$ may be handled in practice using the fact \cite[p.~212]{whitehead_book} that they are free $ \pi_1(X; x)$-modules, with bases labeled by $n$-cells of $X$.

Now let us fix a group $G$, an abelian group $K$ on which $G$ acts by automorphisms and a homomorphism $\bm \alpha : \pi_1(X;X_0) \to G$. Thus for every path $\gamma$ there is an endomorphism $ k \mapsto \alpha_{\gamma} \rhd k$ of $K$, trivial if $\gamma$ is contractible in $X$. It obeys the composition law $ \alpha_{\gamma' \gamma} =  \alpha_{\gamma'} \,  \alpha_{\gamma}$. In our applications we will mostly consider the case $G= \coker(\Delta)$ and $K = \ker (\Delta)$ for some crossed module $\mathbb G$, with $\bm \alpha = \overline{\bm \epsilon}$. This is not relevant for the discussion in this appendix.

\begin{figure}[h!tb]
\centering
\begin{tikzcd}[column sep=-1.2cm, row sep=1cm]
&& \pi_{p+1}(X_{p+1},X_p;X_0) \arrow[drr, "\partial",blue, dashed]\arrow[dddr,blue,"\overline{\partial}",pos=0.6]&& \\
{\color{red}\pi_{p+1}(X_{p+1};X_0)}\arrow[urr,red,dashed]\arrow[rrrr,dashed]&&&&{\color{blue}\pi_p(X_p;X_0)}\arrow[ddl,blue,dashed] \\
&&&&\\
& \pi_{p+2}(X_{p+2},X_{p+1};X_0)\arrow[uul,"\partial",red,dashed]\arrow[uuur,red,"\overline{\partial}",pos=0.4] \arrow[rr]& &\pi_{p}(X_p,X_{p-1};X_0) &
\end{tikzcd}
    \caption{
       The two colored maps factors through the dashed ones marked by the same colors. The composition depictured by the dashed black arrow is the trivial map.
    } \label{fig:pieciokacik}
\end{figure}
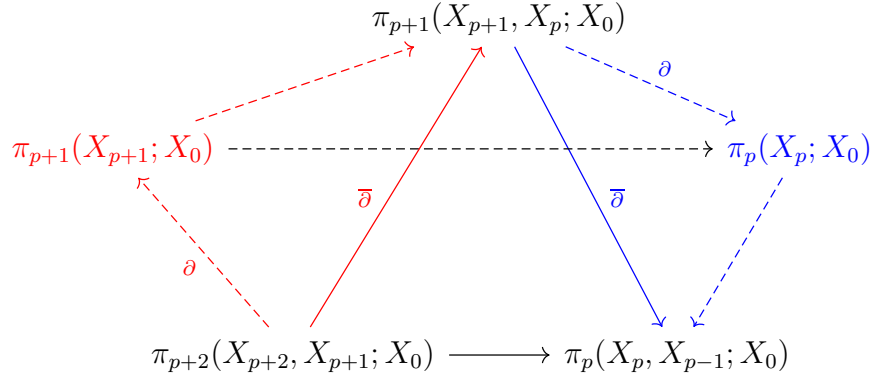

By an $\bm{\alpha}$-twisted $p$-cochain on $X$ valued in $K$ we shall mean:
\begin{itemize}
    \item $p=0$: collection of elements $\rho_v \in K$ labeled by vertices $v$,
    \item $p=1$: assignment of $\psi_{\gamma} \in K$ to every path $\gamma$, subject to the composition law $\psi_{\gamma' \gamma}= \psi_{\gamma'} \, (\alpha_{\gamma'} \rhd \psi_{\gamma})$ whenever $s(\gamma')=t(\gamma)$,
    \item $p \geq 2$: homomorphism $\chi : \pi_p(X_p,X_{p-1};X_0) \to K$ satisfying the equivariance condition $\chi_{\gamma \rhd \tau} = \alpha_{\gamma} \rhd \chi_{\tau}$.
\end{itemize}
The set of all $p$-cochains is a group, which we denote by $C^p(X,K, \bm{\alpha})$. Next we define a differential $\delta : C^p(X, K, \bm{\alpha}) \to C^{p+1}(X, K, \bm{\alpha})$ in the following way:
\begin{itemize}
    \item $p=0$: $(\delta \rho)_{\gamma} = \rho_{t(\gamma)} \, \left( \alpha_{\gamma} \rhd \rho_{s(\gamma)}^{-1} \right)$,
    \item $p=1$: $(\delta \psi)_{\sigma} = \psi_{\partial \sigma}$, where $\partial: \pi_2(X_2,X_1;X_0) \to \pi_1(X_1;X_0)$ is as defined in section \ref{sec:geo_setup},
    \item $p \geq 2$: $(\delta \chi)_{\tau} = \chi_{\overline{\partial} \tau}$, where $\overline{\partial} : \pi_{p+1}(X_{p+1},X_p;X_0) \to \pi_{p}(X_p, X_{p-1};X_0)$ is the composition of homomorphisms $\partial : \pi_{p+1}(X_{p+1},X_p;X_0) \to \pi_p(X_p; X_0)$ and $\pi_p(X_p; X_0) \to \pi_p(X_p, X_{p-1} ;X_0)$.
\end{itemize}
With this differential, $C^{\bullet}(X, K, \bm{\alpha})$ is a cochain complex, whose cohomology we denote by $H^{\bullet}(X,K, \bm{\alpha})$ and call the twisted cohomology. Another popular name is cohomology with local coefficients. To see that $\delta$ is nilpotent, first note that for a~$0$-cochain $\rho$ we have $(\delta^2 \rho)_{\sigma} = (\delta \rho)_{\partial \sigma} = \rho_{b(\sigma)} \, \left( \alpha_{\partial \sigma} \rhd \rho_{b(\sigma)}^{-1} \right) =1$, as $\alpha_{\partial \sigma} =1$. For a~$p$-cochain $\bm \chi$ with $p \geq 2$ we have $(\delta^2 \chi)_{\tau} = \chi_{\overline \partial^2 \tau}$. Homomorphism $\overline \partial^2$ fits in the commutative diagram shown on figure \ref{fig:pieciokacik}, so it factors through the (trivial) composition of two subsequent homomorphisms in the long exact sequence of relative homotopy groups \cite[Thm.~4.3]{hatcher} of the pair $(X_{p+1},X_{p})$. For $p=1$ one needs triviality of $\partial \overline \partial$, for which an analogous reasoning applies.

Now let us assume that $l : Y \to X$ is a cellular map of CW-complexes. Given \hbox{$\bm {\alpha} \in \mathrm{Hom}(\pi_1(X,X_0), G)$}, its pullback $l^{\ast} \bm{\alpha} \in \mathrm{Hom}(\pi_1(Y, Y_0),G)$ is defined as the composition of $\bm {\alpha}$ with the pushforward map $\pi_1(Y,Y_0) \to \pi_1(X,X_0)$ induced by~$l$. Furthermore, the pullback $l^{\ast} : C^{\bullet}(X,K,\bm{\alpha}) \to C^{\bullet}(Y,K,l^{\ast} \bm{\alpha})$ may be defined in an analogous way. It intertwines between the differentials, so there is an induced pullback map of cohomology $l^{\ast} : H^{\bullet}(X, K, \bm{\alpha}) \to H^{\bullet}(Y, K , l^{\ast} \bm{\alpha})$.

We close this appendix with a remark that twisted cohomology may be defined also without reference to a cell structure on $X$. They depend only on the topology of $X$ and another datum called a local system of abelian groups on $X$. The latter may be (non-canonically) encoded by a single abelian group $K$ and a~homomorphism $\pi_1(X;\ast) \to \mathrm{Aut}(K)$ for some base point $\ast$.

\section{Classifying spaces} \label{sec:classifying}

Due to the length of this appendix, we divided it into several parts. In \ref{appendix:csg} we recall the basic properties of classifying spaces of groups. Appendix \ref{appendix:cscmp} is devoted to the definition and the proof of the fundamental property of classifying spaces of crossed modules, which relates field configurations on a space $X$ valued in a crossed module $\mathbb G$ with maps $X \to B \mathbb G$. In \ref{appendix:postnikov} we explain the relation of the so-called Postnikov class with the problem of constructing field configurations (or equivalently, maps to $B \mathbb G$). In \ref{appendix:cswecm} we construct maps between classifying spaces corresponding to homomorphisms of crossed modules and obtain the corollary that weakly equivalent crossed modules have homotopy equivalent classifying spaces. A~simple proof of existence of classifying spaces is given in \ref{appendix:existence}.

\subsection{Classifying spaces of groups}
\label{appendix:csg}

We begin with a short review of the classifying space $BG$ of a group $G$. One way to define it\footnote{There is a more general notion of a classifying space of a topological group \cite{husemoller}, for which this definition is not suitable. Here only discrete groups are considered.} is as a connected CW-complex with fundamental group $G$ and trivial higher homotopy groups. It is well known \cite[Thm.~7.1]{whitehead_book} that such space exists and is determined uniquely up to a homotopy equivalence. One may also assume that $B G$ has exactly one $0$-cell $\ast$, which we take as its base point.

We claim that gauge orbits of $G$-valued lattice gauge fields on $X$ are in one-to-one correspondence with homotopy classes of maps $X_1 \to BG$. Flatness of a gauge field is equivalent to existence of an extension of the corresponding map to $X_2$. If~this condition is satisfied, extending to the whole $X$ is automatic, and furthermore this extension is unique up to homotopy. There is also a correspondence between flat gauge fields (rather than gauge equivalence classes) on $X$ and homotopy classes of maps of pairs\footnote{Map of pairs $(X,Y) \to (X',Y')$ with $Y \subseteq X$ and $Y' \subseteq X'$ is a map $X \to X'$ which takes $Y$ into $Y'$. Definition of homotopy classes of maps of pairs allows only homotopies for which $Y$ is mapped to $Y'$ at all times. Maps of triples and their homotopy classes are defined analogously.} $(X,X_0) \to (BG, \ast)$. Again, flatness condition may be lifted by considering maps $(X_1,X_0) \to (BG, \ast)$.

To prove the above claims, let us first note that any mapping $X \to BG$ is homotopic to one which sends the whole $X_0$ to $\ast$, by the homotopy extension property \cite[p.~15]{hatcher} of the pair $(X,X_0)$. Such map sends every edge of $X$ to a loop in $BG$ based at the base point $\ast$. As a result it determines a homomorphism $\pi_1(X_1,X_0) \to \pi_1(BG, \ast) \cong G$, i.e.\ a lattice gauge field on $X$. Two maps $h_{\bm \alpha},h_{\bm \alpha'}$ are homotopic if and only if they determine gauge-equivalent fields $\bm \alpha$ and $\bm \alpha'$. Indeed, constructing a homotopy between them amounts to constructing an extension to\footnote{Here $I$ is the unit interval.} $I \times X$ of the map $\{ 0, 1 \} \times X \to BG$ given by $h_{\bm \alpha}$ and $h_{\bm \alpha'}$, respectively, on $\{ 0 \} \times X$ and $\{ 1 \} \times X$. This can be done iteratively, cell-by-cell.

\begin{figure}[ht]
\begin{minipage}[c]{0.35\textwidth}
\centering
\begin{tikzpicture}
\draw
(0,0) coordinate (a1) -- ++(0:2) coordinate (a2)
(a2) -- ++(90:2) coordinate (a3)
(a3) -- ++(180:2) coordinate (a4)
(a4) -- (a1)
;
\node at (a4) {$\bullet$};

\draw[very thick, decoration={markings, mark=at position 0.5 with {\arrow{stealth}}},postaction={decorate}](a1) -- node[below] {$\alpha_e^{-1}$}(a2);
\draw[very thick, decoration={markings, mark=at position 0.5 with {\arrow{stealth}}},postaction={decorate}](a2) -- node[right] {$\xi_{s(e)}$}(a3);
\draw[very thick, decoration={markings, mark=at position 0.5 with {\arrow{stealth}}},postaction={decorate}](a3) -- node[above] {$\alpha_e'$}(a4);
\draw[very thick, decoration={markings, mark=at position 0.5 with {\arrow{stealth}}},postaction={decorate}](a4) -- node[left] {$\xi^{-1}_{t(e)}$}(a1);
\end{tikzpicture}
\end{minipage}\hfill
 \begin{minipage}[c]{0.65\textwidth}
    \caption{Extension problem encountered in the construction of a homotopy between two maps $X \to BG$ cell by cell. The bold dot is the chosen base point of the square.} \label{fig:homotopy_gauge}
  \end{minipage}
\end{figure}
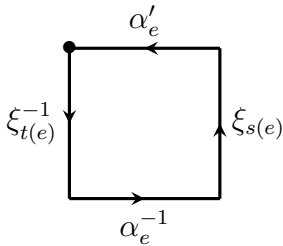

First we consider $1$-cells, which are of the form $I \times \{ v \}$ with $v$ - vertices of $X$. These can be mapped to any loops in $BG$, which determine elements $\xi_v \in \pi_1(BG, \ast)$. Next we extend through $2$-cells, which are products of $I$ and edges of $X$. Considering an edge $e$, we arrive at the problem of extending to the whole square the map on the boundary depictured on figure \ref{fig:homotopy_gauge}. This is possible if and only if the boundary map is null-homotopic, i.e.\ if $ \alpha_e'\,\xi_{s(e)}\, \alpha_e^{-1}\, \xi_{t(e)}^{-1}=1$ in $G$. In other words, if $\bm \alpha'$ and $\bm \alpha$ are gauge equivalent, we have to choose $\bm \xi$ in the previous step which is a~gauge transformation from $\bm \alpha$ to $\bm \alpha'$. Afterwards one has to extend through higher cells. This is always possible since higher homotopy groups of $BG$ vanish. Thus $h_{\bm \alpha}$ and $h_{\bm \alpha'}$ are homotopic. If $\bm \alpha$ and $\bm \alpha'$ are not gauge equivalent, it is impossible to extend the map through $2$-cells regardless of the choice of an~extension through $1$-cells. Hence $h_{\bm \alpha}$ and $h_{\bm \alpha'}$ are not homotopic.

We still have to determine which gauge fields can be realized by some map to~$BG$. On the $1$-skeleton of $X$ we can realize any gauge field, simply by constructing the corresponding map cell-by-cell. An obstruction arises if one attempts to extend the map from $X_1$ to $X_2$. Concretely, extension over a face $f$ is possible if and only if the bounding loop is sent to a trivial loop in $BG$, i.e.~if $\alpha_{\partial f}=1$. Thus a map $h: X_1 \to BG$ extends to $X_2$ if and only if the corresponding gauge field is flat. Further extension from $X_2$ to $X$ is unobstructed, again because higher homotopy groups of $BG$ are trivial.

The only part that remains to be proven is the one concerning homotopy classes of pairs $(X,X_0) \to (BG, \ast)$. Such homotopy class determines a homomorphism $\pi_1(X_1,X_0) \to G$. We already know that every homomorphism is realized by some homotopy class. Furthermore, in the construction of a homotopy between two maps determined by the same homomorphism we may take $\xi_v=1$ for every $v$, and hence the homotopy may be taken to be stationary on $I \times X_0$.

Being done with the proof, notice that there exists a distingushed $G$-valued gauge field $\bm{ \iota}$ on $B G$, corresponding to the tautological (identity) homomorphism \hbox{$\pi_1(BG, \ast) \to G$.} It is universal in the sense that one has $h_{\bm \alpha}^{\ast} \, \bm \iota = \bm \alpha$ for a map $h_{\bm \alpha} : X \to B G$ corresponding to a gauge field $\bm \alpha$ on $X$. Furthermore the twisted cohomology groups $H^{\bullet}(BG,K,\bm \iota)$ are defined for any $G$-module $K$. They are also called the cohomology groups of the group $G$ and can be constructed in a purely algebraic manner, see \cite{kbrown}. The universal character of the field $\bm \iota$ implies that pullback through $h_{\bm \alpha}$ maps $H^{\bullet}(BG,K, \bm \iota)$ to $H^{\bullet}(X,K, \bm \alpha)$.

\subsection{Classifying spaces of crossed modules}
\label{appendix:cscmp}

Here we will describe classifying spaces of crossed modules. For our purposes, the~following definition is suitable: $B \mathbb G$ is a connected CW-complex which contains $B \mathcal E$, the classifying space of the group $\mathcal E$, as a subcomplex and has homotopy groups
\begin{equation}
\pi_n(B \mathbb G; \ast) =
\begin{cases}
\coker(\Delta) & \text{for } n=1, \\
\ker(\Delta) & \text{for } n=2, \\
0 & \text{for } n \geq 3.
\end{cases}
\end{equation}
Furthermore, it is required that $\Pi_2(B \mathbb G, B \mathcal E ; \ast) \cong \mathbb G$. Again, we may assume that $B \mathbb G$ has exactly one $0$-cell $\ast$, which we choose as the base point.

It is known that such space $B \mathbb G$ exists and is determined uniquely up to a~homotopy equivalence by the above properties \cite{maclane_whitehead,loday,brown_higgins,brown01}. The latter fact is also obtained as a simple corollary from the discussion in the appendix \ref{appendix:cswecm}, while the former is proven in the appendix \ref{appendix:existence}.

The property of $B \mathbb G$ most important for us is that field configurations $(\bm \epsilon, \bm \varphi)$ on $X$ with flat $\bm \varphi$, modulo vertex and edge transformations, are in one-to-one correspondence with homotopy classes of maps $X \to B \mathbb G$. Clearly the flatness constraint may be lifted by considering maps $X_2 \to B \mathbb G$ instead. We remark also that homotopy classes of maps of triples $(X,X_1,X_0) \to (B \mathbb G, B \mathcal E, \ast)$ correspond to field configurations with flat $\bm \varphi$. Again, the flatness condition may be removed by replacing $X$ with $X_2$. Finally, it will be clear from the proof that a map $(X, X_1, X_0) \to (B \mathbb G, B \mathcal E , \ast)$ is homotopic as a map of triples to one with image in $B \mathcal E$ if and only if the corresponding configuration $(\bm \epsilon, \bm \varphi)$ has trivial $\bm \varphi$, i.e.~$\varphi_f=1$ for every face $f$.

For the purpose of the proof, we may again asssume that all maps $X \to B \mathbb G$ take $X_0$ to a base point $\ast$. Let us consider first homotopy classes of maps of $X_1$ into $B \mathbb G$ and $B \mathcal E$. Proceeding as in the above exposition of classifying spaces of groups one may show that they are in one-to-one correspondence with gauge equivalence classes of $\pi_1(B \mathbb G, \ast) \cong \coker(\Delta)$ and $\pi_1(B \mathcal E, \ast)\cong \mathcal E$-valued gauge fields on $X_1$, respectively. Furthermore, the map $[X_1, B \mathcal E] \to [X_1, B\mathbb G]$ induced by the inclusion of $B \mathcal E$ in $B \mathbb G$ corresponds to reduction modulo $\mathrm{im}(\Delta)$, so in particular it is surjective. Using the homotopy extension property of the pair $(X,X_1)$ we conclude that any map $X \to B \mathbb G$ is homotopic to one which maps $X_1$ to $B \mathcal E$ and $X_0$ to $\ast$. Such map sends every edge $e$ of $X$ to a loop in $B \mathcal E$ based at $\ast$, and hence determines an element $\epsilon_e \in \pi_1(B \mathcal E, \ast) \cong \mathcal E$.

Now consider the problem of extending a map $h_{\bm \epsilon} : X_1 \to BG$ which determines an $\mathcal E$-valued gauge field $\bm \epsilon$ to $X_2$. For every face $f$ we have to extend the map on the boundary whose homotopy class is given by the element $\overline \epsilon_{\partial f} \in \pi_1(B \mathbb G, \ast)$. An~extension exists if and only if $\overline \epsilon_{\partial f}=1$, i.e.~if $\epsilon_{\partial f}$ belongs to $\mathrm{im}(\Delta)$. Homotopy class of this extension, regarded as a map of triples $(I^2,\partial I, \ast) \to (B \mathbb G, B \mathcal E, \ast)$, determines and is determined by an element $\varphi_f \in \pi_2 (B \mathbb G, B \mathcal E, \ast)$ such that $\Delta \varphi_f = \epsilon_{\partial f}$. Summarizing, every homomorphism $\Pi_2(X_2,X_1;X_0) \to \mathbb G$ is realized by some map of triples $(X_2,X_1,X_0) \to (B \mathbb G, B \mathcal E, \ast)$. Conversely, any homotopy class of maps of triples is determined by the corresponding homomorphism. Thus a~bijection $[(X_2,X_1,X_0), (B \mathbb G, B \mathcal E, \ast)] \cong \mathrm{Hom}(\Pi_2(X_2,X_1;X_0), \mathbb G)$ is established.

Next, let us take two maps $h_{\bm \epsilon, \bm \varphi} , h_{\bm \epsilon', \bm \varphi'} : (X_2,X_1,X_0) \to (B \mathbb G, B \mathcal E, \ast)$, labeled by the corresponding field configurations, and consider the problem of deciding if they are homotopic as maps $X_2 \to B \mathbb G$. Thus we ask if the map $\{ 0 , 1 \} \times X_2 \to B \mathbb G$ given by $h_{\bm \epsilon', \bm \varphi'}$ and $h_{\bm \epsilon, \bm \varphi}$ on $\{ 1 \} \times X_2$ and $\{ 0 \} \times X_2$ extends to $I \times X_2$. By using the homotopy extension property of the pair $\Big{(} I \times X_2, (\{ 0, 1 \} \times X_2) \cup (I \times X_0)\Big{)}$ we conclude that every such extension is homotopic to one which sends $I \times X_0$ to $B \mathcal E$. Then cells $I \times \{ v \}$ are sent to loops in $B \mathcal E$ described by elements $\xi_v \in \pi_1(B\mathcal E, \ast)$, which can be chosen at will. Next we extend through $2$-cells. We~encounter a problem analogous to the one illustrated on figure \ref{fig:homotopy_gauge}. An extension exists if $ \epsilon_e' \, \left( \xi_{t(e)} \, \epsilon_e \,  \xi_{s(e)}^{-1} \right)^{-1} \in \mathcal E$ represents a trivial element of $\pi_1(B \mathbb G, \ast) = \coker(\Delta)$. Assuming this is true, homotopy classes of extensions are described by elements $\psi_e \in \pi_2(B \mathbb G, B \mathcal E, \ast) = \Phi$ such that
\begin{equation}
    \epsilon_e' = \Delta \psi_e \, \xi_{t(e)}\,  \epsilon_e \, \xi_{s(e)}^{-1}.
\end{equation}

\begin{figure}[h!tb]
\begin{minipage}[c]{0.35\textwidth}
\centering
\begin{tikzpicture}
\begin{scope}[very thick]
      \draw[ao, fill=ao,opacity=.2,decoration={markings, mark=at position 0.8 with {\arrow[scale=1.5]{stealth}}},
        postaction={decorate}] \boundellipse{0,0}{1.5}{0.6};
        \draw[ao,decoration={markings, mark=at position 0.8 with {\arrow[scale=1.5]{stealth}}},
        postaction={decorate}] \boundellipse{0,0}{1.5}{0.6};
        \node (a1) at (-1.5,0.1) {};
        \node at (-1.5,0) {{\color{ao}$\bullet$}};
        \node (c1) at (0,0) {$\varphi_f'$};
        \node (e1) at (0.7,-0.8){$\epsilon'_{\partial f}$};
        \node (d1) at (1.5,0.1) {};

        \draw[red, fill=red,opacity=.2,decoration={markings, mark=at position 0.8 with {\arrow[scale=1.5]{stealth}}},
        postaction={decorate}] \boundellipse{0,-4}{1.5}{0.6};
        \draw[red,decoration={markings, mark=at position 0.8 with {\arrow[scale=1.5]{stealth}}},
        postaction={decorate}] \boundellipse{0,-4}{1.5}{0.6};
        \node (a2) at (-1.5,-4.1) {};
        \node at (-1.5,-4) {{\color{red}$\bullet$}};
        \node (c2) at (0,-4) {$\varphi_f$};
        \node (e2) at (0.7,-4.8){$\epsilon_{\partial f}$};
        \node (d2) at (1.5,-4.1) {};

        \draw[very thick, decoration={markings, mark=at position 0.5 with {\arrow{stealth}}},postaction={decorate}](a2) -- node[left] {$\xi_{b(f)}$}(a1);
        \draw[dashed, very thick](d2) -- node[left] {}(d1);

        \node[above left] at (-1.5, -0.2) {$(1,b(f))$};
        \node[above left] at (-1.5, -4.2) {$(0,b(f))$};

        \node (f1) at (3, -3) {$\psi_{\partial f}^{(\epsilon '')}$};
        \node (f1a) at (0.5,-2) {};
        \draw[-latex,blue] (f1) to[out=140,in=0] node[midway,font=\scriptsize,above] {} (f1a);
\end{scope}
\end{tikzpicture}
\end{minipage}\hfill
 \begin{minipage}[c]{0.65\textwidth}
    \caption{
       The cylinder is given a cellular structure with two $0$-cells, indicated by bold dots. Three edges, indicated by solid lines, are mapped according to elements $\epsilon_{\partial f}, \epsilon'_{\partial f}, \xi_{b(f)} \in \mathcal E$. Two faces are given by bases of the cylinder and are mapped according to elements $\varphi_f, \varphi_f' \in \Phi$. The last face, based at $(1, b(f))$, forms the lateral surface. It is mapped according to the element $\psi_{\partial f}^{(\epsilon'')}$.
    } \label{fig:cylinderek}
  \end{minipage}
\end{figure}
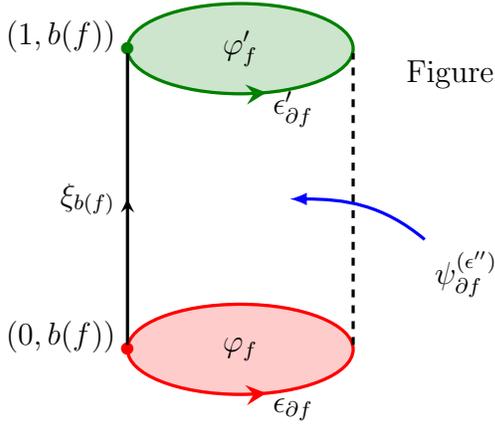

From now we focus on one extension and proceed to extending through $3$-cells. These are products of $I$ and faces of $X_2$. A calculation shows that for every face $f$ one encounters the problem of extending a map from the boundary of a~cylinder, illustrated on figure \ref{fig:cylinderek}, to its interior. This is possible if and only if the corresponding element  $\varphi_f'^{-1} \, \left( \psi_{\partial f}^{( \epsilon'')} \, (\xi_{b(f)} \rhd \varphi_f ) \right) \in \pi_2(B \mathbb G, \ast) = \ker(\Delta)$ is trivial. Here $\bm \epsilon''$ is given by $\epsilon''_e = \xi_{t(e)}\, \epsilon_e \, \xi_{s(e)}^{-1}$.

\begin{figure}[h!tb]
\begin{minipage}[c]{0.45\textwidth}
\centering
\begin{tikzcd}[column sep=0.9cm, row sep=0.9cm](\bm \epsilon,\bm \varphi)\arrow[rr]\arrow[rd,"\bm{\xi}"]& &(\bm{\epsilon'},\bm{\varphi'})\\
&(\bm{\epsilon''},\bm{\varphi''})\arrow[ur, "\bm{\psi}"]&
\end{tikzcd}
\end{minipage}\hfill
 \begin{minipage}[c]{0.55\textwidth}
    \caption{
       Field configuration $(\bm \epsilon', \bm \varphi')$ is obtained from $(\bm \epsilon, \bm \varphi)$ by a vertex transformation $\bm \xi$ followed by an edge transformation $\bm \psi$.
    } \label{fig:triangle_2}
  \end{minipage}
\end{figure}
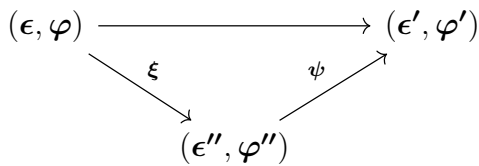

Summarizing, a homotopy between $h_{\bm \epsilon, \bm \varphi}$ and $h_{\bm \epsilon', \bm \varphi'}$ exists if and only if there exist collections $\bm \xi \in \mathcal E^{(0)}_X$ and $\bm \psi \in \Phi^{(1)}_X$ fitting in a diagram of the form presented on the figure \ref{fig:triangle_2}. In other words, configurations $(\bm \epsilon, \bm \varphi)$ and $(\bm \epsilon', \bm \varphi')$ have to be related by the action of vertex and edge transformations. 

We have completed the classification of homotopy classes of maps $X_2 \to B \mathbb G$. Now let us observe that restriction from $X$ to $X_2$ defines maps
\begin{equation*}
[X, B \mathbb G] \to [X_2, B \mathbb G], \qquad [(X, X_1 , X_0),(B \mathbb G, B \mathcal E, \ast)]\rightarrow [(X_2, X_1 , X_0),(B \mathbb G, B \mathcal E, \ast)].
\end{equation*}
We claim that they are injective. Indeed, suppose that $l,l' : X \to B \mathbb G$ are such that their restrictions to $X_2$ are homotopic. Every map $(\{ 0, 1 \} \times X) \cup (I \times X_2) \to B \mathbb G$ extends to $I \times X$, since $\pi_n(B \mathbb G; \ast)$ is trivial for $n \geq 3$. If the initial homotopy was a~homotopy of maps of triples, then so is the extension. This completes the proof of the claim.

Next we ask when does a map $h : X_2 \to B \mathbb G$ extend to $X$. Firstly, the answer depends only on the homotopy class of $h$, by the homotopy extension property of $(X, X_2)$. Thus we may assume that $h$ is a map of triples $(X_2, X_1, X_0) \to (B \mathbb G, B \mathcal E, \ast)$ and write $h=h_{\bm \epsilon, \bm \varphi}$. Secondly, if an extension to $X_3$ exists, then there exists also an extension to $X$, by triviality of higher homotopy groups of $B \mathbb G$. It remains to decide when it is possible to extend through $3$-cells. We consider a ball $q$. Its boundary is mapped to $B \mathbb G$ with homotopy class $\varphi_{\partial q} \in \pi_2(B \mathbb G ; \ast ) = \ker(\Delta)$. Thus an extension to whole $X_3$ exists if and only if $\bm \varphi$ is flat. Hence the proof that homotopy classes of maps $X \to B \mathbb G$ are in one-to-one correspondence with configurations with flat $\bm \varphi$ modulo vertex and edge transformations, as well as of the corresponding statement for the maps of triples, is completed.

\subsection{Postnikov class}
\label{appendix:postnikov}

In this appendix we consider the following question: for which $\bm{\epsilon}$ there exists a~flat $\bm{\varphi}$? First let us observe that the answer depends only on $\bm{\overline \epsilon}$ modulo gauge transformations since flatness of $\bm \varphi$ is invariant under vertex and edge transformations. Now let us choose one representative $\bm {\overline \epsilon}$, its lift to $\bm \epsilon$ and any $\bm \varphi$ satisfying the fake flatness condition. Next we define $\widehat \delta \bm \varphi : \pi_3(X_3, X_2 ; X_0) \to \ker(\Delta)$ by the formula $(\widehat \delta  \varphi)_{\omega} = \varphi_{\partial \omega}$. Let us observe that it has the following properties:
\begin{itemize}
    \item $\pi_1(X; X_0)$-equivariance, i.e.~$(\widehat \delta \varphi)_{\gamma \rhd \omega} = \overline{\epsilon}_{\gamma} \rhd (\widehat \delta \varphi)_{ \omega}$ for a path $\gamma$ starting at the base point of $\omega$. Thus $\widehat \delta \bm \varphi$ is a~twisted $3$-cochain, see appendix \ref{sec:twisted}.
    \item $\widehat \delta \bm \varphi$ is a~twisted cocycle. The proof of this is analogous to the proof of the fact that $\delta^2$ is trivial. Nevertheless, it is not necessarily true that $\widehat \delta \bm \varphi$ is in the image of $\delta$: $\bm \varphi$, being valued in the non-abelian group $\Phi$, is not a $2$-cochain.
    \item The cohomology class of $\widehat \delta \bm \varphi$ does not depend on the choice of a lift of $\bm{ \overline \epsilon}$ to $\bm \epsilon$ nor the choice of $\bm \varphi$. Indeed, edge transformations do not change $\widehat \delta \bm \varphi$ at all, while plaquette transformation $\bm \chi$ merely shifts it by $\delta \bm \chi$.
    \item $\widehat \delta \bm \varphi$ is trivial (i.e. $\varphi_{\partial \omega}=1$ for every $\omega$) if and only if $\bm \varphi$ is flat. Here we are using the fact that balls $q$ generate $\pi_3(X_3,X_2;x)$ as a $\pi_1(X_1;x)$-module for any $x \in X_0$.
\end{itemize}
It is clear from the above properties that the cohomology class $[\widehat \delta \bm \varphi] \in H^3(X, \ker(\Delta), \bm {\overline \epsilon})$ is trivial if and only if $\bm {\overline \epsilon}$ is such that there exists a~compatible configuration $(\bm \epsilon, \bm \varphi)$ with flat $\bm \varphi$.

\begin{figure}[h!tb]
\begin{minipage}[c]{0.5\textwidth}
\centering
\begin{tikzcd}[column sep=1.2cm, row sep=0.9cm](X_2,X_1,X_0)\arrow[r,"h_{\bm \epsilon, \bm \varphi}"]&  (B\mathbb G,B\mathcal E,\ast)\\
(Y_2,Y_2,Y_0)\arrow[u,"l"]\arrow[ur, "h_{l^\ast \bm\epsilon , l^\ast \bm \varphi}", swap]&
\end{tikzcd}
\end{minipage}\hfill
 \begin{minipage}[c]{0.5\textwidth}
    \caption{
       Pullback of field configurations may be defined in terms of the associated maps to the classifying space: $(l^{\ast} \bm \epsilon, l^{\ast} \bm \varphi)$ corresponds to the map $l \circ h_{\bm \epsilon, \bm \varphi}$.
    } \label{fig:trojkacik_P}
  \end{minipage}
\end{figure}


Cocyle $\widehat \delta \bm \varphi$ satisfies an important naturality property. Namely, let us consider a~map of triples $l : (Y_2,Y_1,Y_0) \to (X_2,X_1,X_0)$ for some CW-complex $Y$. Then it makes sense to pull back a field configuration $(\bm \epsilon, \bm \varphi)$ on $X$ to a configuration $l^{\ast} (\bm \epsilon, \bm \varphi) = (l^{\ast} \bm \epsilon, l^{\ast} \bm \varphi)$ on $Y$. One possible description of this pullback operation is through the diagram \ref{fig:trojkacik_P}. Equivalently, $(l^{\ast} \bm \epsilon, l^{\ast} \bm \varphi)$ is given by the composition
\begin{equation*}
\Pi_2(Y_2,Y_1;Y_0) \xrightarrow{l_{\ast}} \Pi_2(X_2,X_1;X_0) \xrightarrow{(\bm \epsilon, \bm \varphi)} \mathbb G.
\end{equation*}
Clearly we have $l^{\ast} [\widehat \delta \bm \varphi] = [ \widehat \delta l^{\ast} \bm \varphi ] \in H^3 (Y, \ker(\Delta), l^{\ast} \bm{\overline \epsilon})$. This innocuous-looking statement allows to relate $[\widehat \delta \bm \varphi]$ to a universal example.

\begin{figure}[h!tb]
\begin{minipage}[c]{0.4\textwidth}
\centering
\begin{tikzcd}[column sep=1.5cm, row sep=1.2cm]X_2\arrow[r,hook]\arrow[d,"h_{\bm \epsilon,\bm \varphi}",swap]&  X\arrow[d,"h_{\overline{\bm \epsilon}}"]\\
B\mathbb G\arrow[r,"\Upsilon"]& B\mathrm{coker}(\Delta)
\end{tikzcd}
\end{minipage}\hfill
 \begin{minipage}[c]{0.6\textwidth}
    \caption{ Given a map $h_{\bm \epsilon, \bm \varphi} : X_2 \to B \mathbb G$ we may compose it with $\Upsilon$ and then extend to a~map $ X \to B \coker(\Delta)$, uniquely up to a~homotopy. This extension corresponds to the gauge field $\bm {\overline \epsilon}$.
      }
 \label{fig:kwadracik_P}
  \end{minipage}
\end{figure}

The identity homomorphism $\pi_1(B \mathbb G, \ast) \to \coker(\Delta)$ determines, up to a~homotopy, a map of pairs $\Upsilon : (B \mathbb G, \ast) \to (B \coker(\Delta), \ast )$. Therefore for a~configuration $(\bm \epsilon, \bm \varphi)$ on $X$ (not necessarily with flat $\bm \varphi$) we have a commutative diagram of continuous maps presented on the figure \ref{fig:kwadracik_P}. Suppose that there existed a map $\Xi : B \coker(\Delta) \to B \mathbb G$ such that $\Upsilon \circ \Xi$ was homotopy equivalent to the self-identity map on $B \coker(\Delta)$. Then the map $h = \Xi \circ h_{\bm {\overline \epsilon}} : X \to B \mathbb G$ would be such that $\Upsilon \circ h$ is homotopic to $h_{\bm {\overline \epsilon}}$, yielding a conclusion that some configuration $(\bm \epsilon', \bm \varphi')$ with $\bm {\overline \epsilon}'=\bm {\overline \epsilon}$ and flat $\bm \varphi'$ exists. This is not always true, so the desired $ \Xi$ does not always exist. On the other hand one could attempt to construct it cell-by-cell. The obstruction to do this is a universal example for the cohomology classes $[\widehat \delta \bm \varphi]$, as we will now demonstrate.

Let $\bm {\overline \iota}$ be the tautological $\coker(\Delta)$-valued gauge field on $B \coker(\Delta)$. We may construct its lift to a $\mathbb G$-valued field configuration $(\bm \iota , \bm o )$ on $B \mathbb G$, which determines a~mapping $h_{\bm \iota , \bm o } : (B \coker(\Delta)_2, B \coker(\Delta)_1, \ast) \to (B \mathbb G, B \mathcal E, \ast)$. Thus we may form the cocycle $\widehat \delta \bm o$, which is a representative of the so-called Postnikov class
\begin{equation}
\beta = [\widehat \delta \bm o] \in H^3(B \coker(\Delta), \ker(\Delta), \overline \iota).
\end{equation}
We reiterate the fact that $\beta$ does not depend on the choice of $\bm \iota$ and $\bm o$, although the representative cocycle $\widehat \delta \bm o$ certainly does. The map $h_{\bm \iota, \bm o}$ induces the identity homomorphism between fundamental groups, and conversely, any map with this property is homotopic to one of the form $h_{\bm \iota, \bm o}$ for some $(\bm \iota, \bm o)$. Thus if $\beta$ is nontrivial, a right homotopy inverse $\Xi$ of $\Upsilon$ does not exist. Conversely, if $\beta$ is trivial, then some $h_{\bm \iota, \bm o}$ extends to the whole $B \coker(\Delta)$. Denoting the extension by $\Xi$, we have that $\Upsilon \circ \Xi$ induces the identity map on $\pi_1(B \coker(\Delta),\ast)$ and hence is homotopic to the identity map, by the classification of maps valued in classifying spaces of groups.

We claim that for any field configuration $(\bm \epsilon, \bm \varphi)$ on $X$ one has the relation
\begin{equation}
[ \widehat \delta \bm \varphi ] = h_{\bm{\overline \epsilon}}^{\ast}  \, \beta.
\end{equation}
Indeed, consider the field configuration $(\bm \epsilon', \bm \varphi') = ( h_{\bm {\overline \epsilon}}^{\ast} \, \bm \iota,h_{\bm {\overline \epsilon}}^{\ast} \, \bm o)$. Then $\bm{\overline \epsilon'} = \bm{\overline \epsilon}$, which implies that $\widehat \delta \bm \varphi'= h_{\bm {\overline \epsilon}}^{\ast} \, \widehat \delta \bm o$ and $\widehat \delta \bm \varphi$ are cohomologous. In particular, a configuration $(\bm \epsilon'', \bm \varphi'')$ with flat $\bm \varphi''$ and $\bm {\overline \epsilon''} = \bm{\overline \epsilon}$ exists if and only if the pullback $h_{\bm{\overline \epsilon}}^{\ast} \, \beta$ of the Postnikov class is trivial.

\subsection{Homomorphisms and weak equivalences}
\label{appendix:cswecm}

In this appendix we will assume that the $1$-skeleton of $B \mathbb G$ is contained in $B \mathcal E$. This is possible, because the inclusion of $B \mathcal E$ in $B \mathbb G$ induces an epimorphism of fundamental groups, see \cite[p.~219]{whitehead_book}. With this condition the identity map on $B \mathbb G$ may be regarded as a map of triples $(B \mathbb G, B \mathbb G_1, \ast) \to (B \mathbb G, B \mathcal E, \ast)$. Thus it determines a~$\mathbb G$-valued field configuration $(\bm \kappa, \bm \eta)$ on $B \mathbb G$, called the tautological configuration. This configuration has flat $\bm \eta$. The corresponding map takes $B \mathcal E$ to $B \mathcal E$, so $\bm \eta$ restricted to $B \mathcal E$ is trivial: $\eta_f=1$ for every face $f$ in $B \mathcal E$ (but not necessarily for faces in $B \mathbb G$).

Configuration $(\bm \kappa, \bm \eta)$ is universal: if $h_{\bm \epsilon, \bm \varphi}$ is a cellular map $(X_2,X_1,X_0) \to (B \mathbb G, B \mathcal E, \ast)$ corresponding to a~configuration $(\bm \epsilon, \bm \varphi)$, then $h_{\bm \epsilon, \bm \varphi}^{\ast} \, (\bm \kappa, \bm \eta) =(\bm \epsilon, \bm \varphi)$. This is because in this case the map $l$ on the figure \ref{fig:trojkacik_P} is simply the inclusion of $(B \mathbb G_2, \mathbb G_1, \ast)$ in $(B \mathbb G, B \mathcal E, \ast)$, so $h_{l^{\ast} \bm \epsilon, l^{\ast} \bm \varphi} = h_{\bm \epsilon, \bm \varphi}$.

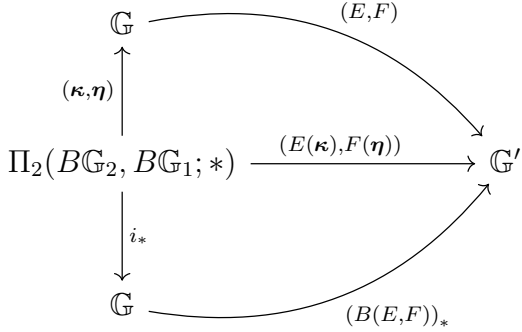
\begin{figure}[h!tb]
\begin{minipage}[c]{0.5\textwidth}
\centering
\begin{tikzcd}[column sep=1.5cm, row sep=1.2cm]
\mathbb G\arrow[drr,bend left, "{(E,F)}"] && \\
\Pi_2(B\mathbb G_2,B\mathbb G_1;\ast)\arrow[rr, "{\left(E(\bm \kappa), F(\bm \eta)\right)}",pos=0.4]\arrow[u,"{(\bm \kappa,\bm \eta)}"]\arrow[d,"i_\ast"]&& \mathbb G'\\
\mathbb G\arrow[urr,bend right, "{\left(B(E,F)\right)_\ast}",swap]&&
\end{tikzcd}
\end{minipage}\hfill
 \begin{minipage}[c]{0.5\textwidth}
    \caption{The upper triangle commutes by definition of the configuration $(E(\bm \kappa), F(\bm \eta))$, while the lower triangle commutes by construction of the map $B(E,F)$.}
 \label{fig:duzy_diagram}
  \end{minipage}
\end{figure}

Now let $\mathbb G'$ be another crossed module and let $(E,F) : \mathbb G \to \mathbb G'$ be a~homomorphism. Then $(E(\bm \kappa), F(\bm \eta))$ is a~$\mathbb G'$-valued configuration on $B \mathbb G$, so it determines a map $(B \mathbb G_2, B \mathbb G_1, \ast) \to (B \mathbb G', B \mathcal E', \ast)$, unique up to a homotopy of maps of triples. Since $\bm \eta$ was flat, so is $F(\bm \eta)$. Thus the corresponding map extends to whole $B \mathbb G$, uniquely up to a homotopy of maps of triples $(B \mathbb G, B \mathbb G_1, \ast) \to (B \mathbb G', B \mathcal E', \ast)$. We denote the extension by $B(E,F)$. Furthermore, $F(\bm \eta)$ is trivial on $B \mathbb G$, so (perhaps after a homotopy of maps of triples) $B(E,F)$ takes $B \mathcal E$ to $B \mathcal E'$. Then $B(E,F)$ induces a homomorphism
\begin{equation*}
(B(E,F))_{\ast} \, : \, \mathbb G = \Pi_2(B \mathbb G, B \mathcal E; \ast) \to \Pi_2(B \mathbb G', B \mathcal E' ; \ast) = \mathbb G'.
\end{equation*}
We claim that $(B(E,F))_{\ast} = (E,F)$. Indeed, let $i : (B \mathbb G_2 , B \mathbb G_1, \ast) \to (B \mathbb G, B \mathcal E, \ast)$ be the inclusion. Since $B(E,F)$ corresponds to the configuration $(E(\bm \kappa),F(\bm \eta))$, we have that the diagram of homomorphisms of crossed modules on figure \ref{fig:duzy_diagram} is commutative. On~the other hand we have $i_{\ast} = (\bm \kappa, \bm \eta)$, by construction of $(\bm \kappa, \bm \eta)$. Therefore
\begin{equation}
 (B(E,F))_{\ast} \circ i_{\ast} = (E,F) \circ i_{\ast}.
 \label{eq:BEFi}
\end{equation}

Maps $\pi_1(B \mathbb G_1 ; \ast) \to \pi_1(B \mathbb G ; \ast)$ and $\pi_2(B \mathbb G_2 , B \mathbb G_1 ; \ast) \to \pi_2(B \mathbb G, B \mathcal E; \ast)$ are epimorphisms, so \eqref{eq:BEFi} implies the validity of the claim. Indeed, surjectivity of the first homomorphism is clear. Secondly, we know that the inclusion $B \mathbb G_2 \to B \mathbb G$ induces an epimorphism of second homotopy groups and that the second homotopy groups of $B \mathbb G_1$ and $B \mathcal E$ are trivial. Hence by naturality of the long exact sequence of relative homotopy groups, the diagram on figure \ref{fig:duzy_diagram_2} is commutative with exact rows. The proof is completed by the four lemma.

\begin{figure}[h!tb]
\centering
\begin{tikzcd}[column sep=0.6cm, row sep=1.2cm]
0& \pi_1(B\mathbb G_2;\ast)\arrow[d,"\cong"]\arrow[l] & \pi_1(B\mathbb G_1;\ast)\arrow[d,twoheadrightarrow] \arrow[l]& \pi_2(B\mathbb G_2, B\mathbb G_1;\ast)\arrow[d]\arrow[l]&\pi_2(B\mathbb G_2;\ast)\arrow[d,twoheadrightarrow] \arrow[l]& 0\arrow[d,"\cong"]\arrow[l]\\
0& \pi_1(B\mathbb G;\ast)\arrow[l] & \pi_1(B\mathcal{E};\ast) \arrow[l]& \pi_2(B\mathbb G, B\mathcal{E};\ast)\arrow[l]&\pi_2(B\mathbb G;\ast)\arrow[l] & 0\arrow[l]
\end{tikzcd}
    \caption{The upper row and the lower row are pieces of long exact sequences of homotopy groups for pointed pairs $(B \mathbb G_2, B \mathbb G_1)$ and $(B \mathbb G, B \mathcal E)$, respectively. Downwards arrows are induced by the inclusion map.}
 \label{fig:duzy_diagram_2}
\end{figure}

We have proven that $(B(E,F))_{\ast}=(E,F)$, so in particular the maps of first and second homotopy groups induced by $B(E,F)$ are $\overline E$ and $\overline F$, respectively. Thus if $(E,F)$ is a weak isomorphism, then $B(E,F)$ is a homotopy equivalence, by Whitehead's theorem \cite[p.~346]{hatcher}. Thus it induces a bijection $[X, B \mathbb G] \cong [X , B \mathbb G']$ for every space $X$, so topological gauge theories based on $\mathbb G$ and $\mathbb G'$ are equivalent. More explicitly, this equivalence is given by mapping a $\mathbb G$-valued configuration $(\bm \epsilon, \bm \varphi)$ on $X$ to a $\mathbb G'$-valued configuration $(E(\bm \epsilon), F(\bm \varphi))$.

We remark that the above result implies that $B \mathbb G$ is determined uniquely up to a homotopy equivalence, a fact which we have never used. Indeed, if $\widetilde{B \mathbb G}$ is another construction of the classifying space of $\mathbb G$, the above construction gives a homotopy equivalence $B \mathbb G \to \widetilde{B \mathbb G}$ induced by the identity homomorphism $\mathbb G \to \mathbb G$.

\subsection{Construction of classifying spaces} \label{appendix:existence}

In this appendix we fix a crossed module $\mathbb G$ and construct a classifying space $B \mathbb G$ together with its subcomplex $B \mathcal E$ by gluing cells. In the process we will repeatedly use standard results \cite[p.~215]{whitehead_book} concerning the effect of attaching cells on homotopy groups, in particular the fact that the $n$-th homotopy group is not changed by attaching cells of dimension greater than $n+1$ (say, by the cellular approximation theorem). The latter is true also for relative homotopy groups.

Firstly, for the $0$-skeleton we take a single point $\ast$. To proceed further, we choose a presentation of $\mathcal E$, i.e.\ a set $\{ \epsilon_i \}_{i \in I}$ and relations $\{ \rho_j \}_{j \in J}$. For each $i \in I$ we attach to $\ast$ a single edge, so that
$B \mathbb G_1 = B \mathcal E_1$ is $\bigvee_{i \in I} S^1$, a bouquet of circles. Now the fundamental group of $B \mathcal E_1$ is free with generators indexed by the set $I$. We~denote the generator corresponding to $i \in I$ by $\epsilon_i$. Each relation $\rho_j$ is a word in the alphabet $\{ \epsilon_i \}_{i \in I}$, so it defines an element of the fundamental group of $B \mathcal E_1$. Space $B \mathcal E_2$ is formed by attaching to $B \mathcal E_1$ a $2$-cell for each $j \in J$, with an attaching map of homotopy class $\rho_j \in \pi_1(B \mathcal E_1 ; \ast)$. Then $\pi_1(B \mathcal E_2; \ast) = \mathcal E$.

Next we choose a set $\{ \varphi_k \}_{k \in K} \subseteq \Phi$ such that the elements $\epsilon \rhd \varphi_k$ with any $\epsilon \in \mathcal E$ generate the group $\Phi$. Space $B \mathbb G_2$ is formed from $B \mathcal E_2$ by attaching a $2$-cell for each $k \in K$, with attaching maps of homotopy classes $\Delta \varphi_k \in \mathcal E = \pi_1(B \mathcal E_2 ; \ast)$. Then the fundamental group of $B \mathbb G_2$ is $\coker(\Delta)$.

Space $B \mathcal E_3$ is formed by attaching $3$-cells to $B \mathcal E_2$ in such a way that $\pi_2(B \mathcal E_3; \ast)$ becomes trivial, e.g. one $3$-cell for each element of a set of generators of $\pi_2(B \mathcal E_3; \ast)$. Then an auxillary space $B \mathbb G_{2 \frac{1}{2}}$ is formed by attaching to $B \mathbb G_2$ the $3$-cells of $B \mathcal E$, or~equivalently \cite[p.~49]{whitehead_book} by gluing to $B \mathcal E_3$ those $2$-cells of $B \mathbb G$ which are not in~$B \mathcal E$.

\begin{figure}[h!tb]
\centering
\begin{tikzcd}[column sep=0.4cm, row sep=0.5cm]
0& \pi_1(B\mathbb G_{2\frac{1}{2}};\ast)\arrow[d,equal]\arrow[l] & \pi_1(B\mathcal{E}_3;\ast)\arrow[d,equal] \arrow[l]& \pi_2(B\mathbb G_{2\frac{1}{2}}, B\mathcal{E}_3;\ast)\arrow[d,twoheadrightarrow,"p",pos=0.3]\arrow[l,"\partial" above, pos=0.4]&\pi_2(B\mathbb G_{2\frac{1}{2}};\ast)\arrow[l]\arrow[d,twoheadrightarrow,"\overline{p}",pos=0.3]&\pi_2(B\mathcal{E}_3;\ast)\arrow[l]\arrow[d,equal]\\
& \mathrm{coker}(\Delta) & \mathcal{E} \arrow[l]& \Phi \arrow[l,"\Delta" above]& \ker(\Delta)\arrow[l]& 0\arrow[l]
\end{tikzcd}
    \caption{Commutative diagram whose upper row is a portion of the long exact sequence of homotopy groups of the pair $(B \mathbb G_{2 \frac{1}{2}}, B \mathcal E_3)$.}
 \label{fig:duzy_diagram_3}
\end{figure}


To proceed further, we need to understand the group $\widetilde \Phi := \pi_2(B \mathbb G_{2 \frac{1}{2}}, B \mathcal E_3 ; \ast)$. Since $B \mathbb G_{2 \frac{1}{2}}$ is obtained from $B \mathcal E_3$ by attaching faces, Whitehead's theorem applies and we have that $\widetilde \Phi$ is generated by elements $\epsilon \rhd \phi_k$ (with $\phi_k$ - the generator corresponding to the $k$-th face), subject only to relations following from Peiffer identities in the crossed module $\widetilde {\mathbb G} := \Pi_2(B \mathbb G_{2 \frac{1}{2}}, B \mathcal E_3; \ast)$. Furthermore, the boundary homomorphism $\pi_2(B \mathbb G_{2 \frac{1}{2}}, B \mathcal E_3 ; \ast) \to \pi_1(B \mathcal E_3 ; \ast) = \mathcal E$ is given by
\begin{equation}
\partial \left( \epsilon \rhd \phi_k \right) \mapsto \epsilon \,  \Delta \varphi_k \, \epsilon^{-1}.
\end{equation}
The assignment $p (\epsilon \rhd \phi_k) = \epsilon \rhd \varphi_k$ defines a group epimorphism $p : \widetilde \Phi \to \Phi$. Furthermore, $(\mathrm{id},p)$ is a homomorphism of crossed modules $\widetilde {\mathbb G} \to \mathbb G$. All that is summarized by the commutative diagram with exact rows presented on figure \ref{fig:duzy_diagram_3}. We let $\{ \lambda_l \}_{l \in L}$ be a set of generators of $\ker(p)$. Then for each $l$ we have that $\partial \lambda_l = \Delta(p(\lambda_l))$ is trivial. On the other hand the kernel of $\partial$ may be identified with $\pi_2(B \mathbb G_{2 \frac{1}{2}}; \ast)$, since $\pi_2(B \mathcal E_3 ; \ast)$ is trivial. Thus we may regard $\lambda_l$ as elements of $\pi_2 (B \mathbb G_{2 \frac{1}{2}};\ast)$. They generate the kernel of $\overline p$. The space $B \mathbb G_3$ is formed from $B \mathbb G_{2 \frac{1}{2}}$ by attaching $3$-cells with attaching maps of homotopy classes $\lambda_l$. Then we have $\Pi_2(B \mathbb G_3 , B \mathcal E_3; \ast) = \mathbb G$.

At this point all homotopy groups up to degree $2$ are as desired. The procedure may be continued inductively: for every $k \geq 4$ space $B \mathcal E_{k}$ is obtained from $B \mathcal E_{k-1}$ by attaching $k$-cells in such a way that $\pi_{k-1}(B \mathcal E_k; \ast)$ becomes trivial. Then $B \mathbb G_{k}$ is formed from $B \mathbb G_{k-1}$ by attaching all $k$-cells of $B \mathcal E$ and possible some additional cells needed to assure that $\pi_{k-1}(B \mathbb G_k; \ast)$ becomes trivial. Finally, we let $B \mathbb G$ (resp. $B \mathcal E$) be the union of all $B \mathbb G_k$ (resp. $B \mathcal E_k$), endowed with the weak topology.

\end{document}